%% file: main.tex
\documentclass[%
 twocolumn,
 showpacs,
 amsmath,amssymb,
 aps,
 pre,
]{revtex4-2}

\usepackage{graphicx}
\usepackage{dcolumn}
\usepackage{bm}
\usepackage{stmaryrd}
\usepackage{subfigure}

\graphicspath{{./figures/}}

\begin{document}


\title{A universal phase-field mixture representation of thermodynamics \\ and shock wave mechanics in
 porous soft biologic continua }

\author{J.D. Clayton}
 \email{john.d.clayton1.civ@army.mil}
\affiliation{Terminal Effects Division, DEVCOM ARL, Aberdeen Proving Ground, MD 21005-5066 USA}%

\date{\today}

\begin{abstract}
A continuum mixture theory is formulated for large deformations, thermal effects, phase interactions, and degradation of
soft biologic tissues.  Such tissues consist of one or more solid and fluid phases and can demonstrate
nonlinear anisotropic elastic, viscoelastic, thermoelastic, and poroelastic physics. Under
 extremely large or rapid deformations, for example impact or shock loading, 
 tissues may fracture, tear, or rupture. Mechanisms are encompassed in a universal, thermodynamically consistent formulation
that combines the continuum theory of mixtures with phase-field mechanics of fracture.
A metric tensor of generalized Finsler space supplies geometric insight on effects
rearrangements of microstructure, for example degrading collagen fibers.
Governing equations are derived, and energy potentials and kinetic laws  
posited, for generic soft porous tissues with solid and liquid or gas phases. 
Shock waves are modeled as singular surfaces; Hugoniot states and shock
decay are studied analytically. 
Suitability of the framework for representing blood, skeletal muscle, and liver is demonstrated.
Insight into physics presently unresolved by experiments is obtained.
\end{abstract}

\keywords{Suggested keywords}
                              
                              \pacs{87.10.Pq, 87.19.R-, 62.20.D-, 46.50.+a, 46.40.Cd}
\maketitle


\input{s1}

\input{s2}

\input{s3a}

\input{s3b}

\input{s4a}

\input{s4b}

\input{s4c}

\input{s4c2}

\input{s5}

\bibliography{refs}

\end{document}

%% file: s1.tex
\section{\label{sec1} Introduction}

Constitutive models describe the complex
physics of soft biological tissue---for example, skin, muscle, connective tissue, blood vessels, and the internal organs---when subject to mechanical and thermal stimuli.
These materials are often simultaneously
nonlinear anisotropic elastic, viscoelastic, thermoelastic, and poroelastic \cite{fung1990,fung1993}.
Large deformations manifesting nonlinear mechanical response
occur even under normal physiological activity. 
Medical events involving disease and surgery incur cutting or tearing (i.e., fracture) of the tissue.
For traumatic scenarios involving dynamic impact or blast, even more extreme deformations and rapid loading rates arise \cite{fung1990,cooper1997,saraf2007,wilgeroth2012aip,claytonAIP2020}.  

Microstructures are often hierarchical and highly complex \cite{fung1990,fung1993}. Materials consist of multiple solid- and fluid-like phases. The liver includes the spongy parenchyma, connective tissue, blood vessels and ducts, plus blood and bile \cite{ricken2010,zheng2021}. Skeletal and cardiac muscles contain fibers (i.e., cells), connective tissue, blood and interstitial fluid \cite{fung1993,holz2009}. Skin includes dermal tissue layers, cells, and interstitial fluid \cite{sachs2021}. Lung includes parenchyma (e.g., alveoli and ducts), air, and surfactant fluid, as well as stiffer elements of the bronchiole structure \cite{fung1990}. Membraneous tissues (e.g., in skin) contain ground substance, elastin, and collagen fibers \cite{yang2015}. Blood has cells immersed in the extracellular plasma \cite{fung1993,zub2013,varchanis2018}. Cells themselves include solid-like walls and internal fluids \cite{lied2010,waugh1979}.

Most constitutive treatments of mechanics of soft tissues focus on their nonlinear elastic response \cite{holz2000,holz2009,chagnon2015}.
Effects of fluids are included implicitly in the energy densities and parameters, perhaps augmented with viscoelasticity or other dissipation elements \cite{rubin2002, gultekin2016}.
Even in the non-dissipative case, sophisticated models are needed 
to account for anisotropy and nonlinearity, for example due to collagen fiber distributions \cite{gasser2006,planas2007,kalh2020}.
If tearing or degradation occur, continuum damage mechanics theories are fairly standard \cite{ito2010,li2016}, whereby a phenomenological kinetic equation is prescribed for an internal state variable measuring local loss of stress-bearing capacity.

In contrast to the former, the phase-field approach has been more recently advocated for
modeling tearing of soft biologic tissues \cite{raina2015,gultekin2019,chit2022}.
Diffuse-interface modeling has witnessed
use over a broad application space; models with
 elasticity include phase changes, twinning, and dislocations in crystals \cite{levitas2009,yang2010,claytonPHYSD2011,acharya2020}.
The method appeared for brittle fracture around 25 years ago \cite{karma2001,eastgate2002,henry2004,marconi2005,spatschek2007}, 
typically for small-strain, isotropic elasticity theory. 
Finite-strain approaches for fracture appeared thereafter \cite{claytonIJF2014,levitas2014}.
Fluid cavitation \cite{levitas2011} and fluid injection \cite{santillan2017} have been
modeled.

Continuum mixture theories and porous media theories emerged in the mid-20th century
\cite{biot1941,truesdell1960,bowen1976,bowen1982}.
Microstructure details
are smeared, but these models distinguish stresses and deformations of each
constituent and capture transfers of mass, momentum, and energy between
phases. Mixture- and porous-media theories introduce length and time scales
that are absent in single-phase viscoelasticity \cite{yang1991,ehlers2002}.
Although porous media models have been used for soft tissues 
\cite{yang1991,huyghe1992,ricken2010,regueiro2014,macminn2016,zheng2021}, they have not been combined with the phase-field fracture approach for the class of soft biologic materials. Finite strains \cite{wilson2016} and thermal exchange effects \cite{suh2021} have, however, been considered previously with phase-field fracture of brittle porous media.

The theory of porous media classically models a two-phase system of 
one solid and one fluid, though models with multiple fluids exist.  The fully dense
solid and fluid phases are usually treated as incompressible, but the mixture overall
is compressible as fluid is locally squeezed out \cite{yang1991,huyghe1992,ricken2010}.
Solid and fluid are usually, but not always \cite{zheng2021}, treated as inviscid individually,
but interactions of viscous origin between phases are captured
by permeability and dissipation from fluid transport (e.g., Darcy's law).
An effective stress principle \cite{terzaghi1943} decomposes total
stress into solid skeleton stress and pore pressure.
Constitutive equations are supplied for the solid skeleton (i.e., drained material) and
fully dense fluid rather than for partial stresses of mixture components.
This method has proven successful for soil and rock mechanics in which
the distinction between solid and fluid(s) is clear (e.g., water flowing through sand or porous rock), and it has been used elsewhere for porous tissues \cite{yang1991,huyghe1992,ehlers2019}.

The effective stress principle with solid skeleton concept is not
implemented here for several reasons.
First, tissue may consist of multiple solid and fluid phases. For
example, the liver has the parenchyma, blood vessels, and connective tissue (which might be tied or displace independently) and distinct fluids of blood and bile. Second, 
designation of a constituent as a ``solid'' or ``fluid'' may be ambiguous, for example, the extracellular matrix of tissue or blood demonstrating both viscous (fluid-like) and viscoelastic (solid-like) physics \cite{zub2013,varchanis2018,fielding2023}. With no single true solid or fluid phase, the soil mechanics analogy is tenuous. 
Cracking in ambiguously soft materials also shows richer physics than in brittle elastic solids due to stochastic bond opening and closure (i.e., healing) \cite{mulla2018}.

A mixture theory similar to that of Refs.~\cite{bowen1972,bowen1976,baer1986,mad2019} is advocated here instead, with constituent energies formulated on a per-unit-mass basis.  Complexity in thermodynamic derivations is reduced since partial stresses rather than effective
stress and pore pressure are used. However, potential difficulty can arise
in ascertaining properties of isolated phases as experiments might measure 
 responses of only some (e.g., fluid) phase(s) and the mixture as a whole. This issue is rectified by example later. The typical assumption of incompressibility of individual phases is abandoned to resolve bulk sound waves and longitudinal shocks. Incompressibility is also unrealistic if materials dilate or undergo cavitation (i.e., tensile damage).

A unified framework is formulated in Secs.~\ref{sec2} and \ref{sec3}, newly synthesizing 
continuum mixture theory, nonlinear anisotropic thermoelasticity, viscosity, viscoelasticity,
and phase-field fracture mechanics for biologic systems.
A universal, thermodynamically consistent theory containing all such features relevant to soft-tissue
mechanics for intense rapid loading does not seem to
 exist. Another unique aspect is generalized Finsler geometry 
describing the material manifold with evolving microstructure \cite{claytonSYMM2023}.

Traditional continuum models are couched in ambient Euclidean 3-space.
Residual stresses from growth and remodeling have been considered
using Riemannian geometry with a metric of non-vanishing curvature 
tensor \cite{taka1990,yavari2010}.
A recent approach \cite{claytonJGP2017,claytonZAMP2017} accounts for effects of microstructure using a
generalized Finsler metric \cite{bejancu1990} as opposed to a Riemannian metric.
In addition to residual stress and growth, the Finsler-geometric approach on the material manifold has been used to describe changes in local configurations of collagen fibers
as tissues degrade and tear  \cite{claytonSYMM2023}.
A Finsler metric was used elsewhere to quantify effects of fiber orientations
on mechanical responses of soft tissues in a discrete bond-based model \cite{takano2017} rather than a continuum physics approach as herein.
The Finsler-geometric theory has similarities to micropolar theory \cite{ikeda1973,*ikeda1981,eremeyev2020}, but the former is more general \cite{claytonJGP2017}. Much past work focused on hard crystalline solids \cite{claytonIJGMMP2018,claytonMMS2022}.

Shock waves in soft tissues are analyzed here with the mixture theory and constitutive frameworks.
Hugoniot solutions and evolution equations are derived for compressive shocks,
extending prior works \cite{coleman1966,chen1971,bowen1974,claytonIJES2022} to simultaneously account for
internal state variables (e.g., order parameters) and the Finsler metric that, in the
present application, is transformed to an osculating Riemannian metric \cite{rund1959,amari1962,claytonSYMM2023}.
Dissipation from viscoelasticity, momentum and energy transfer between phases,
and degradation due to shear-induced tearing all potentially 
affect shock amplitudes over time. Treatment of shocks as singular surfaces of velocity
differs from those of continuous 
 waves in nonlinear materials \cite{destrade2005}.
 Shock structures (e.g., shapes of continuous wave forms) have been
 analyzed elsewhere for nonlinear solids \cite{ortega2011}
 and fluids \cite{taniguchi2014,mad2019}.
  
Modeled in Sec.~\ref{sec4} are three 
soft-tissue systems: skeletal muscle with interstitial fluid, liver with blood, and lung with air. Solutions to physical problems involving tension,
compression, or shock-wave loading grant new understanding of the physics
demonstrated by these materials.  Concluding remarks give closure in Sec.~\ref{sec5}.

%% file: s2.tex
\section{\label{sec2} Governing Equations}

The present theory builds significantly on the finite-strain mixture theories
of Refs.~\cite{bowen1976,bowen1982,claytonIJES2022}.
Two enhancements are incorporated here.  First, time-dependent general internal
state vectors are introduced. Elements of these vectors are later
associated with history dependent mechanisms in the material microstructure,
namely viscoelasticity, active tension, and damage, all generally anisotropic.
A dependence of energy potentials on internal state and the gradient of
internal state is permitted, with terms
associated with internal state gradients properly incorporated in the
balance of energy and boundary conditions.
This enables a phase-field type representation 
when the variables are interpreted as order parameters
\cite{gurtin1996,levitas2009,claytonPHYSD2011}, suitable for modeling
regularized fracture \cite{claytonIJF2014,levitas2014,wilson2016,gultekin2019}.
Second, metric tensors on spatial and material manifolds are permitted
to depend on internal state and can be time-dependent.
Distances measured in the material can include remnant
strains from dissipative processes, or biologic growth and
remodeling if the latter physics are resolved by state variables.
If internal-state dependence of metrics is explicit and
distinct from coordinate dependence,  a generalized Finsler
representation emerges \cite{claytonJGP2017,claytonZAMP2017}.
If internal-state dependence is implicit and state vectors
are (time-dependent) functions of position \cite{rund1975}, then
an osculating Riemannian geometry \cite{amari1962,claytonSYMM2023}
is obtained. In the Riemannian setting,
similarities with Refs.~\cite{taka1990,yavari2010} are noted. 

A mixture of $N \geq 1$ constituents is considered, where at time $t$,
these occupy a shared infinitesimal control volume $\rm{d} \Omega$
centered at spatial position $\bf{x}$. Greek superscripts $\alpha = 1, \ldots, N$
denote constituents having reference coordinates $\bf{X}^\alpha$; constituents
coincident at $\bf{x}$ may occupy different $\bf{X}^\alpha$ due to diffusion processes.
Motions are
\begin{equation}
\label{eq:motions}
{\bf x} = { \bm \chi}^\alpha ( {\bf X}^\alpha, t).
\end{equation}
A spatial manifold comprising the material body, parameterized by coordinate chart(s) $\{ x^k \}$,
is $\mathfrak{m}$. Referential manifold(s) $\mathfrak{M}^\alpha$ are parameterized by
coordinate chart(s) $\{(X^\alpha)^K \}$ corresponding to reference positions of constituent $\alpha$.
Denote by $ \{ {\bm \xi}^\alpha ( {\bf x},t) \}$ and
$\{ \bm \Xi^\alpha ( {\bf X}^\alpha, t) \}$ 
sets of internal state variables viewed as auxiliary coordinates over
$\mathfrak{m}$ and $\mathfrak{M}^\alpha$, respectively.
Metric tensors $\bf {g}$ and $\bf G^\alpha $ with components  $g_{ij}$ and $G^\alpha_{IJ}$ are permitted to be coordinate- and time-dependent, of the forms
\begin{align}
\label{eq:metrics1}
& {\bf g} = {\bf g}( {\bf x}, t) = \tilde{ \bf g} ( {\bf x}, \{ {\bm \xi}^\alpha( {\bf x},t) \} ), \\
& {\bf G}^\alpha   = {\bf G}^\alpha ( {\bf X}^\alpha, t) = \tilde{ \bf G}^\alpha ( {\bf X}^\alpha, \{ {\bm \Xi}^\alpha( {\bf X}^\alpha,t) \} ). \label{eq:metrics2}
\end{align}
The $ \sim $ notation denotes the generalized Finsler description \cite{bejancu1990,claytonJGP2017,claytonZAMP2017} of
metrics on $\mathfrak{m}$ and $\mathfrak{M}^\alpha$, whereas unadorned versions
are interpreted as osculating Riemannian metric tensors \cite{rund1959,amari1962,claytonSYMM2023} at any fixed time $t$.
Determinants are written $g = \det {\bf g}$ and $G^\alpha = \det {\bf G}^\alpha$.
The partial time derivative at fixed $\bf x$ is $\partial_t (\cdot)$; the material time derivative at fixed ${\bf X}^\alpha$ is $D^\alpha_t (\cdot)$, related by
\begin{equation}
\label{eq:MTD}
D^\alpha_t (\cdot) = \partial_t (\cdot) + \nabla (\cdot)  \cdot {\bm \upsilon}^\alpha,
\quad
 ({\upsilon}^\alpha)^k = D^\alpha_t ({ \chi}^\alpha)^k.
\end{equation}
Particle velocity is ${\bm \upsilon}^\alpha$, and $\nabla(\cdot)$ is the covariant derivative
with respect to $\bf x$ where Christoffel symbols are those of the Levi-Civita connection derived from
$g_{ij}$. The covariant derivative with respect to $\bf X^\alpha$ on $\mathfrak{M}^\alpha$ is
$\nabla_0^\alpha (\cdot)$. Spatial and referential gradient operators obey
\begin{equation}
\label{eq:grads}
\nabla (\cdot) = \partial (\cdot) / \partial x^k \otimes {\bf g}^k, \quad
\nabla_0^\alpha (\cdot) = \partial (\cdot) / \partial (X^\alpha)^K \otimes {\bf G}^K.
\end{equation}
When used on typical scalars, vectors, and tensors, 
$\partial_t(\cdot)$ and $D^\alpha_t(\cdot)$ are performed with
 natural basis vectors
${\bf g}_k = \partial {\bf x} / \partial x^k$
and ${\bf G}^\alpha_K = \partial {\bf X} / \partial (X^\alpha)^K$
held fixed with respect to $t$ at ${\bf x}$ and $\bf X^\alpha$, respectively,
so $\partial_t {\bf g}_k$ and $D_t^\alpha {\bf G}^\alpha_K$
vanish in such circumstances. This produces commutation rules:
\begin{equation}
\label{eq:commute}
\nabla [\partial_t (\cdot) ] = \partial_t [ \nabla (\cdot) ],
\quad
\nabla_0^\alpha [ D^\alpha_t (\cdot) ] =D_t^\alpha [\nabla_0^\alpha (\cdot)].
\end{equation}
The deformation gradient ${\bf F}^\alpha$ and Jacobian determinant $J^\alpha$ are defined as follows,
giving a relation between reference and spatial gradient operators:
\begin{align}
\label{eq:defgrad}
& (F^\alpha)^i_J = \frac{ \partial (\chi^\alpha)^i }{ \partial (X^\alpha)^J}, \quad
J^\alpha = \det [(F^\alpha)^i_J] \sqrt{g/G^\alpha}, \\
\label{eq:nabla0}
& {\bf F}^\alpha = (F^\alpha)^i_J {\bf g}_i \otimes {\bf G}^J, \quad \nabla^\alpha_0 (\cdot) = \nabla(\cdot) {\bf F}^\alpha.
\end{align}
The velocity gradient ${\bf l}^\alpha$ and its trace are
\begin{equation}
\label{eq:velgrad}
{\bf l}^\alpha = \nabla {\bm \upsilon}^\alpha =  D^\alpha_t {\bf F}^\alpha ({\bf F^\alpha})^{-1}, \quad
 \nabla \cdot {\bm \upsilon}^\alpha =  {\rm tr} {\bf l}^\alpha.
\end{equation}
Spatial and material volume elements, $ {\rm d} \Omega$ and ${\rm d} \Omega^\alpha_0$, obey 
\begin{equation}
\label{eq:vols}
{\rm d} \Omega ( {\bf x}( {\bf X}^\alpha, t),t ) = J^\alpha( {\bf X}^\alpha, t) {\rm d} \Omega_0^\alpha ( {\bf X}^\alpha, t).
\end{equation}
Time derivatives of local volume elements allow for time dependence of metric tensors, extending Ref.~\cite{yavari2010} as 
\begin{align}
\label{eq:dotOmega}
& \partial_t ( {\rm d} \Omega ) = {\textstyle {\frac{1}{2}}} { \rm tr} (\partial_t {\bf g})  {\rm d} \Omega  = 
\partial_t (\ln \sqrt{g})  {\rm d} \Omega , \\
& D^\alpha_t ( {\rm d} \Omega_0^\alpha ) = {\textstyle {\frac{1}{2}}} { \rm tr} (D^\alpha_t {\bf G}^\alpha) {\rm d} \Omega_0^\alpha  = 
D^\alpha_t (\ln \sqrt{G ^\alpha})  {\rm d} \Omega_0^\alpha, \\
& D^\alpha_t ( {\rm d} \Omega ) = (D^\alpha_t J^\alpha + J^\alpha D^\alpha_t \ln \sqrt{G^\alpha} ){\rm d} \Omega_0^\alpha \nonumber \\ & \qquad \quad \, \, \, =
( \nabla \cdot {\bm \upsilon}^\alpha + \partial_t \ln \sqrt{g}) {\rm d} \Omega . \label{eq:Jdot}
\end{align}
In \eqref{eq:Jdot}, $D^\alpha_t \sqrt{g/G^\alpha}$
is included in $D_t^\alpha J^\alpha$,
and $\partial_t g = D^\alpha_t g$ since $\nabla g$ vanishes
for a Levi-Civita connection \cite{claytonDGKC2014}. 

Define the following, all generally fields of $( {\bf x},t)$:
partial Cauchy stress tensor $ {\bm \sigma}^\alpha$,
traction vector ${ \bf t}^\alpha = {\bm \sigma}^\alpha \cdot {\bf n}$,
body force per unit mass $ {\bf b}^\alpha$,
partial internal energy per unit mass $u^\alpha$,
heat source per unit mass $r^\alpha$,
heat flux vector ${\bf q}^\alpha$,
mass supply rate $c^\alpha$,
linear momentum exchange $ {\bf h}^\alpha$,
and energy exchange rate $\epsilon^\alpha$.
The spatial mass density of constituent $\alpha$ is $\rho^\alpha({\bf x},t)$.
Referential mass density at a fixed time $t = t_0$
is $\rho_0^\alpha({\bf X^\alpha})$ with metric ${\bf G}_0^\alpha({\bf X^\alpha})$
and mass element ${\rm d} m_0^\alpha ({\bf X}^\alpha) = \rho_0^\alpha \sqrt{G_0^\alpha / G^\alpha} {\rm d} \Omega^\alpha_0$. 
At arbitrary $t$, the mass element is ${\rm d} m^\alpha ({\bf X}^\alpha,t) = \rho^\alpha {\rm d} \Omega =  \rho^\alpha J^\alpha {\rm d} \Omega_0^\alpha$.

\subsection{\label{sec2a}Continuous processes}
A global energy balance for each constituent is posited on a region of $\mathfrak{m}$ occupying
control volume $\Omega$,
enclosed by oriented boundary $\partial \Omega $ with unit outward normal ${\bf n}$.
Velocity and heat flux normal to $\partial \Omega$ are $\upsilon^\alpha_n = {\bm \upsilon^\alpha} \cdot {\bf n}$ and $q^\alpha_n = {\bf q^\alpha} \cdot {\bf n}$.
The global balance per constituent extends that of Ref.~\cite{bowen1976} to account for working of generalized tractions $\{ \bf z^\alpha \} = \{ {\bm \zeta^\alpha} \cdot {\bf n} \} $ work-conjugate to rates of auxiliary variables $ \{ D_t^\alpha {\bm \xi}^\alpha \}$ on $ \partial \Omega$,
similarly to Ref.~\cite{gurtin1996}:
\begin{align}
\nonumber
& \partial_t \int_\Omega
\rho^\alpha ( u^\alpha +  {\textstyle {\frac{1}{2}}}|{\bm \upsilon}^\alpha|^2)
{\rm{d} \Omega} \nonumber \\
& \qquad \qquad + \oint_{ \partial \Omega} \rho^\alpha ( u^\alpha +  {\textstyle {\frac{1}{2}}}|{\bm \upsilon}^\alpha|^2) \upsilon^\alpha_n {\rm{d} \partial \Omega} \nonumber \\ & =
\oint_{ \partial \Omega}
({\bf t}^\alpha \cdot {\bm \upsilon}^\alpha
+ \{ {\bf z}^\alpha \} \cdot \{ D^\alpha_t {\bm \xi}^\alpha \}
 - q_n^\alpha)
 {\rm{d} \partial \Omega} \nonumber
\\ & \qquad \qquad +
 \int_\Omega [ \rho^\alpha( {\bf b}^\alpha \cdot {\bm \upsilon}^\alpha + r^\alpha)  
 + {\bf h}^\alpha \cdot {\bm \upsilon}^\alpha + \epsilon^\alpha \nonumber
\\ & \qquad \qquad \qquad \, \,
+ c^\alpha(u^\alpha +  {\textstyle {\frac{1}{2}}}|{\bm \upsilon}^\alpha|^2) ]
  {\rm{d} \Omega}.
 \label{eq:ebalglob}
\end{align}
Angular momentum exchange between constituents \cite{bowen1976} is omitted herein.
Field variables in \eqref{eq:ebalglob} are assumed sufficiently smooth on $\Omega$ such that the  divergence theorem applies on ${\mathfrak m}$. Energy conservation under rigid motions (i.e., various time-dependent translations and rigid-body rotations \cite{green1964,marsden1983}) of the entire mixture then furnishes local conservation laws for mass, momenta, and energy: 
\begin{align}
\label{eq:massbal}
& \partial_t \rho^\alpha + \nabla \cdot (\rho^\alpha { \bm \upsilon}^\alpha) + \rho^\alpha \partial_t \ln \sqrt{g}= c^\alpha, \\
\label{eq:mombal}
& \nabla \cdot {\bm \sigma}^\alpha + \rho^\alpha {\bf b}^\alpha + {\bf h}^\alpha= \rho^\alpha D^\alpha_t {\bm \upsilon}^\alpha , \quad {\bm \sigma}^\alpha = ({\bm \sigma}^\alpha)^{\mathsf T}, 
\end{align}
\begin{align}
\label{eq:ebal}
& \rho^\alpha D_t^\alpha u^\alpha = {\bm \sigma}^\alpha: \nabla {\bm \upsilon}^\alpha
+ \nabla \cdot ( \{ { \bm \zeta}^\alpha \} \cdot  \{ D_t^\alpha {\bm \xi}^\alpha \} ) \nonumber \\
& \qquad  \qquad \qquad \qquad \quad \, - \nabla \cdot {\bf q}^\alpha + \rho^\alpha r^\alpha + \epsilon^\alpha. 
 \end{align}
 
 Denote by $\eta^\alpha$ the local entropy per unit mass and $\theta^\alpha > 0$ the absolute temperature of constituent $\alpha$. As in Refs.~\cite{bowen1976,claytonIJES2022}, an entropy inequality for each constituent is avoided in lieu of an inequality for the whole mixture:
 \begin{align}
 \label{eq:entglob}
 &\partial_t  \int_\Omega   \sum_\alpha\rho^\alpha \eta^\alpha {\rm{d} \Omega}
 +  \oint_{ \partial \Omega} \sum_\alpha\rho^\alpha \eta^\alpha  \upsilon_n^\alpha {\rm{d} \partial \Omega } \nonumber \\
  & \qquad \qquad \geq \int_\Omega \sum_\alpha \frac{\rho^\alpha r^\alpha}{\theta^\alpha} {\rm{d} \Omega}
 -  \oint_{ \partial \Omega} \sum_\alpha \frac{q_n^\alpha}{\theta^\alpha} {\rm{d} \partial \Omega }.
 \end{align}
 From the divergence theorem, \eqref{eq:massbal}, and localization
  \footnote{Misprints in eqs.~(2.15), (4.1), (4.19)$_1$, (4.30)$_3$, (4.45), (4.63), (5.2)$_2$, (5.13)$_2$ of Ref.~\cite{claytonIJES2022} are corrected herein.},
 \begin{equation}
 \label{eq:entbal}
 \sum_\alpha [\rho^\alpha D^\alpha_t \eta^\alpha +  \frac{ \nabla \cdot {\bf q}^\alpha} { \theta^\alpha} -
 \frac{ {\bf q^\alpha } \cdot \nabla \theta^\alpha}{(\theta^\alpha)^2}
  - \frac{ \rho^\alpha r^\alpha }{ \theta^\alpha} + c^\alpha \eta^\alpha] \geq 0. 
 \end{equation}
 Let $\psi^\alpha$ be Helmholtz free energy per unit mass, whereby substitution of \eqref{eq:ebal}
 into \eqref{eq:entbal} yields
 \begin{align}
 \label{eq:psi}
 & \psi^\alpha = u^\alpha - \theta^\alpha \eta^\alpha, \\
 \label{eq:entbal2}
&  \sum_\alpha ({1}/{\theta^\alpha})[ {\bm \sigma}^\alpha: \nabla {\bm \upsilon}^\alpha
+ \nabla \cdot ( \{ { \bm \zeta}^\alpha \} \cdot  \{ D_t^\alpha {\bm \xi}^\alpha \} ) \nonumber
\\ & \qquad \qquad - ( { {\bf q^\alpha } \cdot \nabla \theta^\alpha}) /{\theta^\alpha}
+ \epsilon^\alpha + c^\alpha \theta^\alpha \eta^\alpha \nonumber
\\ & \qquad \qquad \qquad - \rho^\alpha ( D^\alpha_t \psi^\alpha + \eta^\alpha D^\alpha_t \theta^\alpha) ] \geq 0.
  \end{align}
  
  \subsection{\label{sec2b}Singular surfaces}
  Now consider a propagating singular surface $\Sigma(t)$ in $\mathfrak{m}$, with image $\Sigma^\alpha(t)$ in 
  $\mathfrak{M}^\alpha$. The Eulerian function $\phi$ defines this surface, from which the unit normal $\bf n$ and 
  Eulerian speed $\mathcal{U} > 0 $ in the direction of $\bf n$ follow:
  \begin{equation}
  \label{eq:shockvel}
  \phi( {\bf x}, t) = 0, \quad {\bf n} = \nabla \phi / | \nabla \phi |, \quad \mathcal{U} = - \partial_t \phi / | \nabla \phi |.
  \end{equation}
  Let $(\cdot)^+$ and $(\cdot)^-$ label limiting values of $(\cdot)$ as $\Sigma$ is approached
  from either side;  $\bf n$ is directed from the $(\cdot)^-$ side (behind $\Sigma$) to the $(\cdot)^+$ side.  The jump across $\Sigma$ is
  \begin{equation}
  \label{eq:jump}
  \llbracket (\cdot) \rrbracket = (\cdot)^- - (\cdot)^+.
  \end{equation}
   
    Across a shock front $\Sigma$, each ${\bm \chi}^\alpha$ is continuous, but space-time
  derivatives of ${\bm \chi}^\alpha$ are not necessarily so. Nor always are other field
  variables such as $\rho^\alpha$, ${\bm \sigma}^\alpha$, $\theta^\alpha$, etc.
  With $\bf n$ defined per \eqref{eq:shockvel}, the normal velocity, 
  heat flux, and tractions are $\upsilon_n^\alpha = {\bm \upsilon}^\alpha \cdot {\bf n}$,
  $q_n^\alpha = {\bf q}^\alpha \cdot {\bf n}$, ${ \bf t}^\alpha = {\bm \sigma}^\alpha \cdot {\bf n}$,
   and $\{ \bf z^\alpha \} = \{ {\bm \zeta^\alpha} \cdot {\bf n} \} $. Conservation laws for mass, linear momentum, and energy, and the entropy imbalance,
  across $\Sigma$ are derived using principles set forth in Refs.~\cite{truesdell1960,casey2011}.
  Specifically, a closed region $\Omega$ of $\mathfrak{m}$ is
  partitioned by $\Sigma$ at an instant in time into two sub-bodies, within which
  the local continuum balance laws \eqref{eq:massbal}--\eqref{eq:ebal} and \eqref{eq:entbal} hold.
  These continuum laws are integrated over $\Omega$ and then over each sub-body, the latter accounting for possible boundary contributions on $\Sigma$ from tractions and heat flux.
  Integrals are assumed to be physically meaningful even if $\mathfrak{m}$ is non-Euclidean.
  The total mass rate, linear momentum rate, energy rate, and entropy production rate
  are required to be equal for $\Omega$ and the summed contributions of each sub-body with boundary $\Sigma$.
  Application of a form of Reynolds' transport theorem  \cite{casey2011,claytonIJF2017} derived from the divergence theorem and \eqref{eq:Jdot} then gives analogs of \eqref{eq:massbal},
 \eqref{eq:mombal}, and \eqref{eq:ebal} across $\Sigma$ for each $\alpha$:
 \begin{align}
 \label{eq:massjump}
 & \llbracket \rho^\alpha (\upsilon_n^\alpha - \mathcal{U}) \rrbracket = 0, \\
 \label{eq:momjump}
 & \llbracket \rho^\alpha {\bm \upsilon^\alpha} (\upsilon_n^\alpha - \mathcal{U}) \rrbracket = \llbracket { \bf t}^\alpha \rrbracket , \\
 \label{eq:ejump} 
 & \llbracket \rho^\alpha ( u^\alpha +  {\textstyle {\frac{1}{2}}}|{\bm \upsilon}^\alpha|^2)
 (\upsilon_n^\alpha - \mathcal{U}) \rrbracket  \nonumber
 \\ & \qquad \quad = \llbracket {\bf t}^\alpha \cdot {\bm \upsilon}^\alpha
 + \{ {\bf z}^\alpha \} \cdot \{ D^\alpha_t {\bm \xi}^\alpha \}
 - q_n^\alpha \rrbracket.
 \end{align}
 A jump equation for angular momentum can be derived, but it encompasses nothing beyond \eqref{eq:momjump}. Similarly, arguments in Ref.~\cite{casey2011} applied to \eqref{eq:entbal}
 furnish a local entropy inequality across $\Sigma$ for the mixture:
 \begin{equation}
 \label{eq:entjump}
 \sum_\alpha \llbracket \rho^\alpha \eta^\alpha (\mathcal{U}- \upsilon_n^\alpha) - q_n^\alpha / \theta^\alpha \rrbracket \geq 0.
 \end{equation}
 
 Now consider one-dimensional (1-D) loading conditions:
 $n_k \rightarrow n_1 = 1$,
 $x^k \rightarrow x^1 = x$,
  $(\chi^\alpha)^k \rightarrow (\chi^\alpha)^1 = \chi^\alpha$,
  $(F^\alpha)^i_J \rightarrow (F^\alpha)^1_1 = \partial \chi^\alpha / \partial X^\alpha = F^\alpha$,
  $(\upsilon^\alpha)^k \rightarrow (\upsilon^\alpha)^1 = \upsilon^\alpha_n
=  \upsilon^\alpha $,
 $(t^\alpha)_k \rightarrow (t^\alpha)_1 = (\sigma^\alpha)^1_1
 = t^\alpha$,
 $\{ (z^\alpha)_k \} \rightarrow \{ (z^\alpha)_1 \} = \{ (\zeta^\alpha)^1_1 \} 
 =\{ z^\alpha \}$,
  $\{ (\xi^\alpha)^k \} \rightarrow \{ (\xi^\alpha)^1 \} = \{ \xi^\alpha \} $, and
   $(q^\alpha)^k \rightarrow (q^\alpha)^1 =q^\alpha_n
=  q^\alpha $.
The shock is planar, with Eulerian speed $\mathcal U$.
Eulerian forms of Rankine-Hugoniot equations 
 \eqref{eq:massjump}--\eqref{eq:entjump} reduce to
\begin{align}
 \label{eq:massjump1}
 & \llbracket \rho^\alpha (\upsilon^\alpha - \mathcal{U}) \rrbracket = 0, \\
 \label{eq:momjump1}
 & \llbracket \rho^\alpha { \upsilon^\alpha} (\upsilon^\alpha - \mathcal{U}) \rrbracket = \llbracket {  t}^\alpha \rrbracket , 
\\
 \label{eq:ejump1} 
 & \llbracket \rho^\alpha ( u^\alpha +  {\textstyle {\frac{1}{2}}}|{ \upsilon}^\alpha|^2)
 (\upsilon^\alpha - \mathcal{U}) \rrbracket  \nonumber
 \\ & \qquad \quad = \llbracket { t}^\alpha  { \upsilon}^\alpha
 + \{ { z}^\alpha \}  \{ D^\alpha_t { \xi}^\alpha \}
 - q^\alpha \rrbracket , 
  \end{align}
 \begin{align}
 \label{eq:entjump1}
& \sum_\alpha \llbracket \rho^\alpha \eta^\alpha (\mathcal{U}- \upsilon^\alpha) - q^\alpha / \theta^\alpha \rrbracket \geq 0.
 \end{align}
   
A Lagrangian description of a singular surface, denoted $\Sigma^\alpha(t)$ for
each constituent $\alpha$, is, analogously to
\eqref{eq:shockvel},
  \begin{align}
  \label{eq:shockvel0}
 &  \Phi^\alpha ( {\bf X^\alpha}, t) = 0, 
  \qquad {\bf N}^\alpha = \nabla^\alpha_0 \Phi^\alpha / | \nabla^\alpha_0 \Phi^\alpha |, \nonumber
  \\ 
  & \mathcal{U}^\alpha = - D^\alpha_t \Phi^\alpha / | \nabla_0^\alpha \Phi^\alpha |.
  \end{align}
In 1-D, Eulerian and Lagrangian shock speeds are \cite{bowen1974}
\begin{equation}
\label{eq:shockspd1D}
{\mathcal U}(t) = {\rm d} \Sigma (t) / {\rm d} t, \quad
{\mathcal U}^\alpha(t) = {\rm d} \Sigma^\alpha (t) / {\rm d} t.
\end{equation}
From 1-D inverse motion $X^\alpha (\chi^\alpha,t) = (\chi^\alpha)^{-1}(x,t)$ \cite{bowen1974},
\begin{equation}
\label{eq:velrelate}
\Sigma^\alpha (t) = (\chi^\alpha)^{-1}( \Sigma(t),t)
\Rightarrow
{\mathcal U}^\alpha = (F^\alpha)^{-1} ( {\mathcal U} - \upsilon^\alpha),
\end{equation}
with $ (F^\alpha)^{-1} = \partial (\chi^\alpha)^{-1} / \partial x$.
Now assume a continuous referential density field
$\rho_0^\alpha (X^\alpha)$ exists and can be related in 1-D 
to any other spatial density field $\rho^\alpha(x,t)$ via
$\rho_0^\alpha = F^\alpha \rho^\alpha$.
Sufficient conditions for the latter consistent
with mass conservation are $c^\alpha = \rho^\alpha \partial_t \ln \sqrt{g}$
with $g( \chi^\alpha(X^\alpha,t),t) = G^\alpha (X^\alpha,t)$.
In this case, \eqref{eq:massjump1} with \eqref{eq:velrelate}
and $\llbracket \rho^\alpha_0 \rrbracket = 0$ produces
$\llbracket \mathcal U^\alpha \rrbracket = 0$.
Thus, noting that $\mathcal U^\alpha$ is intrinsic
and \eqref{eq:velrelate} should hold at all $(\cdot)^\pm$
states, the following Lagrangian forms of the 1-D
Rankine-Hugoniot equations are derived:
\begin{align}
 \label{eq:massjump0}
 & \rho^\alpha_0 {\mathcal U}^\alpha \llbracket 1 / \rho^\alpha \rrbracket = - \llbracket \upsilon^\alpha \rrbracket , \\
 \label{eq:momjump0}
 &  \rho^\alpha_0  {\mathcal U}^\alpha \llbracket \upsilon^\alpha \rrbracket = -  \llbracket {  t}^\alpha \rrbracket , \\
 \label{eq:ejump0} 
 &   \rho^\alpha_0  {\mathcal U}^\alpha \llbracket  u^\alpha +  {\textstyle {\frac{1}{2}}}|{ \upsilon}^\alpha|^2
 \rrbracket  \nonumber
 \\ & \qquad \quad = - \llbracket { t}^\alpha  { \upsilon}^\alpha
 + \{ { z}^\alpha \}  \{ D^\alpha_t { \xi}^\alpha \}
 - q^\alpha \rrbracket , \\
 \label{eq:entjump0}
& \sum_\alpha (\rho^\alpha_0  {\mathcal U}^\alpha \llbracket  \eta^\alpha \rrbracket - \llbracket q^\alpha / \theta^\alpha \rrbracket ) \geq 0.
 \end{align}
Routine algebra produces velocity jumps and Lagrangian shock speeds
in terms of jumps in stress and mass density:
\begin{equation}
\label{eq:veljumps}
\llbracket \upsilon^\alpha \rrbracket = ( \llbracket t^\alpha \rrbracket \llbracket 1/ \rho^\alpha \rrbracket)^{1/2}, \quad
\rho^\alpha_0 \mathcal{U}^\alpha =  ( \llbracket t^\alpha \rrbracket / \llbracket 1/ \rho^\alpha \rrbracket)^{1/2}.
\end{equation}
Also, since $\rho_0^\alpha / \rho^\alpha = F^\alpha$ by construction,
\begin{equation}
\label{eq:jumps3}
\llbracket \upsilon^\alpha \rrbracket = - {\mathcal U}^\alpha \llbracket F^\alpha \rrbracket, \qquad
\llbracket t^\alpha \rrbracket = \rho_0^\alpha ({\mathcal U}^\alpha )^2 \llbracket F^\alpha \rrbracket.
\end{equation}
The Rankine-Hugoniot energy balance follows by eliminating
particle and shock velocities from \eqref{eq:ejump0}:
\begin{equation}
\label{eq:RHe0}
\llbracket u^\alpha \rrbracket = \langle t^\alpha \rangle \llbracket 1/ \rho^\alpha \rrbracket
- \frac{ \llbracket \{ { z}^\alpha \}  \{ D^\alpha_t { \xi}^\alpha \}
 - q^\alpha \rrbracket } {( \llbracket t^\alpha \rrbracket / \llbracket 1/ \rho^\alpha \rrbracket)^{1/2}}.
\end{equation}
where $\langle (\cdot) \rangle = \frac{1}{2} [ (\cdot)^+ + (\cdot)^-]$ is the average
across $\Sigma^\alpha$.

The displacement derivative $\delta_t (\cdot)$ is defined as follows in 1-D \cite{truesdell1960}, consistently with \eqref{eq:MTD} and \eqref{eq:shockspd1D}:
\begin{equation}
\label{eq:deltaderiv}
\delta_t (\cdot) = \partial_t (\cdot) + {\mathcal U}\partial (\cdot) / \partial x 
= D^\alpha_t (\cdot) + {\mathcal U}^\alpha \partial (\cdot) / \partial X^\alpha.
\end{equation}
This is the time derivative of a quantity measured by an observer moving with
the shock front.

Jump equations can be derived for conservation of mass, momentum, and energy between any two points in a structured
steady wave form \cite{claytonNEIM2019,claytonJMPS2021}.  In the 1-D Lagrangian description,
let $\mathcal{D}^\alpha$ be a constant steady wave speed, such that for
a differentiable function $f(X^\alpha,t)$, within the steady waveform,
\begin{align}
\label{eq:fsteady1}
& f(X^\alpha, t) = f(X^\alpha - {\mathcal D}^\alpha t) = f(Y^\alpha), \\
& \partial f / \partial X^\alpha = {\rm d } f / {\rm d} Y^\alpha, 
\quad
D^\alpha_t f = {-\mathcal D}^\alpha {\rm d } f / {\rm d} Y^\alpha.
\label{eq:fsteady2}
\end{align}
Applying \eqref{eq:fsteady2} to 1-D equations $D^\alpha_t F^\alpha = \partial \upsilon^\alpha / \partial X^\alpha$
and 
\begin{equation}
\label{eq:linmom1D}
 \rho_0^\alpha D_t^\alpha \upsilon^\alpha = \partial t^\alpha / \partial X^\alpha
 + \rho^\alpha_0 b^\alpha + (\partial \chi ^\alpha / \partial X^\alpha )h^\alpha,
\end{equation}
which is the first of \eqref{eq:mombal}, gives
\begin{align}
\label{eq:sw1}
& { \rm d} \upsilon^\alpha / { \rm d} Y = {- \mathcal D}^\alpha { \rm d} F^\alpha / { \rm d} Y, 
\\
& { \rm d} t^\alpha / { \rm d} Y = - \rho_0^\alpha {\mathcal D}^\alpha { \rm d} \upsilon^\alpha / { \rm d} Y - \rho^\alpha_0 b^\alpha -  F^\alpha h^\alpha.
\label{eq:sw2}
\end{align}
Select two points at steady-wave coordinates $Y^\pm$, and define the jump in a quantity between these points as in \eqref{eq:jump}: $ \llbracket f(Y) \rrbracket = f(Y^-) - f(Y^+)$.
Direct integration of \eqref{eq:sw1} over $Y^- \rightarrow Y^+$ and substitution into \eqref{eq:sw2} 
gives
\begin{align}
\label{eq:sw4}
& \llbracket \upsilon^\alpha \rrbracket = - {\mathcal D}^\alpha \llbracket F^\alpha \rrbracket, \\
& \llbracket t^\alpha \rrbracket = \rho_0^\alpha ({\mathcal D}^\alpha )^2 \llbracket F^\alpha \rrbracket + \int_-^+ (\rho^\alpha_0 b^\alpha +  F^\alpha h^\alpha) {\rm d} Y.
\label{eq:sw5}
\end{align}
Using the same procedure for 1-D continuum laws of energy conservation and
entropy production, \eqref{eq:ebal} and \eqref{eq:entbal}, 
\begin{align}
\label{eq:ebal1D}
& \rho_0^\alpha D_t^\alpha u^\alpha = t^\alpha \partial \upsilon^\alpha / \partial X^\alpha 
+ \partial ( \{  z \}^\alpha \}   \{ D_t^\alpha { \xi}^\alpha \} ) / \partial X^\alpha \nonumber \\
& \qquad  \qquad  \quad -  \partial q^\alpha / \partial X^\alpha + \rho_0^\alpha r^\alpha +  (\partial \chi ^\alpha / \partial X^\alpha ) \epsilon^\alpha,
 \\
 & \sum_\alpha [\rho_0^\alpha D^\alpha_t \eta^\alpha + \partial ( q^\alpha / \theta^\alpha ) /
 \partial X^\alpha - \rho_0^\alpha r^\alpha / \theta^\alpha 
  \nonumber \\
    &  \qquad  \qquad   \qquad  \qquad  \quad
     + (\partial \chi ^\alpha / \partial X^\alpha ) c^\alpha \eta^\alpha ] \geq 0,
\label{eq:ent1D}
\end{align}
produces jumps between two points $Y^\pm$ in a steady wave:
\begin{align}
&   \rho^\alpha_0  {\mathcal D}^\alpha \llbracket  u^\alpha +  {\textstyle {\frac{1}{2}}}|{ \upsilon}^\alpha|^2
 \rrbracket  =
  - \llbracket { t}^\alpha  { \upsilon}^\alpha
 + \{ { z}^\alpha \}  \{ D^\alpha_t { \xi}^\alpha \}
 - q^\alpha \rrbracket \nonumber \\
 & \qquad  + \int_-^+ \{ \rho_0^\alpha (r^\alpha + b^\alpha \upsilon^\alpha )
+ F^\alpha (\epsilon^\alpha + h^\alpha \upsilon^\alpha) \}
 {\rm d} Y,
\label{eq:sw6} \\
& \sum_\alpha ( \rho^\alpha_0  {\mathcal D}^\alpha \llbracket  \eta^\alpha \rrbracket - \llbracket q^\alpha / \theta^\alpha \rrbracket ) \nonumber \\
& \qquad \qquad - \int_-^+ (\rho^\alpha_0 r^\alpha / \theta^\alpha - F^\alpha c^\alpha \eta^\alpha) {\rm d} Y
 \} \geq 0.
\label{eq:sw7}
\end{align}
Relation \eqref{eq:sw4} is identical to the mass conservation law for a singular surface in the first of \eqref{eq:jumps3} when ${\mathcal D}^\alpha \rightarrow {\mathcal U}^\alpha$. 
Relation \eqref{eq:sw5} is identical to the momentum conservation law in the second of \eqref{eq:jumps3} when the integral on the right of \eqref{eq:sw5} vanishes (e.g., constant body force and null drag).
Relation \eqref{eq:sw6} is identical to the energy conservation law in \eqref{eq:ejump0}
when its integral terms vanish, and \eqref{eq:sw7} is identical to entropy inequality \eqref{eq:entjump0} when its integral terms involving heat and mass supplies vanish.
In such cases, the Eulerian jump conditions in \eqref{eq:massjump1} and \eqref{eq:momjump1} can be recovered for a steady wave form of Eulerian speed ${\mathcal D}$ where $f(x,t) = f(x - {\mathcal D} t)$ when ${ \mathcal D} = F^\alpha { \mathcal D}^\alpha + \upsilon^\alpha = {\rm constant}$.  

\subsection{\label{sec2c} Total expressions for mixture}

The spatial mass density of the mixture $\rho$,
mean velocity of the mixture $ {\bm \upsilon}$ , and diffusion velocities $ {\bm \mu}^\alpha$ are
\begin{align}
\label{eq:rho}
 \rho = \sum_\alpha \rho^\alpha, \quad
{ \bm \upsilon} = \frac{1}{ \rho} \sum_\alpha \rho^\alpha {\bm \upsilon}^\alpha, \quad
 {\bm \mu}^\alpha = {\bm \upsilon}^\alpha -  { \bm \upsilon}.
\end{align}
Note $\sum_\alpha \rho^\alpha {\bm \mu}^\alpha = {\bf 0}$. Define the material time derivative of $\square$ with respect to the mixture
as
\begin{equation}
\label{eq:MTDT}
\dot{ \square} = \partial_t (\square) + \nabla (\square)  \cdot {\bm \upsilon}
\, \Rightarrow \,
  D^\alpha_t (\square) = \dot {\square} + (\nabla \square) \cdot {\bm \mu}^\alpha.
\end{equation}
Summing \eqref{eq:massbal} over $\alpha$ then gives the total mass balance
for the mixture at space-time location $( {\bf x},t)$:
\begin{equation}
\label{eq:massmix}
\dot{\rho} + \rho \nabla \cdot  {\bm \upsilon}  = \hat{C},
\qquad \hat{C} = \sum_\alpha c^\alpha - \rho \, \partial_t \ln \sqrt{g}.
\end{equation}
In classical mixture theory \cite{bowen1976}, rates $c^\alpha$ account for
exchange of mass between phases but not mass production from extrinsic sources, thus locally summing to zero. To accommodate mass production from biologic (e.g., cellular growth) processes even in a single-phase material, $\sum_\alpha c^\alpha({\bf x},t)$ could be nonzero \cite{yavari2010,volokh2006}.

Define, respectively, the total Cauchy stress tensor,
total body force vector, total internal energy density, total entropy density, total heat supply,
 total heat flux, total internal state vector, and total conjugate force vector:
\begin{align}
\label{eq:stresstot}
& {\bm \sigma } = \sum_\alpha ( {\bm \sigma}^\alpha - \rho^\alpha {\bm \mu}^\alpha \otimes {\bm \mu}^\alpha), \quad  {\bf b} = \frac{1}{\rho} \sum_\alpha \rho^\alpha {\bf b}^\alpha, \\
\label{eq:etot}
&  u = \frac{1}{\rho} \sum_\alpha \rho^\alpha \left[ {u}^\alpha + {\textstyle{\frac{1}{2}}} | {\bm \mu}^\alpha |^2 \right],
\\ \label{eq:etatot}
& \eta = \frac{1}{\rho} \sum_\alpha \rho^\alpha {\eta}^\alpha, 
\qquad  r = \frac{1}{\rho} \sum_\alpha \rho^\alpha {r}^\alpha, \\
 \label{eq:qtot}
 & {\bf q } = \sum_\alpha ( {\bf q}^\alpha - {\bm \sigma}^\alpha \cdot {\bm \mu}^\alpha + \rho^\alpha u^\alpha  {\bm \mu}^\alpha + \frac{\rho^\alpha}{2} |{\bm \mu}^\alpha |^2 {\bm \mu}^\alpha ),
 \\
 \label{eq:ztot}
 & \{ { \bm \xi } \} = ( \{ {\bm \xi} \}^1, \ldots, \{ {\bm \xi}  \}^N), \quad
  \{ { \bm \zeta } \} = ( \{ {\bm \zeta} \}^1, \ldots, \{ {\bm \zeta}  \}^N).
\end{align}
Define $\hat{c}^\alpha = c^\alpha - \rho^\alpha \partial_t \ln \sqrt{g}$
and impose the constraints 
\begin{align}
\label{eq:sums}
& \sum_\alpha \hat{c}^\alpha = 0, \qquad
\sum_\alpha ({\bf h}^\alpha + \hat{c}^\alpha {\bm \mu^\alpha}) = {\bf 0 }, \\
\label{eq:sums2}
& \sum_\alpha [\epsilon^\alpha + {\bf h}^\alpha \cdot {\bm \mu}^\alpha +  \hat{c}^\alpha(u^\alpha +  {\textstyle {\frac{1}{2}}}|{\bm \mu}^\alpha|^2) ]= 0.
\end{align}
 Ensuring that the second of \eqref{eq:sums} holds, net energy rates from biologic growth processes can be included in $r^\alpha$.
Given \eqref{eq:rho}--\eqref{eq:ztot} with constraints \eqref{eq:sums} and \eqref{eq:sums2},
the total balances of mass, linear momentum, angular momentum, and energy and the dissipation
inequality are obtained
by accumulating \eqref{eq:massbal}, \eqref{eq:mombal}, \eqref{eq:ebal}, and \eqref{eq:entbal} over $\alpha = 1, \ldots, N$, following steps detailed by Bowen~\cite{bowen1976}, with the addition of 
work contributions from \eqref{eq:ztot} whose individual entries are defined as orthogonal:
\begin{align}
\label{eq:momix}
& \dot{\rho} + \rho \nabla \cdot  {\bm \upsilon}  = 0, \quad \nabla \cdot {\bm \sigma} + \rho {\bf b} = \rho \dot { {\bm \upsilon} }
, \quad {\bm \sigma} = {\bm \sigma}^{\mathsf T}, \\
\label{eq:emix}
&  \rho \dot{ u}  = {\bm \sigma}: \nabla {\bm \upsilon}
+ \nabla \cdot ( \{ { \bm \zeta} \} \cdot  \{ {\bm \xi}^\prime \} ) \nonumber 
- \nabla \cdot {\bf q} + \rho r 
\\ &  \qquad  + \sum_\alpha  \rho^\alpha {\bf b}^\alpha \cdot {\bm \mu}^\alpha,
\quad \{ {\bm \xi}^\prime \} =  \{ \dot{\bm \xi} \} + \sum_\alpha  \{ \nabla  {\bm \xi }^\alpha  \cdot {\bm \mu}^\alpha \} , 
\\ 
\label{eq:entmix}
& \rho \dot{\eta}  + \nabla \cdot \sum_\alpha ( \frac{ {\bf q}^\alpha} {\theta^\alpha} + \rho^\alpha \eta^\alpha { \bm \mu}^\alpha ) - \sum_\alpha \frac{ \rho^\alpha r^\alpha}{\theta^\alpha} \geq 0.
\end{align}
For the particular case when $\theta^\alpha = \theta$ is uniform among constituents at $({\bf x},t)$,
then \eqref{eq:entmix} simplifies to
\begin{equation}
\label{eq:entmixsimp}
 \rho \dot{\eta}  + \nabla \cdot (\hat{ \bf q} / \theta) - \rho r/ \theta \geq 0,
 \quad \hat{\bf q} = \sum_\alpha ({\bf q}^\alpha + \theta \rho^\alpha \eta^\alpha { \bm \mu}^\alpha ). 
\end{equation}

Jump conditions across singular surfaces can be derived from \eqref{eq:momix}, \eqref{eq:emix}, and \eqref{eq:entmixsimp} using the methods already discussed for constituent $\alpha$.
Let $\mathcal{U}$ now be the Eulerian shock speed for the mixture as a whole.
Analogs of \eqref{eq:massjump}--\eqref{eq:entjump} are, with an obvious change of notation, 
\begin{align}
\label{eq:massjumpmix}
&    \llbracket \rho (\upsilon_n - \mathcal{U}) \rrbracket = 0, \qquad
  \llbracket \rho {\bm \upsilon} (\upsilon_n - \mathcal{U}) \rrbracket = \llbracket { \bf t} \rrbracket , \\
 \label{eq:ejumpmix} 
 & \llbracket \rho ( u +  {\textstyle {\frac{1}{2}}}|{\bm \upsilon}|^2)
 (\upsilon_n - \mathcal{U}) \rrbracket 
 = \llbracket {\bf t} \cdot {\bm \upsilon}
 + \{ {\bf z} \} \cdot \{  {\bm \xi}^\prime \}
 - \hat{q}_n \rrbracket, \\
  \label{eq:entjumpmix}
 &  \llbracket \rho \eta (\mathcal{U}- \upsilon_n) - \hat{q}_n / \theta \rrbracket \geq 0.
 \end{align}
 Eulerian equations for the 1-D case follow trivially.
 Lagrangian equations for the 1-D case are obtained by defining Lagrangian
 speed $\mathcal{U}_0$, deformation mapping $F$,
 and space-time continuous reference mass density $\rho_0$ to obey
 \begin{equation}
 \label{eq:lag1Ddef}
 \mathcal{U}_0  = F^{-1} ( \mathcal{U} - \upsilon),
 \qquad F = \rho_0 / \rho.
 \end{equation}
 Given $\rho_0$ and $\mathcal{U}$, $F$ and $\mathcal{U}_0$ are well-defined.
 Analogs of \eqref{eq:massjump0}--\eqref{eq:entjump0} are then derived as,
 with $t_k \rightarrow t_1 = \sigma$,
\begin{align}
 \label{eq:massjump0m}
 & \rho_0 {\mathcal U}_0 \llbracket 1 / \rho \rrbracket = - \llbracket \upsilon \rrbracket 
 , \qquad \rho_0  {\mathcal U}_0 \llbracket \upsilon \rrbracket = -  \llbracket {  \sigma} \rrbracket , \\
 \label{eq:ejump0m} 
 &   \rho_0  {\mathcal U}_0 \llbracket  u +  {\textstyle {\frac{1}{2}}}{ \upsilon}^2
 \rrbracket  = - \llbracket { \sigma }  { \upsilon}
 + \{ { z} \}  \{ { \xi}^\prime \}
 - \hat{q} \rrbracket , \\
 \label{eq:entjump0m}
&  \rho_0  {\mathcal U}_0 \llbracket  \eta \rrbracket - \llbracket \hat{q} / \theta \rrbracket  \geq 0.
 \end{align}
 Equations fully analogous to \eqref{eq:veljumps}--\eqref{eq:RHe0} can be derived summarily with the obvious
 substitutions.
 
Presuming invertibility and integrability of the second of \eqref{eq:lag1Ddef}, deformation gradient $F = \partial x / \partial X $ and Lagrangian coordinate $X$ for the mixture exist whereby
\begin{align}
& \dot{F} = (\partial \upsilon / \partial x) F = \partial \upsilon / \partial X, \label{eq:Fmix0} \\
& X(x,t) = \chi^{-1}(x,t) = \int_{x_0}^x F^{-1}(\tilde{x},t) {\rm d} \tilde{x},
\label{eq:Fmix1}
\end{align}
 then the analog of the displacement derivative \eqref{eq:deltaderiv} for the mixture is
\begin{equation}
\label{eq:deltaderiv2}
\delta_t (\square) = \partial_t (\square) + {\mathcal U}\partial (\square) / \partial x 
= ( \dot \square ) + {\mathcal U_0} \partial (\square) / \partial X.
\end{equation}

Given \eqref{eq:momix}, \eqref{eq:emix}, \eqref{eq:entmixsimp}, \eqref{eq:Fmix0}, and \eqref{eq:Fmix1}, the treatment of steady Lagrangian waves can be applied to the mixture as a whole. Let $f(X,t) = f(X-{\mathcal D} t) = f(Y)$ in a steady wave, with $\mathcal{D}$ the constant speed. Using relations akin to \eqref{eq:fsteady1} and \eqref{eq:fsteady2}, the local compatibility and linear momentum 
equations \eqref{eq:Fmix0} and \eqref{eq:momix} (in 1-D) are
\begin{align}
\label{eq:sw1mix}
& { \rm d} \upsilon / { \rm d} Y = {- \mathcal D} { \rm d} F^\alpha / { \rm d} Y, 
\\
& { \rm d} \sigma / { \rm d} Y = - \rho_0 {\mathcal D} { \rm d} \upsilon / { \rm d} Y - \rho_0 b.
\label{eq:sw2mix}
\end{align}
Integrating from $Y^+ \rightarrow Y^-$ gives conditions like \eqref{eq:massjump0m}:
\begin{align}
\label{eq:sw4mix}
& \llbracket \upsilon \rrbracket = - {\mathcal D} \llbracket F \rrbracket =
- {\mathcal D} \rho_0 \llbracket 1/ \rho \rrbracket, \\
& \llbracket \sigma \rrbracket = - \rho_0 {\mathcal D} \llbracket \upsilon \rrbracket + \int_-^+ \rho_0 b \, {\rm d} Y.
\label{eq:sw5mix}
\end{align}
Energy conservation and entropy production in \eqref{eq:emix} and \eqref{eq:entmix} can be addressed similarly, omitted here for brevity.

Returning to the general 3-D case, diffusion problems
are often analyzed
using dimensionless measures of local amounts of
each constituent. 
Recall $\rho^\alpha$ is the local mass of $\alpha$
per unit total spatial volume of mixture.
The spatial volume fraction $n^\alpha$ is
the ratio of volume occupied by $\alpha$ to
that of the mixture, while the real mass density
$\rho^\alpha_{\rm R}$ is the local mass of $\alpha$
per unit spatial volume occupied by $\alpha$ (i.e., $\rho^\alpha_{\rm R}$ is mass density of the isolated, fully dense constituent).
The spatial mass concentration $m^\alpha$ is the mass of $\alpha$
per total mass of the mixture. Equations are
\begin{align}
\label{eq:vfracs}
n^\alpha({\bf x},t) = \frac{\rho^\alpha ( {\bf x},t) }{\rho^\alpha_{\rm R} ( {\bf x},t)},
\qquad \sum_\alpha n^\alpha = 1, \\
\label{eq:massfracs}
m^\alpha({\bf x},t) = \frac{ \rho^\alpha ( {\bf x},t) }{ \rho ( {\bf x},t)},
\qquad \sum_\alpha m^\alpha = 1.
\end{align}
At a reference $t = t_0$ when all particles occupy positions ${\bf X^\alpha}$, reference volume and mass fractions are
$n^\alpha_0 ({\bf X}^\alpha) = \rho^\alpha_0 ({\bf X}^\alpha)/ \rho^\alpha_{ {\rm R} 0} ( {\bf X}^\alpha)$ and
$m^\alpha_0 ({\bf X}^\alpha) = \rho^\alpha_0 ({\bf X}^\alpha)/ \rho_0 ( {\bf X}^\alpha)$.

%% file: s3a.tex
\section{\label{sec3} Constitutive Theory}

Thermodynamic identities
are derived in Sec.~\ref{sec3a}
by appealing to
the local balance of energy and entropy inequality.
Pragmatic and thermodynamically
admissible energy functions and kinetic relations are posited in respective
Secs.~\ref{sec3b} and ~\ref{sec3c}.

\subsection{\label{sec3a}Thermodynamics}
Helmholtz free energy per unit mass and entropy density are of the following functional forms,
each depending only on state variables for its particular constituent $\alpha$
and not others $\beta$ with $\beta \neq \alpha$:
\begin{align}
\label{eq:helm}
&\psi^\alpha = \psi^\alpha( { \bf F}^\alpha, \theta^\alpha, \{ {\bm \xi}^\alpha \}, \{ \nabla^\alpha_0 {\bm \xi}^\alpha \},  {\bf X}^\alpha ), \\
\label{eq:ent}
& \eta^\alpha = \eta^\alpha( { \bf F}^\alpha, \theta^\alpha, \{ {\bm \xi}^\alpha \}, \{ \nabla^\alpha_0 {\bm \xi}^\alpha \},  {\bf X}^\alpha ).
\end{align}
Partial stress consists of elastic $\bar{\bm \sigma}^\alpha$ and
viscous $\hat{\bm \sigma}^\alpha$ parts:
\begin{equation}
\label{eq:stress}
\begin{split}
{\bm \sigma} ^\alpha & = \bar{\bm \sigma}^\alpha ( { \bf F}^\alpha, \theta^\alpha, \{ {\bm \xi}^\alpha \}, \{ \nabla^\alpha_0 {\bm \xi}^\alpha \}, {\bf X}^\alpha ) \\
& \qquad  + \hat{\bm \sigma}^\alpha ( { \bf F}^\alpha, \theta^\alpha, \{ {\bm \xi}^\alpha \}, \{ \nabla^\alpha_0 {\bm \xi}^\alpha \}, D^\alpha_t { \bf F}^\alpha,  {\bf X}^\alpha ).
\end{split}
\end{equation}
Arguments in \eqref{eq:metrics1} and \eqref{eq:metrics2} are linked at time $t$ via \cite{claytonZAMP2017,claytonSYMM2023}
\begin{equation}
\label{eq:isvrelate}
 \{ {\bm \Xi}^\alpha \} (t) = \{ {\bm \Xi}^\alpha (\{ \bm \xi ^\alpha \},  {\bf X}^\alpha, {\bf x} )\} (t).
\end{equation}
Thus, all dependence of response functions on metrics $({\bf g}, \bf{G}^\alpha)$ and states  $\{ {\bm \Xi}^\alpha \} $ is implicitly included via arguments $ \{ {\bm \xi}^\alpha \} $.
Kinetic equations for heat flux, internal state, and interphase mass, momentum, and energy exchange are more general, allowing for dependence on all constituents
$\beta = 1, \ldots, N$ including $\alpha = \beta$ and $\alpha \neq \beta$:
\begin{align}
\label{eq:qflux}
&{\bf q}^\alpha = {\bf q}^\alpha( { \bf F}^\beta, \nabla{\bf F}^\beta, \theta^\beta, \nabla \theta^\beta,  \bm \xi ^\beta,  \nabla^\beta_0 {\bm \xi}^\beta ,  {\bm \upsilon}^\beta, {\rho}^\beta ), \\
\label{eq:hflux}
&{\bf h}^\alpha = {\bf h}^\alpha( { \bf F}^\beta, \nabla{\bf F}^\beta, \theta^\beta, \nabla \theta^\beta,  \bm \xi ^\beta,  \nabla^\beta_0 {\bm \xi}^\beta ,  {\bm \upsilon}^\beta, {\rho}^\beta ), \\
\label{eq:cflux}
&{ {c}}^\alpha = {c}^\alpha( { \bf F}^\beta, \nabla{\bf F}^\beta, \theta^\beta, \nabla \theta^\beta,  \bm \xi ^\beta,  \nabla^\beta_0 {\bm \xi}^\beta ,  {\bm \upsilon}^\beta,{\rho}^\beta ), \\
\label{eq:epsflux}
&{ {\epsilon}}^\alpha = {\epsilon}^\alpha( { \bf F}^\beta, \nabla{\bf F}^\beta, \theta^\beta, \nabla \theta^\beta,  \bm \xi ^\beta,  \nabla^\beta_0 {\bm \xi}^\beta ,  {\bm \upsilon}^\beta, {\rho}^\beta ), 
\\
\nonumber
&{ D^\alpha_t \{ {\bm \xi } ^\alpha \} } = { D^\alpha_t \{ {\bm \xi } ^\alpha \} } (  { \bf F}^\beta, \nabla{\bf F}^\beta, \theta^\beta, \nabla \theta^\beta,  \cdots 
\\ 
\label{eq:isvdot}
& \qquad \qquad \qquad \qquad \qquad \quad \cdots \bm \xi ^\beta,  \nabla^\beta_0 {\bm \xi}^\beta ,  {\bm \upsilon}^\beta, {\rho}^\beta ).
\end{align}
Notation $\{ \cdot \}$  on ${\bm \xi}^\alpha$ and admissible explicit dependence on ${\bf X}^\alpha$ for heterogeneous phases $\alpha$ are omitted in arguments of \eqref{eq:qflux}--\eqref{eq:isvdot} for brevity.
Particular forms of \eqref{eq:qflux}--\eqref{eq:epsflux} must satisfy principles of spatial invariance
for objective spatial vectors ${\bf q}^\alpha$ and ${\bf h}^\alpha$ and scalars
$c^\alpha$ and $\epsilon^\alpha$. Evolution equations \eqref{eq:isvdot} must also be objective.
Invariance under rigid translation of the mixture as a whole necessitates that
dependence on velocities ${\bm \upsilon}^\alpha$ is at most only on the $N-1$ velocity differences
${\bm \upsilon}^1 - {\bm \upsilon}^N, \ldots , {\bm \upsilon}^{N-1} - {\bm \upsilon}^N$.
See Ref.~\cite{bowen1976}.  Diffusion velocities ${\bm \mu}^\alpha = {\bm \mu}^\alpha ({\bm \upsilon}^\beta,\rho^\beta) $ of \eqref{eq:rho} fulfill this requirement. Spatial invariance of \eqref{eq:helm} and \eqref{eq:ent}
is obtained via dependence on ${\bf F}^\alpha$ through symmetric deformation tensor
 ${\bf C}^\alpha$:
\begin{align}
\label{eq:Ctensor}
& {\bf C}^\alpha =( {\bf F}^\alpha)^\mathsf{T} {\bf F}^\alpha, \,
(C^\alpha)^K_J = (G^\alpha)^{KI} (F^\alpha)^i_I g_{ij} (F^\alpha)^j_J;
\\
\label{eq:C2}
& \partial \psi^\alpha / \partial  {\bf F}^\alpha = 2  {\bf F}^\alpha \partial \psi^\alpha / \partial  {\bf C}^\alpha, \quad
 J^\alpha = \sqrt{ \det  {\bf C}^\alpha}; \quad
\\
\label{eq:C3}
& D^\alpha_t  {\bf C}^\alpha = 2 ( {\bf F}^\alpha)^\mathsf{T} {\bf d}^\alpha {\bf F}^\alpha,
 \quad 2 {\bf d}^\alpha = {\bf l}^\alpha + ({\bf l}^\alpha )^\mathsf{T}.
\end{align}
The spatial deformation rate is ${\bf d}^\alpha$, and $D^\alpha_t {\bf C}^\alpha$
is taken with $(G^\alpha)^{IJ}$ and $g_{ij}$ fixed with respect to $t$ in \eqref{eq:C3}.

 Expanding 
$D^\alpha_t \psi$ using the chain rule on \eqref{eq:helm} and inserting the result and \eqref{eq:stress} into \eqref{eq:entbal2} gives
 \begin{align}
 \label{eq:entbal3}
&  \sum_\alpha \frac{1}{\theta^\alpha} \{ [
 \bar{\bm \sigma}^\alpha ({\bf F}^\alpha)^{-\mathsf{T}}
 - 2 \rho^\alpha {\bf F}^\alpha \frac{ \partial \psi^\alpha}{\partial {\bf C}^\alpha}]
 : D_t^\alpha {\bf F}^\alpha
 \nonumber
  \\
  & - \rho^\alpha [ \eta^\alpha + \partial \psi^\alpha / \partial \theta^\alpha ] D_t^\alpha \theta^\alpha
  \nonumber
  \\
&
 + [ \{ ({\bf F}^\alpha)^{-1} {\bm \zeta}^\alpha \} - \rho^\alpha \frac{\partial \psi^\alpha}{\partial \{ \nabla_0 { \bm \xi}^\alpha \} } ]:  \{ D^\alpha_t  (\nabla_0 {\bm \xi}^\alpha ) \}
 \nonumber
 \\  
&  + [ \{ ({\bf F}^\alpha)^{-1}:\nabla_0^\alpha {\bm \zeta}^\alpha \} - 
\rho^\alpha \frac{\partial \psi^\alpha} { \partial \{ {\bm \xi}^\alpha \} } ] \cdot \{ D^\alpha_t { \bm \xi}^\alpha \}  \nonumber 
\\ &  + \hat{\bm \sigma}^\alpha:{\bf d}^\alpha - { ({\bf q^\alpha } \cdot \nabla \theta^\alpha ) }/{\theta^\alpha}
+ \epsilon^\alpha + c^\alpha \theta^\alpha \eta^\alpha \}
\geq 0.
  \end{align}
  Identities from \eqref{eq:commute} and \eqref{eq:nabla0} have been used to obtain
  \begin{equation}
  \label{eq:derivxi}
  \begin{split}
    \nabla \cdot ( \{ { \bm \zeta}^\alpha \} & \cdot  \{ D_t^\alpha {\bm \xi}^\alpha \} )
    = ( \nabla  \cdot  \{ { \bm \zeta}^\alpha \} ) \cdot  \{ D_t^\alpha {\bm \xi}^\alpha \} \\ &
     \qquad \qquad +  \{ ( {\bf F}^\alpha)^{-1} {\bm \zeta}^\alpha \} : \{ D^\alpha_t  (\nabla^\alpha_0 {\bm \xi}^\alpha ) \}.
   \end{split}
  \end{equation}
  From standard arguments \cite{coleman1963,coleman1967,gurtin1996,levitas2014} and
  \eqref{eq:helm}--\eqref{eq:isvdot},
  the first three sets of terms in \eqref{eq:entbal3} should vanish for admissibility under general
  thermodynamic processes, leading to the following constitutive equalities:
  \begin{align}
  \label{eq:consteq1}
  & \bar{\bm \sigma}^\alpha = 2 \rho^\alpha {\bf F}^\alpha \frac{ \partial \psi^\alpha}{\partial {\bf C}^\alpha} ({\bf F}^\alpha)^{\mathsf T},
  \qquad \eta^\alpha = - \frac{\partial \psi^\alpha}{\partial \theta^\alpha},
  \\
 \label{eq:consteq2}
 &  \{ {\bm \zeta}^\alpha \} = \rho^\alpha\{  {\bf F}^\alpha { \partial \psi^\alpha}/
 { \partial \nabla_0^\alpha {\bm \xi}^\alpha \} } = \rho^\alpha \frac { \partial \psi^\alpha}{\partial \{\nabla {\bm \xi}^\alpha \} }, \\
 &  \{ {\bm \pi}^\alpha \} = \rho^\alpha \partial \psi^\alpha / \partial \{ {\bm \xi}^\alpha \}, \label{eq:bmpi}
  \end{align}
  where \eqref{eq:bmpi} defines a conjugate force to internal state variables or order parameters. Then \eqref{eq:entbal3} reduces to
\begin{align}
 \label{eq:entbal4}
&  \sum_\alpha \frac{1}{\theta^\alpha} [ (
 \{ \nabla \cdot {\bm \zeta}^\alpha \} - 
 \{ {\bm \pi}^\alpha \} )
 \cdot \{ D^\alpha_t { \bm \xi}^\alpha \}  \nonumber 
\\ &  + \hat{\bm \sigma}^\alpha:{\bf d}^\alpha - { ({\bf q}^\alpha  \cdot \nabla \theta^\alpha )}/{\theta^\alpha}
+ \epsilon^\alpha + c^\alpha \theta^\alpha \eta^\alpha ]
\geq 0.
  \end{align}
  Applying the Legendre transformation from \eqref{eq:psi} with
  \begin{align}
  \label{eq:legendre}
& u^\alpha = u^\alpha ({\bf F}^\alpha,\eta^\alpha,\{ {\bm \xi}^\alpha \},
\{ \nabla_0^\alpha { \bm \xi }^\alpha \}, \bf{X}^\alpha ), \\
& \theta^\alpha = \theta^\alpha  ({\bf F}^\alpha,\eta^\alpha,\{ {\bm \xi}^\alpha \},
\{ \nabla_0^\alpha { \bm \xi }^\alpha \}, \bf{X}^\alpha ), \label{eq:legendre2}
  \end{align}
 in conjunction with \eqref{eq:consteq1} and \eqref{eq:consteq2}, gives
 \begin{align}
  \label{eq:consteq3}
  & \bar{\bm \sigma}^\alpha = 2 \rho^\alpha {\bf F}^\alpha \frac{ \partial u^\alpha}{\partial {\bf C}^\alpha} ({\bf F}^\alpha)^{\mathsf T},
  \qquad \theta^\alpha = \frac{\partial u^\alpha}{\partial \eta^\alpha},
  \\
 \label{eq:consteq4}
 & \{ {\bm \pi}^\alpha \} = \rho^\alpha \frac{\partial u^\alpha }{ \partial \{ {\bm \xi}^\alpha \} }, \quad
   \{ {\bm \zeta}^\alpha \} 
 = \rho^\alpha \frac { \partial u^\alpha}{\partial \{\nabla {\bm \xi}^\alpha \} }.
  \end{align}
  Define specific heat per unit mass at constant strain $c_\epsilon^\alpha$,
   thermal stress coefficients $ {\bm \beta}^\alpha$, and Gr\"uneisen tensor ${\bm \gamma}^\alpha$:
   \begin{align}
   \label{eq:specht}
  & c_\epsilon^\alpha  
  = \theta^\alpha \partial \eta^\alpha / 
  \partial \theta^\alpha 
   = - \theta^\alpha \, { \partial^2 \psi^\alpha}/{\partial (\theta^\alpha)^2},
\\
\label{eq:grunparam}
 &  {\bm \beta}^\alpha = \rho^\alpha c^\alpha_\epsilon {\bm \gamma}^\alpha = 
  -2 \rho^\alpha \, {\partial^2 \psi^\alpha} / {\partial \theta^\alpha \partial {\bf C}^\alpha } .
   \end{align}
   Define the intrinsic dissipation for constituent $\alpha$:
   \begin{equation}
   \label{eq:dissi}
   {\mathfrak D}^\alpha = (
 \{ \nabla \cdot {\bm \zeta}^\alpha \} - 
 \{ {\bm \pi}^\alpha \} )
 \cdot \{ D^\alpha_t { \bm \xi}^\alpha \} +  \hat{\bm \sigma}^\alpha:{\bf d}^\alpha.
   \end{equation}
 Expand the rate of $\eta^\alpha$ using \eqref{eq:ent}, \eqref{eq:consteq1}, \eqref{eq:specht},
 and \eqref{eq:grunparam}:
 \begin{equation}
 \label{eq:etadot}
 \begin{split}
 \rho^\alpha \theta^\alpha D_t^\alpha \eta^\alpha & =   \rho^\alpha c_\epsilon^\alpha D^\alpha_t \theta^\alpha + {\textstyle{\frac{1}{2}}} \theta^\alpha {\bm \beta}^\alpha: D_t^\alpha {\bf C}^\alpha \\
 &  -\rho^\alpha \theta^\alpha [ (\partial^2 \psi/ \partial \theta^\alpha \partial \{ {\bm \xi}^\alpha \})
 \cdot \{  D_t^\alpha {\bm \xi}^\alpha \}
 \\& + (\partial^2 \psi/ \partial \theta^\alpha \partial \{ \nabla {\bm \xi}^\alpha \})
 : \{  \nabla (D_t^\alpha {\bm \xi}^\alpha) \}].
  \end{split}
 \end{equation}
 From \eqref{eq:ebal}, time differentiation of \eqref{eq:psi}, and
 \eqref{eq:consteq1} and \eqref{eq:consteq2}:
 \begin{equation}
 \label{eq:etadot2}
 \begin{split}
 \rho^\alpha \theta^\alpha D_t^\alpha \eta^\alpha  = {\mathfrak D}^\alpha - \nabla \cdot {\bf q}^\alpha + \rho^\alpha r^\alpha + \epsilon^\alpha.
 \end{split}
 \end{equation}
Temperature rates then are, combining \eqref{eq:etadot} and \eqref{eq:etadot2},
\begin{equation}
\label{eq:temprate}
\begin{split}
\rho^\alpha c_\epsilon^\alpha D^\alpha_t \theta^\alpha  = & \, {\mathfrak D}^\alpha 
- {\textstyle{\frac{1}{2}}} \theta^\alpha {\bm \beta}^\alpha: D_t^\alpha {\bf C}^\alpha
\\
 &  + \rho^\alpha \theta^\alpha [ (\partial^2 \psi/ \partial \theta^\alpha \partial \{ {\bm \xi}^\alpha \})
 \cdot \{  D_t^\alpha {\bm \xi}^\alpha \}
 \\& + (\partial^2 \psi/ \partial \theta^\alpha \partial \{ \nabla {\bm \xi}^\alpha \})
 : \{  \nabla (D_t^\alpha {\bm \xi}^\alpha) \}]
\\ & - \nabla \cdot {\bf q}^\alpha + \rho^\alpha r^\alpha + \epsilon^\alpha.
\end{split}
\end{equation}

Decompositions of metrics 
of \eqref{eq:metrics1} and \eqref{eq:metrics2} into symmetric position-dependent (i.e., classical) and dimensionless, invertible, space-time dependent parts are \cite{claytonSYMM2023,yavari2010} 
\begin{align}
\label{eq:metrics3}
&  {\bf g}( {\bf x}, t) = \bar{ \bf g} ( {\bf x}) \, \hat{ \bf g}( \{ {\bm \xi}^\alpha( {\bf x},t) \} ), \\
& {\bf G}^\alpha  ( {\bf X}^\alpha, t) = \bar{\bf G}^\alpha  ( {\bf X}^\alpha ) \, \hat{ \bf G}^\alpha 
(  \{ { \bm \Xi }^\alpha ( {\bf X}^\alpha, t) \} ). 
\label{eq:metrics4}
\end{align}
A deformation $\bar{\bf C}^\alpha$ and Jacobian $\bar{J}^\alpha$ based on $(\bar{\bf g},\bar{\bf G}^\alpha)$ are
\begin{align}
\label{eq:Cbar1}
& (\bar{C}^\alpha)^K_J = \bar{G}^{KI} (F^\alpha)^i_I \bar{g}_{ij} (F^\alpha)^j_J ({\bf G}^\alpha)_K \otimes ({\bf G}^\alpha)^J,
\\ 
\label{eq:Jbar}
& \bar{J}^\alpha = \sqrt{ \det \bar{\bf C}^\alpha} = J^\alpha \sqrt{ \hat{G}^\alpha / \hat{g}},
\end{align}
with dimensionless $\hat{g} = \det \hat{\bf g}$ and $\hat{G}^\alpha = \det \hat{\bf G}^\alpha$.
Alternative constitutive equations \cite{claytonSYMM2023}, also energetically objective, are obtained by
positing dependence of $\psi^\alpha$, $\eta^\alpha$, $u^\alpha$, and $\theta^\alpha$
through $\bar{\bf C}^\alpha (\bf{F}^\alpha)$ rather than ${\bf C}^\alpha$, whereby
\begin{equation}
\label{eq:Cbar2}
\begin{split}
& \partial \psi^\alpha / \partial  {\bf F}^\alpha = 2  {\bf F}^\alpha \partial \psi^\alpha / \partial  \bar{\bf C}^\alpha,\\
& \quad \leftrightarrow \,
\partial \psi^\alpha / \partial (F^\alpha)^i_J = 2 \bar{g}_{ik} (F^\alpha)^k_L 
\partial \psi^\alpha / \partial (\bar{C}^\alpha)_{JL}.
\end{split}
\end{equation}
Derivations in \eqref{eq:entbal3}--\eqref{eq:temprate} continue to apply for $\psi^\alpha( \bar{ \bf C}^\alpha, \cdot)$, $\eta^\alpha( \bar{ \bf C}^\alpha, \cdot)$, etc.~with several
 changes manifested by \eqref{eq:Cbar2}:
\begin{align}
\label{eq:alt1}
& (\bar{\sigma}^\alpha)^j_i  = 2 {\rho}^\alpha \bar{g}_{ik} (F^\alpha)^k_L (F^\alpha)^j_J
\partial \psi^\alpha / \partial (\bar{C}^\alpha)_{JL} \nonumber
\\
&\qquad \, \, \,  = 2 {\rho}^\alpha \bar{g}_{ik} (F^\alpha)^k_L (F^\alpha)^j_J
\partial u^\alpha / \partial (\bar{C}^\alpha)_{JL},
\end{align}
\begin{align}
\label{eq:alt2}
  & \bar{\bm \beta}^\alpha  = \rho^\alpha c^\alpha_\epsilon \bar{\bm \gamma}^\alpha = 
  -2 \rho^\alpha \, {\partial^2 \psi^\alpha} / {\partial \theta^\alpha \partial \bar {\bf C}^\alpha },
  \\
  \label{eq:alt3}
&  \rho^\alpha c_\epsilon^\alpha D^\alpha_t \theta^\alpha  =  \, {\mathfrak D}^\alpha 
- {\textstyle{\frac{1}{2}}} \theta^\alpha \bar{\bm \beta}^\alpha: D_t^\alpha \bar{\bf C}^\alpha \nonumber
\\
 &  \qquad \qquad + \rho^\alpha \theta^\alpha [ (\partial^2 \psi/ \partial \theta^\alpha \partial \{ {\bm \xi}^\alpha \})
 \cdot \{  D_t^\alpha {\bm \xi}^\alpha \} \nonumber
 \\& \qquad \qquad + (\partial^2 \psi/ \partial \theta^\alpha \partial \{ \nabla {\bm \xi}^\alpha \})
 : \{  \nabla (D_t^\alpha {\bm \xi}^\alpha) \}] \nonumber
\\ & \qquad \qquad - \nabla \cdot {\bf q}^\alpha + \rho^\alpha r^\alpha + \epsilon^\alpha.
  \end{align}
 Mixed-variant $\bar{\bm \sigma}^\alpha$ in \eqref{eq:alt1} excludes $\hat{\bf g}({\bf x},t)$. Contravariant stress 
 defined as $(\bar{\sigma}^\alpha)^{ij} = 
 \frac{1}{2}[ g^{ik} (\bar{\sigma}^\alpha)^j_k + g^{jk} (\bar{\sigma}^\alpha)^i_k$]
 or $(\bar{\sigma}^\alpha)^{ij} = \bar{g}^{ik} (\bar{\sigma}^\alpha)^j_k$
  must be symmetric. The former depends on $\hat{\bf g}({\bf x},t)$ implicitly from $g^{ik}$. This choice presumes, a priori, that skew contributions from \eqref{eq:alt1} perform no work in the energy balance so
  can thus be redefined as zero. The latter prescription either redefines raising/lowering indices on Cauchy stress or presumes that ${\bf d}^\alpha$ is defined in covariant form by lowering of ${\bf l}^\alpha $ with $\bar{g}_{ij}$, rather than the typical $g_{ij}$ prior to symmetrization.
  
\subsection{\label{sec3b}Energy functions}

Internal state variables $\{ {\bm \xi}^\alpha \}$ consist of three sets:
configurational variables associated with
viscoelastic processes $\{ { \bm \Gamma }^\alpha  \}$,
damage variables associated with
degradation processes $\{ { \bf D}^\alpha \}$,
and electrochemical activation (e.g., muscle contraction) variables
$\{ {\bm \Delta}^\alpha \}$ \cite{ito2010,gultekin2016,gultekin2019,claytonSYMM2023}:
\begin{equation}
\label{eq:ISVs}
\{ {\bm \xi}^\alpha \} ({\bf x},t) = ( \{ { \bm \Gamma }^\alpha  \}, \{ { \bf D}^\alpha \}, \{ {\bm \Delta}^\alpha \}) ({\bf x},t).
\end{equation}
In the present work, as in phase-field and related theories
\cite{gultekin2019,claytonSYMM2023}, free and internal energy functions
can depend on spatial gradients of damage variables, which are viewed
as order parameters, but not on spatial gradients of viscoelastic
and tissue activation variables. Denote
by $\varsigma^\alpha_{\rm V}$ and $\varsigma^\alpha_{\rm S}$
degradation functions associated with loss of strength due to
changes in bulk and deviatoric strain energies, respectively. 
These scalar functions obey
\begin{align}
\label{eq:degrade}
& \varsigma^\alpha_{\rm V} = \varsigma^\alpha_{\rm V} ( \{ { \bf D}^\alpha \}, {\bf C}^\alpha) \in [0,1], \, \, 
\varsigma^\alpha_{\rm S} = \varsigma^\alpha_{\rm S} ( \{ { \bf D}^\alpha \}) \in [0,1], \\
\label{eq:degradederiv}
& \partial  \varsigma^\alpha_{\rm V}  / \partial {\bf C}^\alpha ( \{ { \bf D}^\alpha \} , { \bf C}^\alpha  ) = { \bf 0} \, \forall  \, { \bf C}^\alpha \neq { \bf 1}.
\end{align}
A degradation operator for
fibrous energy contributions with similar
properties is $ \varsigma^\alpha_{\rm F}(\{ { \bf D}^\alpha \})$. Let $\Psi^\alpha =   \rho^\alpha_{ {\rm R} 0 } \psi^\alpha $
and $U^\alpha =  \rho^\alpha_{ {\rm R} 0 } u^\alpha $ 
be free and internal energies per unit reference volume of individual phases.
Pragmatic functional forms consist of the following sums:
\begin{align}
\label{eq:Psitot}
 \Psi^\alpha  ({\bf C}^\alpha, & \, \theta^\alpha, \{ {\bm \xi}^\alpha \}, 
 \{ \nabla_0^\alpha {\bf D}^\alpha \}) = \nonumber
 \\ & \quad
 \varsigma^\alpha_{\rm V} ( \{ { \bf D}^\alpha \}, {\bf C}^\alpha)
 \Psi^{ \alpha}_{  {\rm V} } (J^\alpha, \theta^\alpha)  \nonumber 
 \\ 
 & 
  + \varsigma^\alpha_{\rm S} ( \{ { \bf D}^\alpha \}) [ \Psi^{\alpha}_{ {\rm S} } 
  ( {\bf C}^\alpha) +   \Psi^{\alpha}_{ \Gamma} 
  ({\bf C}^\alpha, \{ {\bm \Gamma}^\alpha \} ) ] \nonumber
   \\
 &  +   \varsigma^\alpha_{\rm F}(\{ { \bf D}^\alpha \}) \circ  [\Psi^\alpha_{\rm F} ({\bf C}^\alpha) + \Psi^{\alpha}_{\Phi} 
  ({\bf C}^\alpha, \{ {\bm \Gamma}^\alpha \} )]
 \nonumber \\
 & + \Psi^\alpha_{\rm A} ({ \bf C}^\alpha,\{ {\bm \Delta}^\alpha \}) + \Psi^\alpha_{\rm \theta}(\theta^\alpha)
   \nonumber  \\ 
 & 
 + \Psi^\alpha_{\sigma } (J^\alpha) +  \Psi^\alpha_{\rm D} ( \{ {\bm \xi}^\alpha \}, 
 \{ \nabla_0^\alpha {\bf D}^\alpha \}), 
 \end{align}
 \begin{align}
 \label{eq:Utot}
U^\alpha  ({\bf C}^\alpha, & \, \eta^\alpha, \{ {\bm \xi}^\alpha \}, 
 \{ \nabla_0^\alpha {\bf D}^\alpha \}) = \nonumber
 \\ & \quad
 \varsigma^\alpha_{\rm V} ( \{ { \bf D}^\alpha \}, {\bf C}^\alpha)
 U^{ \alpha}_{  {\rm V} } (J^\alpha, \eta^\alpha)  \nonumber 
 \\ 
 & 
  + \varsigma^\alpha_{\rm S} ( \{ { \bf D}^\alpha \}) [U^{\alpha}_{ {\rm S} } 
  ( {\bf C}^\alpha) +   U^{\alpha}_{ \Gamma} 
  ({\bf C}^\alpha, \{ {\bm \Gamma}^\alpha \} ) ] \nonumber
   \\
 &  + \varsigma^\alpha_{\rm F}(\{ { \bf D}^\alpha \}) \circ [U^\alpha_{\rm F} ({\bf C}^\alpha) + U^{\alpha}_{\Phi} 
  ({\bf C}^\alpha, \{ {\bm \Gamma}^\alpha \} )]
   \nonumber \\ &
 + U^\alpha_{\rm A} ({ \bf C}^\alpha,\{ {\bm \Delta}^\alpha \}) + U^\alpha_{\rm \theta}(\eta^\alpha)
   \nonumber  \\  & 
 + U^\alpha_{\sigma } (J^\alpha) + U^\alpha_{\rm D} ( \{ {\bm \xi}^\alpha \}, 
 \{ \nabla_0^\alpha {\bf D}^\alpha \}).
  \end{align}
  In $\Psi^\alpha$, volumetric equilibrium free energy for the 
  entire constituent $\alpha$ is $\Psi^\alpha_{\rm V}$,
  including isotropic thermoelastic coupling.
  Deviatoric equilibrium energy of the isotropic matrix
  is $\Psi^\alpha_{\rm S}$.
  Viscoelastic configurational energy of the isotropic matrix is
  $\Psi^\alpha_\Gamma$.
  Anisotropic deviatoric equilibrium free energy from fibrous microstructures
  is $\Psi^\alpha_{\rm F}$.
  Configurational energy, often but not always anisotropic, from fibers is $\Psi^\alpha_{\Phi}$.
  Energy from fiber activation is $\Psi^\alpha_{\rm A}$.
  Thermal energy of specific heat is $\Psi^\alpha_\theta$.
  Energy from a nonzero reference pressure is $\Psi^\alpha_\sigma$.
  Surface energy from fractures, tears, and other damage
  is contained in $\Psi^\alpha_{\rm D}$.
  Fully analogous descriptors apply to internal energy contributions
  in $U^\alpha$.
  Forms \eqref{eq:Psitot} and \eqref{eq:Utot} are not the most
   mathematically and physically general, but they are sufficient
   for soft tissue materials of present interest given the scope of
   available data on their properties and response.
   A few, or even most, terms vanish for certain classes of materials
   (e.g., isotropic solids, viscoelastic fluids, gas phases, and so forth).
  All functions in \eqref{eq:Psitot} and \eqref{eq:Utot} in materials with heterogeneous
  properties can further depend explicitly on ${\bf X}^\alpha$, omitted in the arguments for brevity. Dependence of state-dependent metric tensors is implicit,
   for example in $J^\alpha$ and
   scalar functions of certain vectors and tensors.
   \\
   
   \noindent {\bf{Ideal gas EOS}}. 
   For gaseous fluids such as air in the lung, an ideal
   gas model \cite{claytonIJES2022} is sufficient. For the ideal
   gas, $\psi^\alpha = (\Psi^\alpha_{ \rm V} + \Psi^\alpha_\theta) / \rho^\alpha_{ {\rm R} 0 } $
   and  $u^\alpha = (U^\alpha_{\rm V} + U^\alpha_\theta)/ \rho^\alpha_{ {\rm R} 0 }$.
   At a reference state, the following conditions hold: $J^\alpha =1$, $\theta^\alpha = \theta^\alpha_0$, $\eta^\alpha = \eta^\alpha_0$,
   $\rho^\alpha_0 = n^\alpha_0 \rho^\alpha_{ {\rm R} 0 }$, 
   $p^\alpha_{\rm V} = p^\alpha_0 = n^\alpha_0 p^\alpha_{{\rm R} 0}$, and $c^\alpha_\epsilon = c^\alpha_{\epsilon 0}$.
   Quantities with zero subscripts are constants;
   $p^\alpha_{\rm V} = - \frac{1}{3} {\rm tr} \bar{\bm \sigma}^\alpha$ is the partial inviscid pressure.
   From the identity $ \partial J^\alpha / \partial {\bf C}^\alpha = \frac{1}{2}J^\alpha
   ( {\bf C}^\alpha)^{-1}$, the stress contribution is spherical:
    $\bar{\bm \sigma}^\alpha = - p^\alpha_{\rm V} {\bf 1}$.
    The ideal gas constant is ${\mathfrak R}^\alpha$. Equation of state (EOS)
    and internal energy function are
    \begin{equation}
    \label{eq:idealeos}
    p^\alpha_{\rm V} = \rho^\alpha {\mathfrak R}^\alpha \theta^\alpha, \qquad
    u^\alpha = c^\alpha_{\epsilon 0} \theta^\alpha.
    \end{equation}
    From \eqref{eq:idealeos} and $\psi^\alpha = u^\alpha + \theta^\alpha (\partial \psi^\alpha / \partial \theta^\alpha)$, it follows that
    \begin{align}
    \label{eq:idealhelm}
    & \psi^\alpha(J^\alpha, \theta^\alpha) = - {\mathfrak R}^\alpha \theta^\alpha \ln J^\alpha
    - c^\alpha_{\epsilon 0} \theta^\alpha [ \ln (\theta^\alpha / \theta^\alpha_0) -1 ], \\
     \label{eq:idealu}
    & u^\alpha(J^\alpha, \eta^\alpha) = c^\alpha_{\epsilon 0} \theta^\alpha_0
    (J^\alpha)^{-\gamma^\alpha_0} \exp (\eta^\alpha / c^\alpha_{\epsilon 0}).
    \end{align}
    noting $\eta^\alpha_0 = 0$ and $\gamma^\alpha_ 0 = {\mathfrak R}^\alpha / c^\alpha_{\epsilon 0}$. Thermal stress tensor is ${\bm \beta}^\alpha = \rho^\alpha {\mathfrak R}^\alpha ({ \bf C}^\alpha)^{-1}$ in \eqref{eq:grunparam}, and $c^\alpha_\epsilon = c^\alpha_{\epsilon 0}$  in \eqref{eq:specht}.
    \\
    
    \noindent {\bf{Condensed matter EOS}.}
  For solid and liquid tissue phases, an EOS combining
  the third-order logarithmic form used
  for high-pressure physics 
  \cite{poirier1998,claytonIJES2014}
  with
  an exponential form for tissue
  mechanics \cite{claytonMOSM2020} is
  sufficiently general for the present applications.
  Thermoelastic coupling is linear and isotropic
  with constant volumetric expansion coefficient $A^\alpha$,
  and specific heat $c^\alpha_\epsilon$ is constant.
  Reference temperature is $\theta^\alpha_0$,
  and reference pressure is $p^\alpha_{{\rm R} 0}$.
  The reference isothermal bulk modulus is $B^\alpha_\theta$, and
  the pressure derivative of the isothermal bulk modulus in the
  reference state is $B^\alpha_{ \theta {\rm p}}$.
  Analogously, the isentropic bulk modulus and pressure
  derivative are $B^\alpha_\eta$ and $B^\alpha_{ \eta {\rm p}}$.
  Denote a constant controlling exponential stiffening
  by $k^\alpha_{\rm V}$.
  Free energies per unit initial volume of
  constituent $\alpha$ are
  \begin{align}
  \label{eq:EOSf1}
 &  \Psi^\alpha_{\rm V} =\frac{B^\alpha_{\theta}}{2}
 \left[ \frac {\exp \{ k^\alpha_{\rm V} (\ln J^\alpha)^2 \} - 1 } {k^\alpha_{\rm V}}
 - \frac{ (B^\alpha_{ \theta {\rm p}} - 2) (\ln J^\alpha)^3}{3} \right] \nonumber \\
 & \quad - A^\alpha B^\alpha_{\theta}(\theta^\alpha - \theta^\alpha_0)  \ln J^\alpha ,
 \quad \Psi^\alpha_\sigma = -p^\alpha_{{\rm R}0} \ln J^\alpha,
 \\
  \label{eq:EOSf2}
 & \Psi^\alpha_{\theta} = - \rho^\alpha_{ {\rm R} 0} c^\alpha_\epsilon [ \theta^\alpha \ln (\theta^\alpha / \theta^\alpha_0)
 - (\theta^\alpha - \theta^\alpha_0)].
  \end{align}
  The contribution to stress $\bar{\bm \sigma}^\alpha$ from $\Psi^\alpha_{\rm V}$
  and $\Psi^\alpha_\sigma$ is spherical, with Cauchy pressure
  \begin{align}
  \label{eq:pEOS}
&   p^\alpha_{ \rm V } = 
 - \frac { \rho^\alpha } { \rho^\alpha_{ {\rm R} 0}  }
  \frac{ \partial ( \Psi^\alpha_{\rm V} + \Psi^\alpha_\sigma)} {\partial \ln J^\alpha}
   \nonumber \\
  & =  - \frac { \rho^\alpha } { \rho^\alpha_{ {\rm R} 0}  }
  B^\alpha_\theta \ln J^\alpha [ \exp \{ k^\alpha_{\rm V} (\ln J^\alpha)^2 \}
  - {\textstyle{\frac{1}{2}}} (B^\alpha_{ \theta {\rm p}} - 2) \ln J^\alpha ] \nonumber \\
 &  \qquad + \frac { \rho^\alpha } { \rho^\alpha_{ {\rm R} 0}  }  A^\alpha B^\alpha_\theta
 (\theta^\alpha - \theta^\alpha_0)  + \frac { \rho^\alpha } { \rho^\alpha_{ {\rm R} 0}  } p^\alpha_{{\rm R}0} .
  \end{align}
  From \eqref{eq:massbal}, if $c^\alpha = \rho^\alpha \partial_t \ln \sqrt{g}$,
$\partial_t  \ln \sqrt{g} = D^\alpha_t  \ln \sqrt{G^\alpha}$,
and $\rho_0^\alpha = \rho_0^\alpha({\bf X}^\alpha)$,
 then $\rho^\alpha_0 = \rho^\alpha J^\alpha \Rightarrow 
 { \rho^\alpha } / { \rho^\alpha_{ {\rm R} 0}  } = n^\alpha_0 / J^\alpha$.
  The thermal stress tensor and Gr\"uneisen tensor are
  \begin{equation}
  \label{eq:betaEOS}
  {\bm \beta}^\alpha = \frac { \rho^\alpha } { \rho^\alpha_{ {\rm R} 0}  } A^\alpha B^\alpha_\theta 
  ({\bf C}^\alpha)^{-1},
  \quad
  {\bm \gamma}^\alpha = \frac{ A^\alpha B^\alpha_\theta }{ \rho^\alpha_{ {\rm R} 0} c^\alpha_\epsilon }   ({\bf C}^\alpha)^{-1}.
  \end{equation}
  The scalar Gr\"uneisen constant is $\gamma^\alpha_0 = 
  A^\alpha B^\alpha_\theta / ( \rho^\alpha_{ {\rm R} 0} c^\alpha_\epsilon )$.
  Internal energy complementary to \eqref{eq:EOSf1} and \eqref{eq:EOSf2} is
   \begin{align}
  \label{eq:EOSi1}
 &  U^\alpha_{\rm V} =\frac{B^\alpha_{\eta}}{2}
 \left[ \frac {\exp \{ k^\alpha_{\rm V} (\ln J^\alpha)^2 \} - 1 } {k^\alpha_{\rm V}}
 - \frac{ (B^\alpha_{ \eta {\rm p}} - 2) (\ln J^\alpha)^3}{3} \right] \nonumber \\
 & \qquad - \rho^\alpha_{ {\rm R} 0} \theta_0^\alpha \gamma_0^\alpha \eta^\alpha  \ln J^\alpha ,
 \quad U^\alpha_\sigma = -p^\alpha_{{\rm R}0} \ln J^\alpha,
 \\
  \label{eq:EOSi2}
 & U^\alpha_{\theta} = \rho^\alpha_{ {\rm R} 0} \theta^\alpha_0 \eta^\alpha
 [1 + \eta^\alpha / (2 c^\alpha_\epsilon)].
  \end{align}
  Pressure and temperature from \eqref{eq:EOSi1} and \eqref{eq:EOSi2} are
  \begin{align}
  \label{eq:pEOSi}
 &   p^\alpha_{ \rm V } = 
 - \frac { \rho^\alpha } { \rho^\alpha_{ {\rm R} 0}  }
  \frac{ \partial ( U^\alpha_{\rm V} + U^\alpha_\sigma)} {\partial \ln J^\alpha}
   \nonumber \\
  & =  - \frac { \rho^\alpha } { \rho^\alpha_{ {\rm R} 0}  }
  B^\alpha_\eta \ln J^\alpha [ \exp \{ k^\alpha_{\rm V} (\ln J^\alpha)^2 \}
  - {\textstyle{\frac{1}{2}}} (B^\alpha_{ \eta {\rm p}} - 2) \ln J^\alpha ] \nonumber \\
 &  \qquad \qquad + { \rho^\alpha } \theta^\alpha_0 \gamma^\alpha_0 \eta^\alpha + 
 \rho^\alpha p^\alpha_{{\rm R}0} /\rho^\alpha_{ {\rm R} 0}, 
 \end{align}
 \begin{align}
 &
 \theta^\alpha = \frac{1} {\rho^\alpha_{ {\rm R} 0} } \frac{ \partial (U^\alpha_{ \rm V} + U^\alpha_\theta)}{ \partial \eta^\alpha } = \theta_0^\alpha [ 1 
  + \frac{\eta^\alpha}{ c^\alpha_\epsilon} - \gamma_0^\alpha \ln J^\alpha ].
 \label{eq:tEOSi}
  \end{align}
 Bulk moduli $B^\alpha_\theta$ and $B^\alpha_\eta$ are nonnegative,
 and $k^\alpha_{\rm V}$ should be nonnegative for stiffening under large strain typical of soft tissues,
 compressive or tensile. If $B^\alpha_{\theta {\rm p}} > 2$, the material
 stiffens in compression and softens in tension, and vice-versa for $B^\alpha_{\theta {\rm p}} < 2$.
 Similar statements hold for $B^\alpha_{\eta {\rm p}}$.
 Energy functions \eqref{eq:EOSf1} and \eqref{eq:EOSi1} are not (poly)convex in $J^\alpha$.  Polyconvexity is appealing for existence of unique solutions to boundary value problems \cite{balzani2006} but is not essential.
 If $\varsigma^\alpha_{\rm V} < 1$, $p^\alpha_{\rm V}$ contributions from
 $\Psi^\alpha_{\rm V}$ and $U^\alpha_{\rm V}$ (i.e., all terms except rightmost in \eqref{eq:pEOS} and \eqref{eq:pEOSi} with $p^\alpha_{{\rm R} 0}$) require multiplication by $\varsigma^\alpha_{\rm V}$,
 as do ${\bm \beta}^\alpha$, ${\bm \gamma}^\alpha$, and $\gamma_0^\alpha$.
 \\
 
 \noindent {\bf{Deviatoric matrix equilibrium.}} Deviatoric deformation gradient and deformation tensor are, with  $f = f( \tilde{\bf C}^\alpha)$
 a generic differentiable function of its argument,
 \begin{align}
 \label{eq:devF}
 & \tilde{\bf F}^\alpha = (J^\alpha)^{-1/3} {\bf F}^\alpha, \qquad
 \tilde{\bf C}^\alpha = (J^\alpha)^{-2/3}{\bf C}^\alpha, \\
 \label{eq:devF2}
&  \frac{\partial f }{ \partial {\bf C}^\alpha} = 
(J^\alpha)^{-2/3} \left[ \frac {\partial f }{ \partial \tilde{\bf C}^\alpha }
- \frac{1}{3} \left( \frac {\partial f} { \partial \tilde{\bf C}^\alpha }: {\bf C}^\alpha \right)
({\bf C}^\alpha)^{-1} \right].
 \end{align}
 Let $\mu^\alpha_{\rm S} \geq 0$ be a shear modulus. Energy is \cite{balzani2006}
 \begin{equation}
 \label{eq:PsiUdev}
 \Psi^\alpha_{\rm S} = U^\alpha_{\rm S} = {\textstyle{\frac{1}{2}}} \mu^\alpha_{\rm S} ({\rm tr} \,\tilde{\bf C}^\alpha - 3) .
 \end{equation}
 From \eqref{eq:devF2}, the contribution of \eqref{eq:PsiUdev} to Cauchy stress is
 \begin{align}
 \label{eq:devS}
& {\bm \sigma}^\alpha_{\rm S} = 2 \frac{\rho^\alpha}{\rho^\alpha_{ {\rm R} 0} } {\bf F}^\alpha 
 \frac{\partial \Psi^\alpha_{\rm S} }{\partial {\bf C}^\alpha } ({\bf F}^\alpha)^{\mathsf{T}}
 = 2 \frac{\rho^\alpha}{\rho^\alpha_{ {\rm R} 0} } {\bf F}^\alpha 
 \frac{\partial U^\alpha_{\rm S} }{\partial {\bf C}^\alpha } ({\bf F}^\alpha)^{\mathsf{T}} \nonumber
 \\
 & \quad = \frac{\rho^\alpha}{\rho^\alpha_{ {\rm R} 0} } \mu^\alpha_{\rm S}
 [ \tilde{\bf B}^\alpha - {\textstyle{\frac{1}{3}}} ({\rm tr} \tilde{\bf B}^\alpha) {\bf 1} ],
 \quad 
  \tilde{\bf B}^\alpha =  \tilde{\bf F}^\alpha ( \tilde{\bf F}^\alpha)^{\mathsf{T}}.
  \end{align}
  This contribution is linear in spatial deformation tensor
  $\tilde{\bf B}^\alpha$, traceless, and ultimately scaled by $\varsigma^\alpha_{\rm S}$.
  Further nonlinearity can be furnished by $\Psi^\alpha_{\rm F}$ and $U^\alpha_{\rm F}$.
  Function \eqref{eq:PsiUdev} is polyconvex \cite{balzani2006} and isotropic.
  \\
 
 \noindent {\bf{Fiber equilibrium.}} Let index $k$ denote
 a fiber family of reference alignment by unit vector ${\bm \iota}^\alpha_k$.
 Let $\kappa^\alpha_k \in [0,\frac{1}{3}]$ be dispersion constants.
 Structure tensors \cite{gasser2006} are
 \begin{equation}
 \label{eq:structens}
 {\bf H}_k^\alpha = \kappa^\alpha_k {\bf 1} + (1-3 \kappa^\alpha_k) {\bm \iota}^\alpha_k \otimes {\bm \iota}^\alpha_k.
 \end{equation}
 Strain energy contributions are of functional forms
 \begin{align}
 \label{eq:PsiF1}
 \Psi^\alpha_{\rm F} = \Psi^\alpha_{\rm F}( \tilde{\bf C}^\alpha, {\bf H}_k^\alpha ({\bf X}^\alpha))
 = U^\alpha_{\rm F}( \tilde{\bf C}^\alpha, {\bf H}_k^\alpha ({\bf X}^\alpha)),
 \end{align}
 with ${\bf H}_k^\alpha$ time-independent at ${\bf X}^\alpha$ (i.e., not transient state variables). Depending on the number of fiber families $k$ and their orientations,
 different scalar invariants entering \eqref{eq:PsiF1} are
 possible. For the current presentation, one invariant per fiber family
 is sufficient: $I^\alpha_k = \tilde{\bf C}^\alpha:{\bf H}^\alpha_k$.
 The particular form of \eqref{eq:PsiF1} is polyconvex \cite{holz2000,balzani2006,holz2009,claytonSYMM2023}:
 \begin{align}
 \label{eq:PsiFexp}
& \Psi^\alpha_{\rm F} = \sum_k \Psi^\alpha_{{\rm F}k} \nonumber \\ & \quad =
 \sum_k \frac{\mu^\alpha_k}{4 k^\alpha_k}
 \{ \exp[k_k^\alpha (I^\alpha_k -1)^2] -1\} {\rm H}(I^\alpha_k - 1).
 \end{align}
 A fiber modulus and stiffening coefficient are $\mu^\alpha_k \geq 0$ and
 $k^\alpha_k > 0$. Optional right-continuous Heaviside function is ${\rm H}(\cdot)$; this disables 
  fiber stiffness for buckling in compression along ${\bm \iota}^\alpha_k$.
  Contributions to stress are traceless:
  \begin{align}
  \label{eq:PsiFsig}
  & {\bm \sigma}^\alpha_{\rm F} = 2 \frac{\rho^\alpha}{\rho^\alpha_{ {\rm R} 0} } {\bf F}^\alpha 
 \frac{\partial \Psi^\alpha_{\rm F} }{\partial {\bf C}^\alpha } ({\bf F}^\alpha)^{\mathsf{T}}
 = 2 \frac{\rho^\alpha}{\rho^\alpha_{ {\rm R} 0} } {\bf F}^\alpha 
 \frac{\partial U^\alpha_{\rm F} }{\partial {\bf C}^\alpha } ({\bf F}^\alpha)^{\mathsf{T}} \nonumber
 \\
 & \quad = \frac{\rho^\alpha}{\rho^\alpha_{ {\rm R} 0} } \sum_k \mu^\alpha_k
 (I^\alpha_k -1) \exp[k_k^\alpha (I^\alpha_k -1)^2] {\rm H}(I^\alpha_k - 1)  \tilde{\bf h}^\alpha_k , \nonumber 
 \\
 &  \tilde{\bf h}^\alpha_k = \tilde{\bf F}^\alpha {\bf H}^\alpha_k (\tilde{\bf F}^\alpha)^{\mathsf T}
 -{\textstyle{\frac{1}{3}}} {\rm tr} [\tilde{\bf F}^\alpha {\bf H}^\alpha_k (\tilde{\bf F}^\alpha)^{\mathsf T}] {\bf 1}.
  \end{align}
 Family $k$ is isotropic as  $\kappa_k^\alpha \rightarrow \frac{1}{3}$.  Fiber compressibility
 is encompassed by the condensed matter EOS for phase $\alpha$ rather than distinct energetic terms.
 Stress contributions from \eqref{eq:PsiFsig} are affected by $\varsigma^\alpha_{\rm F}$ if fibers are damaged.
 \\
 
\noindent {\bf Matrix viscoelasticity.} 
The viscoelastic formulation combines features from
prior works \cite{holz1996,holz1996b,holz2002,claytonMOSM2020,claytonACTA2020}
in a thermodynamically consistent manner.
Let $\{ {\bm \Gamma}^\alpha \} \rightarrow \{ {\bm \Gamma}^\alpha_{{\rm V} l},
{\bm \Gamma}^\alpha_{{\rm S} m}, {\bm \Gamma}^\alpha_{ {\Phi} k,n} \}$
be internal strain-like configurational variables for constituent $\alpha$.
Index $l$ spans a set of discrete relaxation time constants $\tau^\alpha_{ {\rm V} l} = \tau^\alpha_{ {\rm V} 1}, \ldots$
for viscoelastic relaxation processes associated with volumetric deformation of
the matrix. Index $m$ spans times $\tau^\alpha_{ {\rm S} m}$ associated with deviatoric (shear)
deformation of the matrix. 
Index $n$ spans times $\tau^\alpha_{ {\Phi} k,n}$ associated with fiber family $k$
 discussed in the next subsection.
 Internal stresses $\{ {\bf Q}^\alpha_{ {\rm V} l },{\bf Q}^\alpha_{ {\rm S} m } \}$
 conjugate to the matrix internal strains, in coordinates
 referred to ${\mathfrak M}^\alpha$, obey \cite{holz1996b,holz2002}
 \begin{align}
 \label{eq:Q1}
 & {\bf Q}^\alpha_{ {\rm V} l } = 
 - \frac{ \partial \Psi^\alpha_{\Gamma} }{\partial {\bm \Gamma}^\alpha_{{\rm V} l} }
 = 2 \frac{\partial \Psi^\alpha_{ {\rm V} l}} {\partial {\bf C}^\alpha},
 \quad
 {\bf Q}^\alpha_{ {\rm S} m } = 
 - \frac{ \partial \Psi^\alpha_{\Gamma} }{\partial {\bm \Gamma}^\alpha_{{\rm S} m} }
 = 2 \frac{\partial \Psi^\alpha_{ {\rm S} m}} {\partial {\bf C}^\alpha},
 \end{align}
 \begin{align}
 \label{eq:Q2}
 & \Psi^\alpha_{\Gamma} = U^\alpha_{\Gamma} = \sum_l \Psi^\alpha_{ {\rm V} l} ({\bm \Gamma}^\alpha_{{\rm V} l},{\bf C}^\alpha)
 + \sum_m \Psi^\alpha_{ {\rm S} m} ({\bm \Gamma}^\alpha_{{\rm S} m},{\bf C}^\alpha)
\nonumber 
\\
 & \quad =  \sum_l \int {\textstyle{\frac{1}{2}}} {\bf Q}^\alpha_{ {\rm V} l }:{\rm d} {\bf C}^\alpha
 + 
  \sum_m \int {\textstyle{\frac{1}{2}}}  {\bf Q}^\alpha_{ {\rm S} m }:{\rm d} {\bf C}^\alpha.
 \end{align}
Indefinite integrals \eqref{eq:Q2} are not needed explicitly.
Evolution equations for internal stresses are
\begin{align}
 \label{eq:Q3}
 & D^\alpha_t {\bf Q}^\alpha_{ {\rm V} l } + {\bf Q}^\alpha_{ {\rm V} l } / \tau^\alpha_{ {\rm V} l }
 = 2 D^\alpha_t (\partial \hat{\Psi}^\alpha_{ {\rm V}l} / \partial {\bf C}^\alpha), \\
 \label{eq:Q4}
 & 
 D^\alpha_t {\bf Q}^\alpha_{ {\rm S} m } + {\bf Q}^\alpha_{ {\rm S} m } / \tau^\alpha_{ {\rm S} m }
 = 2 D^\alpha_t (\partial \hat{\Psi}^\alpha_{ {\rm S}m} / \partial {\bf C}^\alpha), \\
 \label{eq:Q5}
& \hat{\Psi}^\alpha_{ {\rm V} l} = {\textstyle{\frac{1}{2}}} \beta^\alpha_{ {\rm V} l} B^\alpha_\theta (\ln J^\alpha)^2, 
\\
 \label{eq:Q6}
& \hat{\Psi}^\alpha_{ {\rm S} m} = {\textstyle{\frac{1}{2}}} \beta^\alpha_{ {\rm S} m}
 \mu^\alpha_{\rm S} ({\rm tr} \,\tilde{\bf C}^\alpha - 3).
 \end{align}
Dimensionless factors are $\beta^\alpha_{ {\rm V} l} \geq 0,\beta^\alpha_{ {\rm S} m} \geq 0$.
Initial conditions and convolution solutions to \eqref{eq:Q3} and \eqref{eq:Q4} are
\begin{align}
\label{eq:Qics}
& {\bf Q}^\alpha_{ {\rm V} l 0} = 2 \partial \hat{\Psi}^\alpha_{ {\rm V}l} / \partial {\bf C}^\alpha, \quad
{\bf Q}^\alpha_{ {\rm S} m 0} = 2 \partial \hat{\Psi}^\alpha_{ {\rm S}m} / \partial {\bf C}^\alpha, \\
\label{eq:convint1}
&  {\bf Q}^\alpha_{ {\rm V} l }(t) = 
 {\bf Q}^\alpha_{ {\rm V} l 0} \exp [ {-t}/{ \tau^\alpha_{ {\rm V} l}} ]
  \nonumber \\ & \qquad + 
 \int_{0+}^t \exp [ {-(t-s)}/{ \tau^\alpha_{ {\rm V} l}} ]
 D^\alpha_s (2 \partial \hat{\Psi}^\alpha_{ {\rm V} l} / \partial {\bf C}^\alpha)  {\rm d} s,
\\
\label{eq:convint2}
& {\bf Q}^\alpha_{ {\rm S} m }(t) = 
 {\bf Q}^\alpha_{ {\rm S} m 0} \exp [ {-t}/{ \tau^\alpha_{ {\rm S} m}} ]
 \nonumber \\ & \qquad + 
 \int_{0+}^t \exp [ {-(t-s)}/{ \tau^\alpha_{ {\rm S} m}} ]
 D^\alpha_s (2 \partial \hat{\Psi}^\alpha_{ {\rm S}m} / \partial {\bf C}^\alpha)  {\rm d} s.
\end{align}
Cauchy stress contributions are sums over $l,m$:
\begin{align}
\label{eq:QC1}
& {\bm \sigma}^\alpha_{ \Gamma} = 2 \frac{\rho^\alpha}{\rho^\alpha_{ {\rm R} 0} } {\bf F}^\alpha 
 \frac{\partial \Psi^\alpha_{\Gamma} }{\partial {\bf C}^\alpha } ({\bf F}^\alpha)^{\mathsf{T}}
 = 2 \frac{\rho^\alpha}{\rho^\alpha_{ {\rm R} 0} } {\bf F}^\alpha 
 \frac{\partial U^\alpha_{\Gamma} }{\partial {\bf C}^\alpha } ({\bf F}^\alpha)^{\mathsf{T}} \nonumber
 \\
&  = \frac{\rho^\alpha}{\rho^\alpha_{ {\rm R} 0} } \sum_l {\bf F}^\alpha {\bf Q}^\alpha_{ {\rm V} l} 
 ({\bf F}^\alpha)^{\mathsf{T}} 
 + 
  \frac{\rho^\alpha}{\rho^\alpha_{ {\rm R} 0} } \sum_m {\bf F}^\alpha {\bf Q}^\alpha_{ {\rm S} m} 
 ({\bf F}^\alpha)^{\mathsf{T}}. 
 \end{align}
As $t / \tau^\alpha_{{\rm V} l} \rightarrow 0$ and $t / \tau^\alpha_{{\rm S} m} \rightarrow 0$, 
$ {\bm \sigma}^\alpha_{\Gamma}$ in \eqref{eq:QC1} sums to the instantaneous (glassy) viscoelastic stresses
\begin{align}
\label{eq:QC2}
 &
  2 \frac{\rho^\alpha}{\rho^\alpha_{ {\rm R} 0} } \sum_l {\bf F}^\alpha \frac{ \partial \hat{\Psi}^\alpha_{{ \rm V} l} }{\partial {\bf C}^\alpha}  ({\bf F}^\alpha)^{\mathsf{T}}
 +  
  2 \frac{\rho^\alpha}{\rho^\alpha_{ {\rm R} 0} } \sum_m {\bf F}^\alpha \frac{ \partial \hat{\Psi}^\alpha_{ {\rm S} m} }{\partial {\bf C}^\alpha}  ({\bf F}^\alpha)^{\mathsf{T}} \nonumber
  \\ &  =
  \frac{\rho^\alpha}{\rho^\alpha_{ {\rm R} 0} } \sum_l \beta^\alpha_{ {\rm V} l} B^\alpha_\theta
 ( \ln J^\alpha)  {\bf 1} \nonumber
 \\ &
  \qquad + \frac{\rho^\alpha}{\rho^\alpha_{ {\rm R} 0} } \sum_m \beta^\alpha_{{ \rm S} m} 
 \mu^\alpha_{ {\rm S} } 
  [ \tilde{\bf B}^\alpha - \frac{1}{3} ({\rm tr} \tilde{\bf B}^\alpha) {\bf 1} ].
 \end {align}
 As $t / \tau^\alpha_{\rm{V} l} \rightarrow \infty$ and $t / \tau^\alpha_{\rm{S} m} \rightarrow \infty$, 
$ {\bf Q}^\alpha_{ {\rm V} l} \rightarrow {\bf 0} $ and $ {\bf Q}^\alpha_{ {\rm S} m} \rightarrow {\bf 0}$ so that ${\bm \sigma}^\alpha_{ \Gamma} \rightarrow {\bf 0}$ in \eqref{eq:QC1} for  relaxed equilibrium response.
\\

\noindent {\bf Fiber viscoelasticity.} Dissipative response of fiber families $k = 1, \ldots$ is dictated by internal variables ${\bm \Gamma}^\alpha_{ {\Phi} k,n } $ each with $n = 1, \ldots$ relaxation times $\tau^\alpha_{ {\Phi} k,n }$ and conjugate internal stresses ${\bf Q}^\alpha_{ {\Phi} k, n }$. Internal stresses and energies are
 \begin{align}
 \label{eq:Q1f}
 & {\bf Q}^\alpha_{ {\Phi} k,n } = 
 - { \partial \Psi^\alpha_{\Phi} } / { \partial {\bm \Gamma}^\alpha_{ {\Phi} k,n }}
 = 2 {\partial \Psi^\alpha_{ {\Phi} k,n }} / {\partial {\bf C}^\alpha},
 \end{align}
 \begin{align}
 \label{eq:Q2f}
 & \Psi^\alpha_{\Phi} = U^\alpha_{\Phi} 
 = \sum_k \Psi^\alpha_{ {\Phi} k } 
=  \sum_k \sum_n \Psi^\alpha_{ {\Phi} k,n } ({\bm \Gamma}^\alpha_{ {\Phi} k,n } ,{\bf C}^\alpha)
\nonumber 
\\
& 
\qquad \qquad =  \sum_k \sum_n \int {\textstyle{\frac{1}{2}}} {\bf Q}^\alpha_{ {\Phi} k,n }:{\rm d} {\bf C}^\alpha.
 \end{align}
 Evolution equations and stored viscoelastic energies are
 \begin{align}
 \label{eq:Q3f}
 & D^\alpha_t {\bf Q}^\alpha_{ {\Phi} k,n } + {\bf Q}^\alpha_{ {\Phi} k,n } / \tau^\alpha_{ {\Phi} k,n }
 = 2 D^\alpha_t (\partial \hat{\Psi}^\alpha_{ {\Phi} k, n } / \partial {\bf C}^\alpha), \\
 \label{eq:Q5f}
& \hat{\Psi}^\alpha_{ {\Phi} k,n } =  
\frac{\beta^\alpha_{ {\Phi} k,n }  \mu^\alpha_k}{4 k^\alpha_k}
 \{ \exp[k_k^\alpha (I^\alpha_k -1)^2] -1\} {\rm H}(I^\alpha_k - 1),
 \end{align}
 with $\beta^\alpha_{ {\Phi} k,n } \geq 0 $. Initial conditions and solutions (convolution integrals), followed by Cauchy stress terms, are
 \begin{align}
\label{eq:Qicsf}
& {\bf Q}^\alpha_{ {\Phi} k,n 0}  = 2 \partial \hat{\Psi}^\alpha_{ {\Phi} k,n } / \partial {\bf C}^\alpha,  \\
\label{eq:convint1f}
&  {\bf Q}^\alpha_{ {\Phi} k,n } (t) = 
 {\bf Q}^\alpha_{ {\Phi} k,n 0}   \exp [ {-t}/{ \tau^\alpha_{ {\Phi} k,n }}  ]
  \nonumber \\ & \qquad + 
 \int_{0+}^t \exp [ {-(t-s)}/{ \tau^\alpha_{ {\Phi} k,n }}  ]
 D^\alpha_s (2 \partial \hat{\Psi}^\alpha_{ {\Phi} k,n }  / \partial {\bf C}^\alpha)  {\rm d} s ,
 \\
\label{eq:QC1f}
& {\bm \sigma}^\alpha_{\Phi} = 2 \frac{\rho^\alpha}{\rho^\alpha_{ {\rm R} 0} } {\bf F}^\alpha 
 \frac{\partial \Psi^\alpha_{\Phi} }{\partial {\bf C}^\alpha } ({\bf F}^\alpha)^{\mathsf{T}}
 = 2 \frac{\rho^\alpha}{\rho^\alpha_{ {\rm R} 0} } {\bf F}^\alpha 
 \frac{\partial U^\alpha_{\Phi} }{\partial {\bf C}^\alpha } ({\bf F}^\alpha)^{\mathsf{T}} \nonumber
 \\
&  = \frac{\rho^\alpha}{\rho^\alpha_{ {\rm R} 0} } \sum_k \sum_n {\bf F}^\alpha
 {\bf Q}^\alpha_{ {\Phi} k,n }
 ({\bf F}^\alpha)^{\mathsf{T}} .
 \end{align}
 As $t / \tau^\alpha_{ {\Phi} k,n } \rightarrow 0$,
$ {\bm \sigma}^\alpha_{\Phi}$ in \eqref{eq:QC1f} becomes the glassy stress
\begin{align}
\label{eq:QC2f}
 &
  2 \frac{\rho^\alpha}{\rho^\alpha_{ {\rm R} 0} } \sum_k \sum_n {\bf F}^\alpha \frac{ \partial \hat{\Psi}^\alpha_{ {\Phi} k,n }}{\partial {\bf C}^\alpha}  ({\bf F}^\alpha)^{\mathsf{T}}
  =  \frac{\rho^\alpha}{\rho^\alpha_{ {\rm R} 0} } \sum_k \sum_n \{ \beta^\alpha_{ {\Phi} k,n } 
  \nonumber
  \\ &  \quad \times
  \mu^\alpha_k
 (I^\alpha_k -1) \exp[k_k^\alpha (I^\alpha_k -1)^2] {\rm H}(I^\alpha_k - 1)  \tilde{\bf h}^\alpha_k  \}.
 \end {align}
  As $t / \tau^\alpha_{ {\Phi} k,n } \rightarrow \infty$, $ {\bf Q}^\alpha_{ {\Phi} k,n }  \rightarrow {\bf 0} $ leading to ${\bm \sigma}^\alpha_{ \Phi} \rightarrow {\bf 0}$ in \eqref{eq:QC1f} .
\\

\noindent {\bf Active tension.} Electrochemistry of soft tissue cellular activation
\cite{stalhand2016} is beyond the present scope.
A phenomenological approach is instead, generalizing
other continuum models \cite{martins1998,ito2010,franchini2022}.
Let $\{ {\bm \Delta}^\alpha \} \rightarrow \{ \Delta^\alpha_k \} ({\bf X}^\alpha,t)$
be a set of scalar internal variables associated with potentially
active fiber families $k$ in phase $\alpha$.
These variables can include internal strains in contractile elements
and time-dependent switching functions.
Define the fiber orientation tensors ${\bf H}^\alpha_k$ as in \eqref{eq:structens}.
In many models, cells are fully aligned such that $k = 1$ and $\kappa^\alpha_1 = 0$
\cite{martins1998,ito2010},
but this is inessential \cite{franchini2022}.
Stretch in the fiber direction is $\lambda^\alpha_k = \sqrt{I^\alpha_k}$.
A generic energy function is 
\begin{equation}
\label{eq:Psiact}
\Psi^\alpha_{\rm A} = U^\alpha_{\rm A} = \sum_k \Psi^\alpha_{{\rm A} k} =
\sum_k [ \Lambda^\alpha_k (\lambda^\alpha_k, \{ \Delta^\alpha_k \} ) + \chi^\alpha_k (\{ \Delta^\alpha_k \}) ].
\end{equation}
Strain energy functions $\Lambda^\alpha_k$ furnish active stress terms
 \begin{align}
  \label{eq:PsiAsig}
  & {\bm \sigma}^\alpha_{\rm A} = 2 \frac{\rho^\alpha}{\rho^\alpha_{ {\rm R} 0} } {\bf F}^\alpha 
 \frac{\partial \Psi^\alpha_{\rm A} }{\partial {\bf C}^\alpha } ({\bf F}^\alpha)^{\mathsf{T}}
 = \frac{\rho^\alpha}{\rho^\alpha_{ {\rm R} 0} } \sum_k  
 \frac{1}{\lambda^\alpha_k} \frac{\partial \Lambda^\alpha_k}{\partial \lambda^\alpha_k}
 \tilde{\bf h}^\alpha_k.
  \end{align}
  Energy functions $\chi^\alpha_k$ ensure nonnegative net dissipation.
  Stresses ${\bm \sigma}^\alpha_{\rm A}$ should vanish for passive conditions.
\\

\noindent {\bf Damage.} 
Internal variables $\{ {\bf D}^\alpha \}
\rightarrow \{ \bar{D}^\alpha, D^\alpha_k \}$. Scalar damage
measures in the isotropic matrix are  $\bar{D}^\alpha \in [0,1]$,
and $D^\alpha_k \in [0,1]$ are scalar functions
for each fiber family $k$.
All are akin to order parameters in
phase-field fracture theory \cite{claytonIJF2014,gultekin2019}.
Degradation functions in
\eqref{eq:degrade}--\eqref{eq:Utot} are, with
$\bar{\vartheta}^\alpha \in [0,\infty), \vartheta^\alpha_k \in [0,\infty)$ constants,
\begin{align}
\label{eq:degf1}
& \varsigma^\alpha_{ \rm{V}} = [1 - \bar{D}^\alpha {\rm H}( \ln J^\alpha)]^{\bar{\vartheta}^\alpha} ,
\quad \varsigma^\alpha_{ \rm{S}} = (1 - \bar{D}^\alpha)^{\bar{\vartheta}^\alpha},
\\
\label{eq:degf2}
& \varsigma^\alpha_{\rm F} \circ (\cdot) 
=  \varsigma^\alpha_{\rm F} \circ \sum_k (\cdot)_k  \nonumber  = 
\sum_k \varsigma^\alpha_{{\rm F} k} (\cdot)_k
 = \sum_k (1 - D^\alpha_k)^{\vartheta^\alpha_k} (\cdot)_k,
 \\ 
& \varsigma^\alpha_{{\rm F} k} = (1 - D^\alpha_k)^{\vartheta^\alpha_k}.
\end{align}
The Heaviside function in $\varsigma^\alpha_{\rm V}$ prevents degradation in
compression so the bulk modulus is maintained \cite{claytonIJF2014,claytonJMPS2021}.
Operator $\varsigma^\alpha_{\rm F}$ in \eqref{eq:Psitot} and \eqref{eq:Utot} is applied to a sum of energetic contributions (hyperelastic, viscoelastic, and active)
over all families $k$ via \eqref{eq:degf2}.
Partial stress less viscous stress of constituent $\alpha$ is
$\bar{\bm \sigma}^\alpha = {\bm \sigma}^\alpha - \hat{\bm \sigma}^\alpha$.
For an ideal gas, $\bar{D}^\alpha = \bar{ \vartheta}^\alpha = 0 \rightarrow \varsigma^\alpha_{\rm V} = 1$, so $\bar{\bm \sigma}^\alpha = - \rho^\alpha {\mathfrak R}^\alpha \theta^\alpha {\bf 1}$
by \eqref{eq:idealeos}.
For solid and liquid $\alpha$, applying \eqref{eq:Psitot}, this stress is the sum
of \eqref{eq:pEOS}, \eqref{eq:devS}, \eqref{eq:PsiFsig}, 
\eqref{eq:QC1}, \eqref{eq:QC1f}, and \eqref{eq:PsiAsig} scaled by one or more functions in \eqref{eq:degf1} and \eqref{eq:degf2}:
\begin{align}
\label{eq:totstatstress}
& \bar{\bm \sigma}^\alpha = \frac { \rho^\alpha } { \rho^\alpha_{ {\rm R} 0}  }
[1 - \bar{D}^\alpha {\rm H}( \ln J^\alpha)]^{\bar{\vartheta}^\alpha} B^\alpha_\theta \nonumber
\\ &
  \qquad \quad \times \{ \ln J^\alpha [ \exp \{ k^\alpha_{\rm V} (\ln J^\alpha)^2 \}
  - {\textstyle{\frac{1}{2}}} (B^\alpha_{ \theta {\rm p}} - 2) \ln J^\alpha ] \nonumber \\
 & \qquad  \qquad \quad - A^\alpha (\theta^\alpha - \theta^\alpha_0) \} {\bf 1} - ( { \rho^\alpha }  / { \rho^\alpha_{ {\rm R} 0})  p^\alpha_{{\rm R}0}  }  {\bf 1} \nonumber
 \\ & \nonumber
\qquad + \frac { \rho^\alpha } { \rho^\alpha_{ {\rm R} 0}  } (1 - \bar{D}^\alpha)^{\bar{\vartheta}^\alpha}
\mu^\alpha_{\rm S}
 [ \tilde{\bf B}^\alpha - {\textstyle{\frac{1}{3}}} ({\rm tr} \tilde{\bf B}^\alpha) {\bf 1} ] \\
\nonumber 
&
\qquad + \frac { \rho^\alpha } { \rho^\alpha_{ {\rm R} 0}  } 
\sum_k (1 - {D}^\alpha_k )^{\vartheta^\alpha_k} \mu^\alpha_k
 (I^\alpha_k -1) \\
 \nonumber
 & \qquad \qquad \qquad \qquad  \times 
  \exp[k_k^\alpha (I^\alpha_k -1)^2] {\rm H}(I^\alpha_k - 1)  \tilde{\bf h}^\alpha_k
\\
& \nonumber
\qquad + \frac { \rho^\alpha } { \rho^\alpha_{ {\rm R} 0}  } (1 - \bar{D}^\alpha)^{\bar{\vartheta}^\alpha}
 \sum_l {\bf F}^\alpha {\bf Q}^\alpha_{ {\rm V} l} 
 ({\bf F}^\alpha)^{\mathsf{T}} 
 \\
& \nonumber
\qquad + \frac { \rho^\alpha } { \rho^\alpha_{ {\rm R} 0}  } (1 - \bar{D}^\alpha)^{\bar{\vartheta}^\alpha}
\sum_m {\bf F}^\alpha {\bf Q}^\alpha_{ {\rm S} m} 
 ({\bf F}^\alpha)^{\mathsf{T}}
 \\
 & \nonumber
\qquad + \frac { \rho^\alpha } { \rho^\alpha_{ {\rm R} 0}  } 
\sum_k  [(1 - {D}^\alpha_k )^{\vartheta^\alpha_k} 
 \sum_n {\bf F}^\alpha
 {\bf Q}^\alpha_{ {\Phi} k,n }
 ({\bf F}^\alpha)^{\mathsf{T}} ]
\\
& 
\qquad + \frac { \rho^\alpha } { \rho^\alpha_{ {\rm R} 0}  } 
\sum_k  
 \frac{1}{\lambda^\alpha_k} \frac{\partial \Lambda^\alpha_k}{\partial \lambda^\alpha_k}
 \tilde{\bf h}^\alpha_k .
 \end{align}
 To preclude damage induced by normal muscle contraction, the final term is decoupled from $\{ {\bf D}^\alpha\}$
 consistently with \eqref{eq:Psitot} and \eqref{eq:Utot}.
If \eqref{eq:EOSi1} is used instead of \eqref{eq:EOSf1}, spherical terms on the right in \eqref{eq:totstatstress} should
appeal to \eqref{eq:pEOSi} rather than \eqref{eq:pEOS}.

Let $\Psi^\alpha_{\rm D} = U^\alpha_{\rm D}$ comprise cohesive and surface energies
of fracture per unit referential volume scaled by contributions of dimensionless 
Finsler-type metric $\hat{\bf G}^\alpha$ in \eqref{eq:metrics4} \cite{claytonZAMP2017,claytonSYMM2023} where $\hat{G}^\alpha = \hat{G}^\alpha (\{ \bm{\xi}^\alpha \}, {\bf X}^\alpha, {\bf x})$ via \eqref{eq:isvrelate}.
As in phase-field theory,
quadratic forms for matrix and fiber $k$ contributions are
\begin{align}
\label{eq:fracen}
& \hat{\Psi}^\alpha_{\rm D} = \Psi^\alpha_{\rm D} / \sqrt{\hat{G}^\alpha} = 
\bar{E}^\alpha_{\rm C} | \bar{D}^\alpha|^2 + \bar{\Upsilon}^\alpha \bar{l}_{\rm R}^\alpha | \nabla^\alpha_0 \bar{D}^\alpha |^2 \nonumber
\\ 
& \qquad \qquad \quad + \sum_k [ E^\alpha_{{\rm C} k} |D^\alpha_k|^2
+ {\Upsilon}^\alpha_k {l}^\alpha_{{\rm R} k} | \nabla^\alpha_0 {D}^\alpha_k |^2].
\end{align}
In \eqref{eq:fracen}, cohesive energies per unit volume are $\bar{E}^\alpha_{\rm C}$ and $E^\alpha_{{\rm C} k}$, surface energies are $\bar{\Upsilon}^\alpha$ and $\Upsilon^\alpha_k$, and gradient
regularization lengths are $\bar{l}^\alpha_{\rm R} = \bar{\alpha}^\alpha \bar{l}^\alpha$ and 
${l}^\alpha_{{\rm R} k} = \alpha^\alpha_k l^\alpha_k$, all non-negative constants. 
For solids, typically 
$\bar{E}^\alpha_{\rm C} =  \bar{\Upsilon}^\alpha / \bar{l}^\alpha$
and $E^\alpha_{{\rm C} k} = {\Upsilon}^\alpha_k / {l}^\alpha_k$.
Dimensionless factors $ \bar{\alpha}^\alpha \in [0,\infty)$ and $\alpha^\alpha_k l^\alpha_k \in [0,\infty)$ allow independent cohesive energies and gradient regularization lengths
\cite{claytonJMPS2021}.
For cavitation of a fluid, gradient terms can be dropped \cite{levitas2011} (i.e., $\bar{\alpha}^\alpha \rightarrow 0$); cohesion energy $\bar{E}^\alpha_{{\rm C}}$ will capture fracture of the fluid for tensile pressure. 
Isotropic surface energies are assumed for gradient terms of \eqref{eq:fracen}. These could be
extended to anisotropic contributions \cite{gultekin2019} if data exist.
However, $\{ D^\alpha_k \}$ furnish stress-damage anisotropy regardless.

%% file: s3b.tex
\subsection{\label{sec3c}Kinetics}

\noindent {\bf Viscous stress.} Isotropic Newtonian behavior
is usually adequate for each constituent $\alpha$,
with $\hat{B}^\alpha(\theta^\alpha) \geq 0$ and
$\hat{\mu}^\alpha (\theta^\alpha) \geq 0$ possibly temperature-dependent
bulk and shear viscosities.   Viscous stresses and dissipation are
\begin{align}
\label{eq:viscstress1}
& \hat{\bm \sigma}^\alpha = [\hat{B}^\alpha - {\textstyle{\frac{2}{3}}} \hat{\mu}^\alpha]{\rm tr} ({\bf d}^\alpha) {\bf 1} + 2 \hat{\mu}^\alpha {\bf d}^\alpha \\ 
\label{eq:viscsdissi}
& \hat{\mathfrak D}^\alpha =  \hat {\bm \sigma}^\alpha : {\bf d}^\alpha \nonumber
\\ & \quad \, \, = \hat{B}^\alpha |{\rm tr} ({\bf d}^\alpha) |^2 + 
2 \hat{\mu}^\alpha | {\bf d}^\alpha -  {\textstyle{\frac{1}{3}}} {\rm tr} ({\bf d}^\alpha) {\bf 1} |^2 \geq 0.
\end{align}
Viscous pressure and shear are $\hat{p}^\alpha = - \hat{B}^\alpha \nabla \cdot {\bm \upsilon}^\alpha$
and $\hat{\bm \sigma}^\alpha_{\rm S}$, whereby $\hat{\bm \sigma}^\alpha =  -\hat{p}^\alpha {\bf 1} + \hat{\bm \sigma}^\alpha_{\rm S}$.
\\

\noindent {\bf Viscoelasticity.} Viscoelastic
internal state variables
are the subset of $\{ {\bm \xi}^\alpha \}$
consisting of $\{ {\bm \Gamma}^\alpha_{ {\rm V} l}, {\bm \Gamma}^\alpha_{ {\rm S} m}, 
 {\bm \Gamma}^\alpha_{ {\Phi} k,n}\}$.
 Conjugate forces entering \eqref{eq:dissi} are a subset of $\{ {\bm \pi}^\alpha \}$:
 \begin{equation}
 \label{eq:pivisc}
 \begin{split}
  {\bm \pi}^\alpha_{ {\rm V} l } & =  - \varsigma_{\rm S}^\alpha \frac{\rho^\alpha}{\rho^\alpha_{ {\rm R} 0} } {\bf Q}^\alpha_{ {\rm V} l}, \qquad
 {\bm \pi}^\alpha_{ {\rm S} m} = - \varsigma_{\rm S}^\alpha \frac{\rho^\alpha}{\rho^\alpha_{ {\rm R} 0}} {\bf Q}^\alpha_{ {\rm S} m}, \\
  {\bm \pi}^\alpha_{ {\Phi} k,n} & = - \varsigma_{{\rm F}k}^\alpha \frac{\rho^\alpha}{\rho^\alpha_{ {\rm R} 0}} {\bf Q}^\alpha_{ {\Phi} k,n}.
  \end{split}
 \end{equation}
Kinetic laws for internal variables \cite{holz1996,holz1996b,claytonMOSM2020} are
 \begin{equation}
 \label{eq:Gammadots}
 \begin{split}
 D^\alpha_t {\bm \Gamma}^\alpha_{ {\rm V} l} & = \frac{{\bf Q}^\alpha_{ {\rm V} l}}
 { \beta^\alpha_{ {\rm V} l} B^\alpha_\theta  \tau^\alpha_{ {\rm V} l} }, 
 \quad
  D^\alpha_t {\bm \Gamma}^\alpha_{ {\rm S} m} = \frac{{\bf Q}^\alpha_{ {\rm S} m}}
 { \beta^\alpha_{ {\rm S} m} \mu^\alpha_{\rm S}  \tau^\alpha_{ {\rm S} m} },
 \\ 
 D^\alpha_t {\bm \Gamma}^\alpha_{ {\Phi} k,n} & = \frac{{\bf Q}^\alpha_{ {\Phi} k,n}}
 { \beta^\alpha_{ {\Phi} k,n} \mu^\alpha_k  \tau^\alpha_{ {\Phi} k,n}}.
 \end{split}
 \end{equation}
 Dissipation from viscoelasticity is nonnegative in \eqref{eq:dissi}:
\begin{equation}
\label{eq:viscodissi}
\begin{split}
{\mathfrak D}^\alpha_{\Gamma}  = & \quad \frac{\rho^\alpha}{\rho^\alpha_{ {\rm R} 0}}
\sum_l \frac{\varsigma_{\rm S}^\alpha  {\bf Q}^\alpha_{ {\rm V} l} : {\bf Q}^\alpha_{ {\rm V} l} }
 { \beta^\alpha_{ {\rm V} l} B^\alpha_\theta  \tau^\alpha_{ {\rm V} l} }
\\ &  +  \frac{\rho^\alpha}{\rho^\alpha_{ {\rm R} 0}} \sum_m \frac{ \varsigma_{\rm S}^\alpha  {\bf Q}^\alpha_{ {\rm S} m}:{\bf Q}^\alpha_{ {\rm S} m}}
 { \beta^\alpha_{ {\rm S} m} \mu^\alpha_{\rm S}  \tau^\alpha_{ {\rm S} m} } \\
 & 
  +  \frac{\rho^\alpha}{\rho^\alpha_{ {\rm R} 0}}  \sum_k \sum_n \frac{\varsigma_{{\rm F}k}^\alpha {\bf Q}^\alpha_{ {\Phi} k,n}:{\bf Q}^\alpha_{ {\Phi} k,n}}
 { \beta^\alpha_{ {\Phi} k,n} \mu^\alpha_k  \tau^\alpha_{ {\Phi} k,n}} \geq 0.
\end{split}
\end{equation}
  Initial conditions for state variables are
$ {\bm \Gamma}^\alpha_{ {\rm V} l 0} = {\bf 0}$, ${\bm \Gamma}^\alpha_{ {\rm S} m } = {\bf 0}$, and
${\bm \Gamma}^\alpha_{ {\Phi} k,n} = {\bf 0}$. For energies,
$\Psi^\alpha_{{\rm V} l } ({\bm 0},{\bf C}^\alpha) = \hat{\Psi}^\alpha_{{\rm V} l } ({\bf C}^\alpha)$,
$\Psi^\alpha_{{\rm S} m } ({\bm 0},{\bf C}^\alpha)= \hat{\Psi}^\alpha_{{\rm S} m } ({\bf C}^\alpha)$,
$\Psi^\alpha_{{\Phi} k,n } ({\bm 0},{\bf C}^\alpha)= \hat{\Psi}^\alpha_{{\Phi} k,n } ({\bf C}^\alpha)$. Configurational energies are integrated over time as
\begin{align}
\label{eq:econfig}
&  \Psi^\alpha_{\Gamma} 
=  \sum_l  \left[ \hat{\Psi}^\alpha_{{\rm V} l } - \int_{0}^t  {\bf Q}^\alpha_{ {\rm V} l} : D^\alpha_s  {\bm \Gamma}^\alpha_{{\rm V} l } \, {\rm d} s \right] \nonumber \\
&  \qquad \quad
+  \sum_m  \left[ \hat{\Psi}^\alpha_{{\rm S} m } - \int_{0}^t  {\bf Q}^\alpha_{ {\rm S} m} : D^\alpha_s  {\bm \Gamma}^\alpha_{{\rm S} m } \, {\rm d} s \right], \\
&  \Psi^\alpha_{\Phi} 
=  \sum_k \sum_n  \left[ \hat{\Psi}^\alpha_{{\Phi} k,n } - \int_{0}^t  {\bf Q}^\alpha_{ {\Phi} k,n} : D^\alpha_s  {\bm \Gamma}^\alpha_{{\Phi} k,n }  {\rm d} s \right].
\end{align}
%
\noindent {\bf Active tension.} Internal state variables in $\{ {\bm \xi}^\alpha \}$ are
the scalar functions  $\{ \Delta^\alpha_k \}$ with $k$ the fiber family number.
Kinetic equations with initial conditions are imposed directly \cite{martins1998,ito2010,franchini2022}
rather than by more sophisticated electrochemical physics \cite{stalhand2016}
outside the present scope:
\begin{equation}
\label{eq:actkin} 
\begin{split}
D^\alpha_t \{D^\alpha_t \Delta^\alpha_k \} & = \{ D^\alpha_t \Delta^\alpha_k \} ( {\bf X}^\alpha,t), 
\\
 \{ \Delta^\alpha_k \}( {\bf X}^\alpha,0)& =  \{ \Delta^\alpha_{k0} \}( {\bf X}^\alpha) .
 \end{split}
\end{equation}
Evolution equations \eqref{eq:actkin} should implicitly be affected by local states; for example, the history of fiber damage $\{D^\alpha_k \}$, if severe, should limit
maximum contractile stress.
Conjugate forces in \eqref{eq:dissi} are the following parts of $\{ {\bm \pi}^\alpha \}$:
\begin{equation}
\label{eq:piactive} 
\begin{split}
& \{ \pi^\alpha_{ {\rm A} k } \} (\lambda^\alpha_k, \{ \Delta^\alpha_k \} )  = 
   \\  & \qquad \varsigma_{{\rm F}k}^\alpha \frac{\rho^\alpha}{\rho^\alpha_{ {\rm R} 0}}
 \frac {\partial} { \partial \{ \Delta^\alpha_k \} } [\Lambda^\alpha_k (\lambda^\alpha_k, \{ \Delta^\alpha_k \} ) + \chi^\alpha_k (\{ \Delta^\alpha_k \})] .
 \end{split}
\end{equation}
Dissipation from activation or passivation should be nonnegative,
to be ensured by storage-release functions $\chi^\alpha_k$:
\begin{equation}
\label{eq:dissiA}
{\mathfrak D}^\alpha_{\rm A} =  - \sum_k  \{ \pi^\alpha_{ {\rm A} k } \} \cdot \{ D^\alpha_t \Delta^\alpha_k \}
\geq 0.
\end{equation}

\noindent {\bf Damage.} From \eqref{eq:Psitot} and \eqref{eq:fracen},
conjugate forces to damage measures for the matrix in
$\{ {\bm \pi}^\alpha \}$ and $\{ {\bm \zeta}^\alpha \}$ are
\begin{align}
\label{eq:piDbar}
& \bar{\pi}^\alpha_{\rm D} = \rho^\alpha \frac{\partial \psi^\alpha}{\partial \bar{D}^\alpha}
= \frac{\rho^\alpha}{\rho^\alpha_{ {\rm R} 0}} \frac{\partial}{\partial \bar{D}^\alpha}
[ \sqrt{\hat{G}^\alpha} \hat{\Psi}^\alpha_{\rm D} + 
\nonumber \\
& \qquad \qquad \qquad \qquad + \varsigma^\alpha_{\rm V} \Psi^\alpha_{\rm V} 
 + \varsigma^\alpha_{\rm S} (\Psi^\alpha_{\rm S}  + \Psi^\alpha_{\Gamma} ) ]
 \nonumber
 \\ & \quad
 =  \frac{\rho^\alpha}{\rho^\alpha_{ {\rm R} 0}}
 [ 2 \sqrt{\hat{G}^\alpha} \bar{E}^\alpha_{\rm C} \bar{D}^\alpha + \Psi^\alpha_{\rm D}
 \frac{\partial}{\partial \bar{D}^\alpha} \ln \sqrt{\hat{G}^\alpha} ] \nonumber
 \\ & \quad \, \, - 
  \frac{\rho^\alpha}{\rho^\alpha_{ {\rm R} 0}}
\bar{\vartheta}^\alpha  [1 - \bar{D}^\alpha {\rm H}( \ln J^\alpha)]^{\bar{\vartheta}^\alpha - 1}
{\rm H}( \ln J^\alpha) \Psi^\alpha_{\rm V} \nonumber
\\ & \quad \, \, - 
  \frac{\rho^\alpha}{\rho^\alpha_{ {\rm R} 0}}
\bar{\vartheta}^\alpha  [1 - \bar{D}^\alpha]^{\bar{\vartheta}^\alpha - 1}
( \Psi^\alpha_{\rm S} +  \Psi^\alpha_{\Gamma}), 
\\
& \bar {\bm \zeta}^\alpha_{\rm D} = \rho^\alpha {\bf F}^\alpha \frac{\partial \psi^\alpha}{\partial \nabla^\alpha_0 \bar{D}^\alpha} = 
\frac{\rho^\alpha}{\rho^\alpha_{ {\rm R} 0}}
\sqrt{\hat{G}^\alpha} {\bf F}^\alpha \frac{\partial \hat{ \Psi}^\alpha_{\rm D}}{\partial \nabla_0 \bar{D}^\alpha} \nonumber \\
& \quad 
= 2 \frac{\rho^\alpha}{\rho^\alpha_{ {\rm R} 0}}
\sqrt{\hat{G}^\alpha} 
\bar{\Upsilon}^\alpha \bar{l}^\alpha_{\rm R} {\bf F}^\alpha [ (\nabla \bar{D}^\alpha) {\bf F}^\alpha ]. \label{eq:zetabarD}
\end{align}
Define total conjugate force to damage in the matrix:
\begin{equation}
\label{eq:totconbar}
\bar{\digamma}^\alpha_{\rm D} = - \bar{\pi}^\alpha_{\rm D} + \nabla \cdot \bar{\bm \zeta}^\alpha_{\rm D}.
\end{equation} 
Define viscosity $\bar{\nu}^\alpha_{\rm D} \geq 0$. A Ginzburg-Landau kinetic law and nonnegative dissipation for the matrix in \eqref{eq:dissi} are
\begin{align}
\label{eq:TDGLbar}
& \bar{\nu}^\alpha_{\rm D} D^\alpha_t \bar{D}^\alpha = \bar{\digamma}^\alpha_{\rm D}, \\
\label{eq:DDis}
&  \bar{\mathfrak D}^\alpha_{\rm D} = \bar{\digamma}^\alpha_{\rm D} D^\alpha_t \bar{D}^\alpha =  \bar{\nu}^\alpha_{\rm D} |D^\alpha_t \bar{D}^\alpha|^2 \geq 0.
\end{align}
To render the damage rate always nonnegative, \eqref{eq:TDGLbar}
can be modified to 
$\bar{\nu}^\alpha_{\rm D} D^\alpha_t \bar{D}^\alpha = \bar{\digamma}^\alpha_{\rm D}
{\rm H}(\bar{\digamma}^\alpha_{\rm D})$.
Damage kinetics are suppressed for $\bar{\nu}^\alpha_{\rm D} \rightarrow \infty$
and rate-independent for $\bar{\nu}^\alpha_{\rm D} \rightarrow 0$
with equilibrium condition $\bar{\digamma}^\alpha_{\rm D} = 0$.
For rate insensitivity, irreversibility is enforced by
setting $\bar{D}^\alpha (t)$ to the maximum of the argument
 of $\bar{\digamma}^\alpha_{\rm D}(\bar{D}^\alpha; \cdot)(t) = 0$
 and $\bar{D}^\alpha (s) \, \forall \, s < t$, where the latter
 renders $D^\alpha_t \bar{D}^\alpha(t^-) \rightarrow 0$.
Usual, but inessential, initial conditions are $\bar{D}^\alpha_0 = 0$.

Damage in fiber families $k$ is treated analogously.
Viscosities are ${\nu}^\alpha_{{\rm D} k} \geq 0$.
Conjugate thermodynamic forces and dissipative Ginzburg-Landau kinetics are
\begin{align}
\label{eq:piDk}
 {\pi}^\alpha_{{\rm D} k} & = \rho^\alpha \frac{\partial \psi^\alpha}{\partial {D}^\alpha_k }
= \frac{\rho^\alpha}{\rho^\alpha_{ {\rm R} 0}} \frac{\partial}{\partial {D}^\alpha_k}
[ \sqrt{\hat{G}^\alpha} \hat{\Psi}^\alpha_{\rm D}  
+ \varsigma^\alpha_{{\rm F} k} (\Psi^\alpha_{{\rm F} k} + \Psi^\alpha_{\Phi k})]
 \nonumber
 \\ & 
 =  \frac{\rho^\alpha}{\rho^\alpha_{ {\rm R} 0}}
 [ 2 \sqrt{\hat{G}^\alpha} {E}^\alpha_{{\rm C} k} {D}^\alpha_k + \Psi^\alpha_{\rm D}
 \frac{\partial}{\partial {D}^\alpha_k} \ln \sqrt{\hat{G}^\alpha} ] \nonumber
 \\ &  \qquad \quad \, \,
 -  \frac{\rho^\alpha}{\rho^\alpha_{ {\rm R} 0}}
{\vartheta}^\alpha_k  [1 - {D}^\alpha_k]^{{\vartheta}^\alpha_k - 1}
 (\Psi^\alpha_{{\rm F} k} + \Psi^\alpha_{\Phi k}),
 \end{align}
 \begin{align}
 \label{eq:zetafiber}
 {\bm \zeta}^\alpha_{{\rm D} k} & = \rho^\alpha {\bf F}^\alpha \frac{\partial \psi^\alpha}{\partial \nabla^\alpha_0 {D}^\alpha_k} = 
\frac{\rho^\alpha}{\rho^\alpha_{ {\rm R} 0}}
\sqrt{\hat{G}^\alpha} {\bf F}^\alpha \frac{\partial \hat{ \Psi}^\alpha_{\rm D}}{\partial \nabla_0 {D}^\alpha_k } \nonumber \\
& 
= 2 \frac{\rho^\alpha}{\rho^\alpha_{ {\rm R} 0}}
\sqrt{\hat{G}^\alpha} 
{\Upsilon}^\alpha_k {l}^\alpha_{{\rm R} k} {\bf F}^\alpha [ (\nabla {D}^\alpha_k) {\bf F}^\alpha ], 
\end{align}
\begin{align}
\label{eq:digammafiber}
& {\digamma}^\alpha_{{\rm D} k} = - {\pi}^\alpha_{{\rm D} k}  + \nabla \cdot {\bm \zeta}^\alpha_{{\rm D} k} ,
\\
\label{eq:TDGLk}
&  {\nu}^\alpha_{{\rm D} k} D^\alpha_t {D}^\alpha_k = {\digamma}^\alpha_{{\rm D} k}, \\
\label{eq:DDisk}
&{\mathfrak D}^\alpha_{\rm D F} = \sum_k  {\mathfrak D}^\alpha_{{\rm D} k} 
= \sum_k {\digamma}^\alpha_{{\rm D} k} D^\alpha_t {D}^\alpha_k 
\nonumber 
\\ & \qquad \qquad \qquad \quad =  \sum_k {\nu}^\alpha_{{\rm D} k} |  D^\alpha_t {D}^\alpha_k |^2 \geq 0.
\end{align}
To forbid healing,
${\nu}^\alpha_{{\rm D} k} D^\alpha_t {D}^\alpha_k = {\digamma}^\alpha_{{\rm D} k}
{\rm H} ({\digamma}^\alpha_{{\rm D} k})$ in lieu of \eqref{eq:TDGLk}.
For rate insensitivity, ${\nu}^\alpha_{{\rm D} k} \rightarrow 0
\Rightarrow {\digamma}^\alpha_{{\rm D} k} = 0$ with possible irreversibility constraints analogous to 
those for $\bar{D}^\alpha(t)$. Usual initial conditions are $D^\alpha_{k 0} = 0$.
\\

\noindent {\bf Heat conduction.} Fourier conduction
is usually sufficient for each bulk constituent $\alpha$,
with isotropic conductivity $\kappa_\theta^\alpha (\theta, \{ {\bm \xi}^\alpha \} ) \geq 0$, often temperature and internal-state dependent.
It could degrade with damage via, e.g., 
$\kappa^\alpha_\theta \approx \varsigma^\alpha_{\rm V} \kappa^\alpha_{\theta 0}$. 
Heat flux and entropy production are
\begin{align}
\label{eq:fourier}
& {\bf q}^\alpha = - \kappa^\alpha_\theta \nabla \theta^\alpha, 
\quad {\mathfrak D}_{\rm q}^\alpha = \frac{ - {\bf q}^\alpha \cdot \nabla \theta^\alpha }{ \theta^\alpha }
= \frac{\kappa^\alpha_\theta |\nabla \theta^\alpha|^2 }{ \theta^\alpha} \geq 0.
\end{align}

\noindent {\bf Momentum transfer.} 
Momentum exchange includes Darcy-like contributions
from velocity differences ${\bm \upsilon}^\alpha - {\bm \upsilon}^\beta$ =
${\bm \mu}^\alpha - {\bm \mu}^\beta$ \cite{macminn2016,mad2019,claytonIJES2022}
and mass exchange to satisfy \eqref{eq:sums}:
\begin{align}
\label{eq:hdef}
& {\bf h}^\alpha = - \sum_\beta [ \lambda^{\alpha \beta} ({\bm \mu}^\alpha - {\bm \mu}^\beta)]
- \hat{c}^\alpha {\bm \mu}^\alpha.
\end{align}
The inverse hydraulic-type conductivity matrix is $\lambda^{\alpha \beta} = \lambda^{\beta \alpha}$; entries $\lambda^{\alpha \beta}  \geq 0$ can depend on temperature and volume of each phase (e.g., to account for changes in interphase viscosity with temperature and in permeability with
 porosity \cite{macminn2016})
and degrade with damage \cite{wilson2016,suh2021}:
\begin{align}
\label{eq:lambdaab}
& \lambda^{\alpha \beta}(J^\alpha,J^\beta,\theta^\alpha, 
\theta^\beta,\bar{D}^\alpha,\bar{D}^\beta) 
\nonumber \\ & \qquad \qquad= \bar{\lambda}^{\alpha \beta} (J^\alpha,J^\beta,\theta^\alpha,\theta^\beta) \sqrt{ \varsigma^\alpha_{\rm V} \varsigma^\beta_{\rm V}}.
\end{align}

\noindent {\bf Energy transfer.} Energy exchange includes heat transfer 
from temperature differences $\theta^\alpha - \theta^\beta$ as well as 
momentum and mass exchange terms to satisfy \eqref{eq:sums2}:
\begin{align}
\label{eq:epsdef}
& \epsilon^\alpha = - \sum_\beta [\omega^{\alpha \beta} (\theta^\alpha - \theta^\beta)]
- m^\alpha \sum_\beta {\bf h}^\beta \cdot {\bm \mu}^\beta \nonumber 
\\ & \qquad \qquad \qquad \quad
- m^\alpha \sum_\beta 
 \hat{c}^\beta (u^\beta +  {\textstyle {\frac{1}{2}}}|{\bm \mu}^\beta|^2),
\end{align}
recalling mass concentration $m^\alpha = \rho^\alpha / \rho$ from \eqref{eq:massfracs}.
The matrix of heat transfer coefficients $\omega^{\alpha \beta} = \omega^{\beta \alpha} \geq 0$
can depend on state variables and damage \cite{suh2021} like \eqref{eq:lambdaab}:
\begin{align}
\label{eq:omegaab}
& \omega^{\alpha \beta}(J^\alpha,J^\beta,\theta^\alpha, 
\theta^\beta,\bar{D}^\alpha,\bar{D}^\beta) 
\nonumber \\ & \qquad \qquad= \bar{\omega}^{\alpha \beta} (J^\alpha,J^\beta,\theta^\alpha,\theta^\beta) \sqrt{ \varsigma^\alpha_{\rm V} \varsigma^\beta_{\rm V}}.
\end{align}
In compression, contact among fully
broken constituents permits momentum and heat transfer in respective  \eqref{eq:lambdaab} and \eqref{eq:omegaab} even if $\bar{D}^\alpha \rightarrow 1$ or $\bar{D}^\beta \rightarrow 1$.
Damage dependence differing from basic illustrations \eqref{eq:lambdaab} and \eqref{eq:omegaab} (e.g., \cite{wilson2016,suh2021}) can be substituted if more appropriate.
\\

\noindent {\bf Mass transfer.} 
Terms $\hat{c}^\alpha$ sum to zero in \eqref{eq:sums};
they account for mass transfer rates between constituents.
 In biology, these could originate from
 growth and remodeling, for example,
 exchange of nutrients dissolved in a fluid phase to support
 growth of new solid tissue.
 Detailed constitutive equations for
$\hat{c}^\alpha$ are beyond the present scope. 
Thermodynamic constraints emerge from \eqref{eq:entbal4}
with \eqref{eq:hdef} and logical stipulation that
the rightmost two sums in \eqref{eq:entbal4} should be nonnegative in concert:
\begin{align}
\label{eq:chatcon}
 \sum_\alpha \left[ \frac{\epsilon^\alpha}{\theta^\alpha} + c^\alpha \eta^\alpha \right]
& = \sum_{\alpha} \sum_\beta \frac{\omega^{\alpha \beta}(\theta^\alpha - \theta^\beta)^2}{2 \theta^\alpha \theta^\beta}
\nonumber
\\ 
& + \sum_\alpha \frac{m^\alpha}{2\theta^\alpha }\sum_\beta \lambda^{\alpha \beta}
| {\bm \mu}^\alpha - {\bm \mu}^\beta|^2 \nonumber 
\\
& +  \sum_\alpha \frac{m^\alpha}{\theta^\alpha }
\sum_\beta \hat{c}^\beta \left[ \frac{ | {\bm \mu}^\beta |^2}{2} - u^\beta \right] \nonumber
\\ &
+ \sum_\alpha \hat{c}^\alpha \eta^\alpha + \rho \eta \,  \partial_t \ln \sqrt{g} \geq 0.
\end{align}
The first two double sums in \eqref{eq:chatcon} are
always nonnegative. When all $\epsilon^\alpha = 0$ and $\hat{c}^\alpha = 0$ (i.e., no phase interactions),
\eqref{eq:chatcon} becomes, with \eqref{eq:metrics3},
$\eta \, \partial_t \hat{g} \geq 0$. Then when $ \rho \eta = \sum_\alpha \rho^\alpha \eta^\alpha > 0$,  metric $\hat{\bf g}$ should only be dilating.
 
\subsection{Stress, dissipation, and boundary conditions} 
Total stress ${\bm \sigma}^\alpha$ for constituent $\alpha$ is the sum of \eqref{eq:totstatstress} and \eqref{eq:viscstress1}. Total stress for the mixture ${\bm \sigma}$ is \eqref{eq:stresstot}.
Total constituent dissipation $\mathfrak{D}^\alpha$ entering \eqref{eq:dissi} is the sum of \eqref{eq:viscsdissi}, \eqref{eq:viscodissi}, \eqref{eq:dissiA}, \eqref{eq:DDis}, \eqref{eq:DDisk}, and \eqref{eq:fourier}, each individually nonnegative:
\begin{align}
\label{eq:Dissatot}
 {\mathfrak D}^\alpha - ({\bf q}^\alpha \cdot \nabla \theta^\alpha) / \theta^\alpha & = 
\hat{\mathfrak D}^\alpha + {\mathfrak D}^\alpha_\Gamma + {\mathfrak D}^\alpha_{\rm A}
\nonumber \\
& \quad 
+ \bar{\mathfrak D}^\alpha_{\rm D} + {\mathfrak D}^\alpha_{\rm DF} + {\mathfrak D}^\alpha_{\rm q}
\geq 0.
\end{align}
The total dissipation inequality of \eqref{eq:entbal4} is then
\begin{align}
\label{eq:totcdineq}
\sum_\alpha \left[ \frac{ {\mathfrak D}^\alpha + {\mathfrak D}^\alpha_{\rm q}}{\theta^\alpha} \right]
+ \sum_\alpha \left[ \frac{\epsilon^\alpha}{\theta^\alpha} + c^\alpha \eta^\alpha \right] \geq 0.
\end{align}
From \eqref{eq:ebalglob}, boundary conditions are required for each constituent $\alpha = 1, \ldots, N$.
Mechanical conditions are prescribed histories of traction ${\bf t}^\alpha$ or
velocity ${\bm \upsilon}^\alpha$ on $\partial \Omega$.
Thermal conditions are histories of flux $q_n^\alpha$ or temperature $\theta^\alpha$
on $\partial \Omega$. For internal variables $\{ {\bm \xi}^\alpha \}$ with gradient energetic dependence (e.g., order parameters for damage), 
histories of fluxes $ \{ {\bf z}^\alpha \} = (\rho^\alpha \partial \psi^\alpha / \partial \{ \nabla {\bm \xi} \}^\alpha)
\cdot {\bf n}$ or conjugate rates $D^\alpha_t {\bm \xi}^\alpha$ are needed on $\partial \Omega$. 
Histories of body force ${\bf b}^\alpha$ and heat source $r^\alpha$
are prescribed over $\Omega$.

\subsection{\label{sec3e}Finsler metrics}
Coordinate forms of \eqref{eq:metrics3} and \eqref{eq:metrics4} are
\begin{align}
\label{eq:metrics5}
&  {\bf g}( {\bf x}, t) = \bar{g}_{ik} ( {\bf x}) \, \hat{ g}^k_j ( \{ {\bm \xi}^\alpha( {\bf x},t) \} )
{\bf g}^i \otimes {\bf g}^j \\
& {\bf G}^\alpha  ( {\bf X}^\alpha, t) = (\bar{G}^\alpha )_{IK} ( {\bf X}^\alpha ) \nonumber
\\ & \qquad  \qquad  \times  (\hat{  G}^\alpha 
)^K_J (  \{ { \bm \Xi }^\alpha ( {\bf X}^\alpha, t) \} ) ( {\bf G}^\alpha)^I \otimes ( {\bf G}^\alpha)^J.
\label{eq:metrics6}
\end{align}
Canonical transformations \cite{claytonZAMP2017,claytonMMS2022,claytonSYMM2023} between representations of state variables on
$\mathfrak{m}$ and ${\mathfrak M}^\alpha$ are used for \eqref{eq:isvrelate}:
\begin{equation}
\label{eq:isvrelate2}
 \{ {\bm \Xi}^\alpha  ({\bf X}^\alpha, t) \}
 = \{ {\bm \xi}^\alpha ({\bf x}, t )\} \circ {\bm \chi}^\alpha ({\bf X}^\alpha, t).
\end{equation}
Though other relationships are admissible, the analogous transformation law \cite{claytonSYMM2023} between components of $\hat{\bf G}^\alpha$ and $\hat{\bf g}$ is prescribed here, with $\delta^i_J$ Kronecker's delta symbols:
\begin{align}
\label{eq:ghatrelate}
& (\hat{G}^\alpha)^I_J ({\bf X}^\alpha,t) = \delta^I_i \delta^j_J \hat{g}^i_j ( {\bm \chi}^\alpha ({\bf X}^\alpha,t), t),
\end{align}
Only $ \{ {\bm \xi}^\alpha \}$ and $ \hat{g}^i_j( \{ {\bm \xi}^\alpha \})$ are defined constitutively,
with \eqref{eq:isvrelate2} and \eqref{eq:ghatrelate} yielding $ \{ {\bm \Xi}^\alpha \}$ and $ (\hat{G}^\alpha)^I_J( \{ {\bm \Xi}^\alpha \} )$, or vice-versa if referential
versions are defined instead.

Dependence of $\hat{\bf g}$ on $ \{ {\bm \xi}^\alpha \}$
is henceforth restricted to dependence on damage parameters $( \{ \bar{D}^\alpha \}, \{ D^\alpha_k \})$;
 \eqref{eq:isvrelate2} is $\bar{D}^\alpha ( {\bf X}^\alpha, t) = \bar{D}^\alpha ( {\bm \chi}^\alpha ( {\bf X}^\alpha, t), t) \circ {\bm \chi}^\alpha_t$ and so on for $\{ D^\alpha_k \}$.
 If the geometric framework is extended to describe biologic growth and remodeling, then
 $ \{ {\bm \xi}^\alpha \}$ can be expanded with internal state variable(s) associated with
 such processes, for which kinetic equations in \eqref{eq:isvdot} are needed.
 For a theory as in Ref.~\cite{yavari2010}, $\hat{\bf G}^\alpha$ should depend
 on these additional (e.g., growth) function(s), and $\hat{g}^i_j \rightarrow \delta^i_j$, so
 that $\mathfrak{m}$ is Euclidean but $\mathfrak{M}^\alpha$ need not be.
 
In the current application, as tears and commensurate fiber rearrangements arise in constituents of the mixture, the body manifold can expand and shear \cite{claytonSYMM2023}.
Mixed-variant tensor $\hat{\bf g}$ is a product of matrix and fiber terms:
\begin{align}
\label{eq:mprod1}
& \hat{g}^i_j (\{ \bar{D}^\alpha \},\{ D^\alpha_k \}) = \bar{\gamma}^i_k (\{ \bar{D}^\alpha \}) \tilde{\gamma}^k_j (\{ D^\alpha_k \}).
\end{align}
Contributions from isotropic matrix damage $\{ \bar{D}^\alpha \}$ in $\bar{\gamma}^i_j$
are assumed spherical (e.g., Weyl-type scaling \cite{claytonJGP2017}), measured by determinants  $\bar{\gamma}^\alpha = \bar{\gamma}^\alpha ( \bar{D}^\alpha)$.
Spherical contributions of phases $\alpha = 1, \ldots, N$ are merged multiplicatively
in $\hat{\bf g}$ since their sequence is irrelevant. Forms are
\begin{align}
\label{eq:gisohat}
& \bar{\gamma}^i_j =  \delta^i_j \prod_\alpha (\bar{\gamma}^\alpha)^{1/3}, \quad
\bar{\gamma}^\alpha = \exp \left[ \frac{2 n_0^\alpha \bar{\kappa}^\alpha}{\bar{r}^\alpha} (\bar{D}^\alpha)^{ {\bar r}^\alpha} \right].
\end{align}
Recall $n_0^\alpha( {\bf X}^\alpha) \in [0,1] $ is a reference volume fraction of phase $\alpha$,
and $\bar{r}^\alpha > 0$ and $ \bar{\kappa}^\alpha$ are constants, the latter positive for dilatant damage. Remnant volumetric strain \cite{claytonSYMM2023} at $\bar{D}^\alpha = 1$
is the ratio of constants $\bar{\epsilon}^\alpha = n_0^\alpha \bar{\kappa}^\alpha / \bar{r}^\alpha$.
Fiber contributions from constituents $\alpha$ and fiber families $k$ 
are merged additively into $\hat{\bf g}$ since these terms are generally anisotropic \cite{claytonZAMP2017,claytonSYMM2023}. Defining $(H^\alpha_k)^i_j = \delta^i_I \delta^J_j (H^\alpha_k)^I_J$
with $(H^\alpha_k)^I_J = \kappa^\alpha_k \delta^I_J + (1-3 \kappa^\alpha_k) (\iota^\alpha_k)^I (\iota^\alpha_k)_J$
from \eqref{eq:structens},
\begin{align}
\label{eq:ganihat}
& \tilde{\gamma}^i_j = \delta^i_j + \sum_\alpha \sum_k (H^\alpha_k)^i_j \{ \exp \left[ \frac{2 n_0^\alpha \tilde{\kappa}^\alpha_k}{\tilde{r}_k^\alpha} ({D}_k^\alpha)^{ \tilde{r}_k^\alpha} \right] - 1 \}.
\end{align}
Constants $\tilde{r}_k^\alpha > 0$ and $ \tilde{\kappa}_k^\alpha$ measure the
 logarithmic remnant strain contributions 
$\tilde{\epsilon}^\alpha_{k} = n_0^\alpha \tilde{\kappa}^\alpha_k / \tilde{r}^\alpha_k$
at $D^\alpha_k = 1$.
 Noting $ \hat{G}^\alpha = \det \hat{\bf G}^\alpha = \det \hat{\bf g} = \hat{g}$ are related through \eqref{eq:ghatrelate},
derivatives in conjugate forces \eqref{eq:piDbar} and \eqref{eq:piDk} are found from
\eqref{eq:gisohat} and \eqref{eq:ganihat} as
\begin{align}
\label{eq:gderivs}
& {\partial} ( \ln \sqrt{\hat{G}^\alpha} ) / {\partial \bar{D}^\alpha}
= n^\alpha_0 \bar{\kappa}^\alpha (\bar{D}^\alpha)^{ {\bar r}^\alpha - 1}, \\
\label{eq:gderivs2}
&  \frac{\partial   (\ln \sqrt{\hat{G}^\alpha }) }{\partial {D}^\alpha_k }
= \exp \left[ \frac{2 n_0^\alpha \tilde{\kappa}^\alpha_k}{\tilde{r}_k^\alpha} ({D}_k^\alpha)^{ \tilde{r}_k^\alpha} \right]  \nonumber \\
& \qquad \qquad \quad \quad \times (\tilde{\gamma}^{-1})^i_j (H^\alpha_k)^j_i 
n^\alpha_0 \tilde{\kappa}^\alpha_k ({D}_k^\alpha)^{ {\tilde r}^\alpha_k - 1} .
\end{align}
Values $\bar{r}^\alpha \in (0,1)$ and $\tilde{r}^\alpha_k \in (0,1)$ produce non-singular $\hat{\bf g}$
and admit solutions
to some equilibrium problems \cite{claytonSYMM2023}. However, these ranges can result in singularities at $\bar{D}^\alpha = 0$
and $D^\alpha_k = 0$ in \eqref{eq:gderivs} and \eqref{eq:gderivs2}. Such singularities
can be avoided by choosing $\bar{r}^\alpha \geq 1$ and $\tilde{r}^\alpha_k \geq 1$.
Stronger conditions  $\bar{r}^\alpha > 1$ and $\tilde{r}^\alpha_k >1$
usefully ensure  \eqref{eq:gderivs}, \eqref{eq:gderivs2} vanish
 at (e.g., initial) states having $\bar{D}^\alpha = 0$, ${D}_k^\alpha= 0$.

%% file: s4a.tex
\section{\label{sec4} Soft-tissue physics}

\subsection{\label{sec4a} Shock Hugoniot response}

\noindent {\bf Fluids.} The theory is first exercised for three fluids that comprise the majority of soft tissues:
water, extracellular fluid (ECF, representative of blood plasma \cite{fung1993} and interstitial fluid in skeletal muscle and skin), and whole blood, the latter with a realistic hematocrit of 0.4.
Shock responses are modeled via theory of Sec.~\ref{sec2b} with each fluid a single-phase material
($\alpha = N = 1$, superscript henceforth suppressed).
Planar (1-D) impact is along the $x = x^1$ direction. Fluid is quiescent upstream, at density $\rho_0$, temperature $\theta_0$, and ambient pressure $p_0 = p_{{\rm R} 0} = 1$ atm.
Eulerian and Lagrangian shock speeds are identical, labeled ${\mathcal U}$. Particle velocity in Hugoniot states is $\upsilon^- = \upsilon$.  In Hugoniot (i.e., downstream shocked) $(\cdot)^{-}$ states, 
\begin{align}
\label{eq:fluidhug1}
& \rho = \rho_0/J, \quad J = F^1_1 = F = \partial {\chi}/ \partial X,  \\
& u = U/\rho_0, \quad t_1 = \sigma^1_1 = -P = -p, \label{eq:fluidhug2}
\end{align}
where $U$ is energy per unit reference volume and
$P$ is longitudinal Cauchy stress, positive in compression. Since volume fraction $n^1_0 = n^1 = 1$, $\rho_{{\rm R}0} = \rho_0$,
and $\rho_{\rm R} = \rho$.

To calculate the material response, $J$ is decremented from unity. At each decrement, the constitutive model,
here the condensed matter EOS for $U_{\rm V}$ and $p_{\rm V}$ in \eqref{eq:EOSi1}--\eqref{eq:pEOSi}, is solved concurrently
with Hugoniot energy equation \eqref{eq:RHe0} for $P$ and $\eta$.
With macroscopically adiabatic conditions assumed and $\{ z \} = 0$, the
latter reduces to 
\begin{align}
\label{eq:fluidEB}
& U = {\textstyle{\frac{1}{2}}}(P + p_0) ( 1 - J)
\end{align}
since $U_0 = 0 $ and $J = 1$ upstream.
Then $\upsilon$, $\mathcal{U}$, and $\theta$ are found from \eqref{eq:veljumps} and \eqref{eq:tEOSi}.
For single-phase materials, ${\bm \mu}^\alpha = {\bf 0}$, ${\bf h}^\alpha = {\bf 0}$, $\epsilon^\alpha = 0$, and here $\hat{c}^\alpha = 0$. For compression, fluid cavitation is omitted, so no damage is modeled.
Hence, all metrics are Euclidean: $g = G = \hat{g} = \hat{G} = 1$.

\begin{table}[b]
\caption{\label{table1}%
Physical properties or model parameters for water, extracellular fluid (ECF),
human blood, and porcine skeletal muscle (cells and matrix, $\alpha = k = 1$, vol.~fract.~$n_0^1 = 0.9$) } 
\begin{ruledtabular}
\begin{tabular}{lcccc}
Property & Water& ECF & Human Blood & Porcine Muscle \\
\hline
$\rho^\alpha_{{\rm R} 0}$ [g/cm$^3$] & 1.00 &  1.03 & 1.06 & 1.10 \\
$B^\alpha_\eta$ [GPa] & 2.10 & 2.20 & 2.64 & 3.28  \\
$c^\alpha_\epsilon$ [J/g$\cdot$K] & 4.15 & 3.96 & 3.58 & 3.25 \\
$\gamma_0^\alpha$ [-] & 0.120 & 0.132 & 0.160 & 0.313 \\
$B^\alpha_{\eta {\rm p}}$ [-] & 6.96 & 6.96 & 12.0 & 8.0 \\
$k^\alpha_{\rm V}$ [-] & 7.0 & 7.0 & 0.0 & 6.0 \\
$\hat{B}^\alpha$ [mPa$\cdot$s] & 2.1 & $\ldots$ & $\ldots$ & $\ldots$ \\
$\hat{\mu}^\alpha$ [mPa$\cdot$s] & 0.8 & 1.2 & 5.0 & $\ldots$ \\
$\mu^\alpha_{\rm S}$ [kPa] & $\ldots$ & $\ldots$ & $\ldots$ & 1.0 \\
$\mu^\alpha_k$ [kPa] & $\ldots$ & $\ldots$ & $\ldots$ & 1.0 \\
$k^\alpha_k$ [-] & $\ldots$ & $\ldots$ & $\ldots$ & 10.0 \\
$\beta^\alpha_{\rm S}$ [-] & $\ldots$ & $\ldots$ & $\ldots$ & $1.0 \times 10^5$ \\
$\beta^\alpha_{\Phi k}$ [-] & $\ldots$ & $\ldots$ & $\ldots$ & $1.0 \times 10^5$ \\
${\mathsf J}_{\rm C} $ [kJ/m$^2$]  & $\ldots$ & $\ldots$ &  $\ldots$ & 0.84 \\
$l^\alpha$ [mm]   & $\ldots$ & $\ldots$ & $\ldots$ & 0.88 \\
$\vartheta^\alpha$ [-] & $\ldots$ & $\ldots$ & $\ldots$ & 2.0 \\
${\epsilon}^\alpha_{\rm r}$ [-] & $\ldots$ & $\ldots$ & $\ldots$ & 0.2 \\
$\bar{r}^\alpha = \tilde{r}^\alpha_k $ [-]  &$\ldots$ & $\ldots$ & $\ldots$ & 2.0 \\
 \end{tabular}
\end{ruledtabular}
\end{table}

Properties entering the EOS are listed in Table~\ref{table1}. All are obtained or estimated from experimental literature
\cite{kell1975,fung1993,guo2001,nagayama2002,yao2014,jones2018} with the exception
of nonlinear bulk stiffening parameters $B_{\eta {\rm p} }$ and $k_{\rm V}$ that are fit to the experimental Hugoniot data.
The ambient bulk modulus is related to the bulk sound velocity $c_{\rm B} = \sqrt{ (B_\eta + p_0) / \rho_0}$. 
For water, the usual relationship $B_{\eta {\rm p}} = 4S - 1$ is used, where $S$ is the slope of a linear fit to the 
$\mathcal{U}$-$\upsilon$ Hugoniot \cite{claytonNEIM2019}.
Since shock compression data are apparently unavailable for ECF, $B_{\eta {\rm p} }$ and $k_{\rm V}$ for water are assigned.
Newtonian viscosities $\hat{B},\hat{\mu}$ are listed for completeness \cite{fung1993,guo2001}, where $\hat{\mu}$ for blood is for high rates \cite{fung1993} and bulk viscosity is supplied only for water \cite{guo2001}. These do not enter the present analysis: 
Newtonian viscosity \eqref{eq:viscstress1} and Fourier conduction  \eqref{eq:fourier} are necessarily omitted as both are incompatible with treatment of shocks as singular surfaces \cite{morro1980,*morro1980b}.

\begin{figure}[t]
\begin{subfigure}
 (a){\includegraphics[width = 0.36\textwidth]{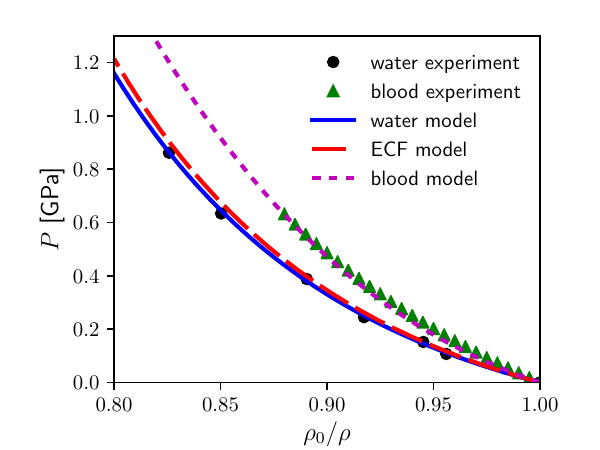} \label{fig1a}}
\end{subfigure} \\
\begin{subfigure}
\,  (b) {\includegraphics[width = 0.36\textwidth]{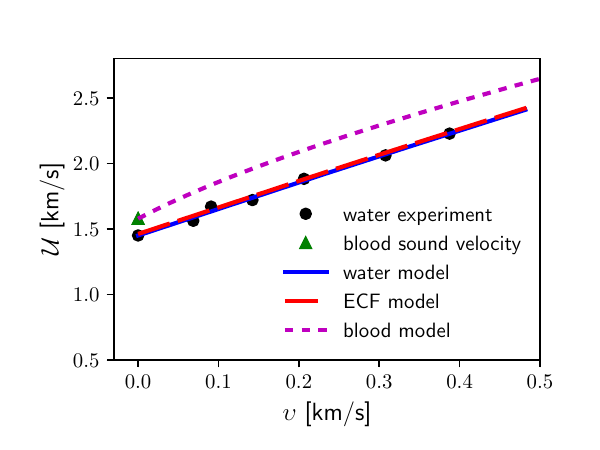} \label{fig1b}} 
\end{subfigure}
\caption{\label{fig1} Model results and shock data \cite{nagayama2002,nagoya1995} for (a) Hugoniot stress vs.~mass density and (b) shock velocity vs.~particle velocity in water, extracellular fluid (ECF), and human blood.}
\end{figure}

Compared in Fig.~\ref{fig1}(a) are $P$-$\rho$ Hugoniot predictions and experimental data on water ($\theta_0  = 297$ K)\cite{nagayama2002} and human blood ($\theta_0 \approx 310$ K) \cite{nagoya1995}. Shock data on ECF do not seem to exist; predictions are for $\theta_0 =310$ K.
Compared in Fig.~\ref{fig1}(b) are $\mathcal{U}$-$\upsilon$ data for water \cite{nagayama2002} and predictions for all three fluids; shock velocity data were not reported for blood in Ref.~\cite{nagoya1995}. The $c_{\rm B}$ value for blood, ${\mathcal U} \rightarrow c_{\rm B}$ as $\upsilon \rightarrow 0$, is from Ref.~\cite{jones2018}.
Results in Fig.~\ref{fig1} confirm validity of the EOS for water and blood. The latter is stiffer than ECF, which is stiffer than water.
\\

\noindent{\bf Skeletal muscle.} Planar shock response of skeletal muscle is predicted next.
Properties and loading conditions replicate impact experiments of Wilgeroth et al.~\cite{wilgeroth2012,wilgeroth2012aip}
on porcine muscle tissue.
The material is modeled as a mixture of two coexisting phases ($\alpha = 1,2$): a ``solid'' tissue of
initial volume fraction $n_0^1 = 0.9$ \cite{neville1979} and
an interstitial fluid depicted by the ECF, comprising remaining fraction $n_0^2 = 0.1$.
The first phase consists of the muscle cells (i.e., fibers), collagenous connective tissues, and ground substance between and encasing the cells (i.e., the extracellular matrix).  Muscle cells contain significant internal fluid whose physical properties are included implicitly in the constitutive model for the first phase.

Experiments \cite{wilgeroth2012} show a single-wave structure with steep shock front (rise time on the order of $\mu {\rm s}$ or smaller) rather than multiple waveforms that would be expected if shock and particle velocities among the phases
differed significantly \cite{bowen1974,claytonIJES2022}.
This suggests inverse hydraulic conductivity is very large (e.g., $\lambda^{\alpha \beta} \tilde{ \rightarrow} \infty$) at
these high-pressure dynamic conditions, and diffusion velocities ${\bm \mu}^\alpha$ are negligible. 
It is thus assumed particle velocity histories match in each phase: $\upsilon^\alpha(x,t) = \upsilon^1(x,t) = \upsilon^2(x,t) = \upsilon(x,t)$.  Therefore, $J(x,t) = F(x,t) = \partial \chi / \partial X$ is identical in each constituent.  Microsecond scales are too brief
for biologic mass exchange: $\hat{c}^\alpha = 0$.
Shock compression is adiabatic: $q^\alpha \rightarrow 0$). Heat transfer in 
$\epsilon^\alpha$ of \eqref{eq:epsdef} is likewise assumed null in Hugoniot states: $\theta^\alpha = \theta^1 = \theta^2 = \theta$.

The solution procedure is similar to that for single fluids, but the constitutive model is now much more
complex.  The shock response is that of the mixture, where governing and jump equations are in \eqref{eq:massjumpmix}--\eqref{eq:Fmix1}. Both phases are quiescent and at reference $\theta_0 = 310$ K
and $p_{{\rm R} 0} = 1$ atm upstream; $\theta_0$ was unreported in Ref.~\cite{wilgeroth2012}, but model results are insensitive to $\theta_0$. From \eqref{eq:stresstot} and \eqref{eq:etot} with ${\bm \mu}^\alpha = {\bf 0}$, mixture 
stress and internal energy are 
\begin{align}
\label{eq:muscstress}
& \bm{\sigma} = \sum_\alpha {\bm \sigma}^\alpha, \quad
U 
= \sum_\alpha n_0^\alpha \rho^\alpha_{ {\rm R} 0} u^\alpha 
= \sum_\alpha n^\alpha_0 U^\alpha.
\end{align}
Because all phases deform equally without mass exchange, mixture density is $\rho = \rho_0/J$.
As explained later in the context of \eqref{eq:mDkinhug}, $\{ {\bf z}^\alpha \} \rightarrow \{ z^\alpha \} = 0$. The analog of \eqref{eq:RHe0} for the mixture reduces to
\eqref{eq:fluidEB} with 
$U_0 = 0$ and $p_0 = \sum_\alpha n^\alpha_0 p^\alpha_{{\rm R} 0}$.
In calculations, $J$ is reduced incrementally from unity. In each decrement, energy equation \eqref{eq:fluidEB} and  constitutive equations for each phase are solved simultaneously and summed, if appropriate, to give mixture values $P$, $U$, 
and $\theta$. Given $\theta$ and $J$, entropy $\eta^\alpha$ of each phase is found by inversion of \eqref{eq:tEOSi}.
Particle and shock velocities are found from mixture analogs of \eqref{eq:veljumps}.
The response of the fluid phase (ECF, $\alpha = 2$) is calculated as before; its energy and pressure contributions are  given fully by $U^\alpha_{\rm V}$ and $p^\alpha_{\rm V}$.

The tissue phase (including intracellular fluids), $\alpha = 1$, has a total internal energy per unit
reference volume $U^\alpha$:
\begin{align}
\label{eq:muscen}
& U^1 = U_{\rm V}^1 + \varsigma^1_{\rm S} \cdot (U^1_{\rm S} +  U^1_{\Gamma}) +  \varsigma^1_{\rm F} \circ (U^1_{\rm F} +  U^1_{\Phi}) + U^1_{\rm D}.
\end{align}
The first term on the right is the EOS (noting $\varsigma_{\rm V} = 1$ for compression), second and third are deviatoric matrix elasticity and viscoelasticity,
third and fourth are deviatoric fiber elasticity and viscoelasticity, and the last is surface energy of soft-tissue degradation (i.e., damage). Only passive states are modeled: $U^1_{\rm A} = 0$.
Thermal variables $\theta^1$ and $\eta^1$ only enter $U_{\rm V}^1$, which fully specifies the partial pressure $p^1$.
Notation $U$ and $\Psi$ is interchangeable for remaining terms that only affect deviatoric response.

Matrix viscoelasticity is limited to the shear response, following typical assumptions for nearly incompressible soft materials \cite{holz1996,holz1996b,gultekin2016}.
For very rapid loading modeled here, viscous relaxation for all ($m$) configurational variables ${\bm \Gamma}^1_{{\rm S} m}$ is assumed negligible with $t/\tau^1_{{\rm S} m} \rightarrow 0$, so 
\begin{align}
\label{eq:muscvisc}
U^1_{\rm S} +  U^1_{\Gamma} \rightarrow U^1_{\rm S} + \sum_m \hat{\Psi}^1_{{\rm S} m} = U^1_{\rm S} ( 1 + \sum_m {\beta}^1_{{\rm S} m}). 
\end{align}
Thus, $\check{\mu}^1_{\rm S}  = \mu^1_{\rm S} (1 + \sum_m \beta^1_{{\rm S} m})$ is the glassy shear modulus of the matrix, with static energy and modulus in \eqref{eq:PsiUdev}.

Muscle fibers comprise one family, $k = 1$, of direction ${\bm \iota}^\alpha_k = {\bm \iota}$ with
$\kappa^1_1 = 0$ in \eqref{eq:structens}.
Viscous relaxation for all ($n$) configurational variables ${\bm \Gamma}^1_{\Phi k,n}$ is assumed negligible:
\begin{align}
\label{eq:muscviscF}
U^1_{\rm F} +  U^1_{\Phi} \rightarrow U^1_{\rm F} + \sum_n \hat{\Psi}^1_{{\Phi}1, n} = U^1_{\rm F} (1 + \sum_n {\beta}^1_{{\Phi}1,n}).
\end{align}
Static strain energy of fibers $U^1_{\rm F}$ is \eqref{eq:PsiFexp} with $\mu^\alpha_k = \mu^1_1$ and ${\rm H}(\cdot)$ omitted, supporting compressive stress.
A dynamic fiber modulus is $\check{\mu}^1_1 = {\mu}^1_1 (1 + \sum_n {\beta}^1_{{\Phi}1,n})$.
Fiber directions relative to $x = x^1$ are ambiguous in Ref.~\cite{wilgeroth2012}.  Calculations apply loading parallel or transverse to ${\bm \iota}$, both pure mode directions.
In the former, the longitudinal sound speed ${ \mathcal C}$ obeys $\rho_0 {\mathcal C}^2 = B_\eta + p_0+ \frac{4}{3} n_0^1 
(\check{\mu}_{\rm S}^1 + \frac{2}{3} \check{\mu}^1_1)$. In the latter,  
$\rho_0 {\mathcal C}^2 = B_\eta + p_0 + \frac{4}{3} n_0^1 
(\check{\mu}_{\rm S}^1 + \frac{1}{6} \check{\mu}^1_1)$.

Damage order parameters for the matrix, $\bar{D}^\alpha = \bar{D}^1 = \bar{D} \in [0,1]$, and fibers, $D^\alpha_k = D^1_1 = D_1 \in [0,1]$,
degrade respective deviatoric stress contributions from matrix and fibers in \eqref{eq:totstatstress}.
They also supply surface energy $U^1_{\rm D} = \Psi^1_{\rm D}$ in \eqref{eq:fracen}.
Jumps in $\bar{D}$ and $D_k$ are allowed across the shock front.
This necessitates $\bar{\alpha}^\alpha = \alpha^\alpha_{k} = 0 \Rightarrow
\bar{l}^\alpha_{\rm R} = {l}^\alpha_{{\rm R} k } = 0$ in \eqref{eq:fracen} to avoid infinite energy in the front. Gradient energies and conjugate forces
$\bar{\bm \zeta}^\alpha_{\rm D}, {\bm \zeta}^\alpha_{{\rm D} k}$ vanish identically, as do $\{ {\bf z}^\alpha \}$. 
Treatment of shocks as singular surfaces mandates
viscosities $\bar{\nu}^\alpha_{\rm D} = \nu^\alpha_{{\rm D} k } = 0$ for fracture
to avoid infinite dissipation in the shock front if $\bar{D}$ and $D_k$ are discontinuous across the front. Kinetic equations \eqref{eq:TDGLbar} and \eqref{eq:TDGLk} therefore reduce as follows,
with $\bar{\pi}^\alpha_{\rm D}$, $\pi^\alpha_{{\rm D} k}$ in \eqref{eq:piDbar}, \eqref{eq:piDk}:
\begin{align}
\label{eq:mDkinhug}
\bar{\pi}^1_{\rm D} = 0, \qquad \pi^1_{{\rm D} 1} = 0.
\end{align}
For each decrement of $J$, \eqref{eq:mDkinhug} are solved simultaneously for  $\bar{D}$ and $D_1$,
affecting $P$ and $U$ in Hugoniot equation \eqref{eq:fluidEB}.
Damage can be nonzero, so the generalized Finsler metrics of Sec.~\ref{sec3e}
are non-trivial. Here, $\bar{g}_{ij} = \delta_{ij}, \bar{G}^\alpha_{IJ} = \delta_{IJ}$,
with $\hat{g}^i_j$ and $(\hat{G}^\alpha)^I_J$ in \eqref{eq:ghatrelate}--\eqref{eq:ganihat}.
Isotropic matrix damage gives \eqref{eq:gisohat}, anisotropic fiber damage \eqref{eq:ganihat}.
Determinants and their derivatives, \eqref{eq:gderivs} and \eqref{eq:gderivs2},
enter \eqref{eq:fracen} and \eqref{eq:mDkinhug}.
For the present loading and material symmetries, with \eqref{eq:ghatrelate}, 
$\hat{g}^i_j$ and $(\hat{G}^\alpha)^I_J$ do not affect $J$ or ${\rm tr} \, \tilde{\bf C}$.
Finsler or osculating Riemannian metrics enter the analysis only through \eqref{eq:fracen} and \eqref{eq:mDkinhug}.

Properties for the ECF and tissue phase used in calculations are in Table~\ref{table1}.
Experimental data on hydrated muscle (e.g., \cite{wilgeroth2012}) furnish properties of the mixture as a whole, 
not the isolated $\alpha = 1$ phase. Given \eqref{eq:muscstress}, the mixture density, isentropic bulk modulus,
bulk sound speed, volumetric thermal expansion coefficient, specific heat, and Gr\"uneisen parameter
are, respectively, 
\begin{align}
\label{eq:muscprops1}
& \rho_0 = \sum_\alpha n_0^\alpha \rho_{{\rm R}0}^\alpha, \quad B_\eta = 
\sum_\alpha n_0^\alpha B^\alpha_\eta,\\
& c_{\rm B} = \sqrt{ (B_\eta + p_0) / \rho_0}, \quad A =  \sum_\alpha n_0^\alpha A^\alpha, \label{eq:muscprops2}
\\
& \rho_0 c_{\rm p} = \sum_\alpha n_0^\alpha \rho_{{\rm R}0}^\alpha c_{\rm p}^\alpha = 
 \rho_0 c_{\epsilon} ( 1 + A \gamma_0 \theta_0), \label{eq:muscprops3}
 \\  & \gamma_0 ={A  B_\eta}/({\rho_0  c_{\rm p}}) = {A  B_\theta}/({\rho_0  c_{\epsilon}}) ,
 \label{eq:muscpropsfin}
\end{align}
where $c_{\rm p}^\alpha$ is specific heat at constant pressure of phase $\alpha$.
Given properties of the mixture \cite{jones2018,wilgeroth2012} and ECF ($\alpha = 2$), \eqref{eq:muscprops1}--\eqref{eq:muscpropsfin} are inverted and solved for thermoelastic properties of the tissue phase ($\alpha = 1$).
Ultimately, experimental values \cite{wilgeroth2012} of $\rho_0$ and $c_{\rm B}$
yield tissue density and bulk modulus.  Nonlinear bulk stiffening parameters $B^1_{\eta {\rm p} }$ and $k^1_{\rm V}$
are fit to the shock Hugoniot data \cite{wilgeroth2012}.

\begin{figure}[t]
\begin{subfigure}
 (a){\includegraphics[width = 0.36\textwidth]{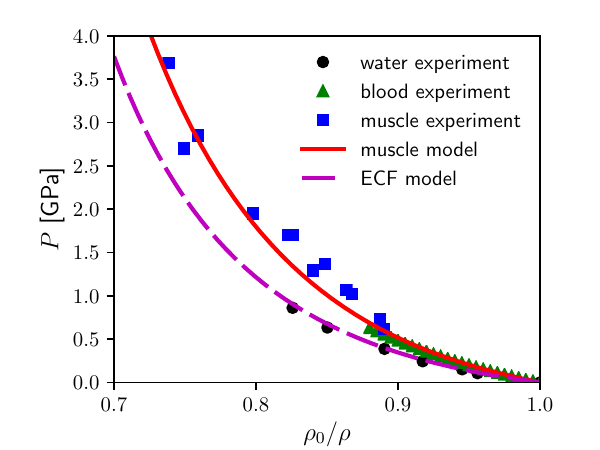} \label{fig2a}}
\end{subfigure} \\
\begin{subfigure}
  (b) {\includegraphics[width = 0.36\textwidth]{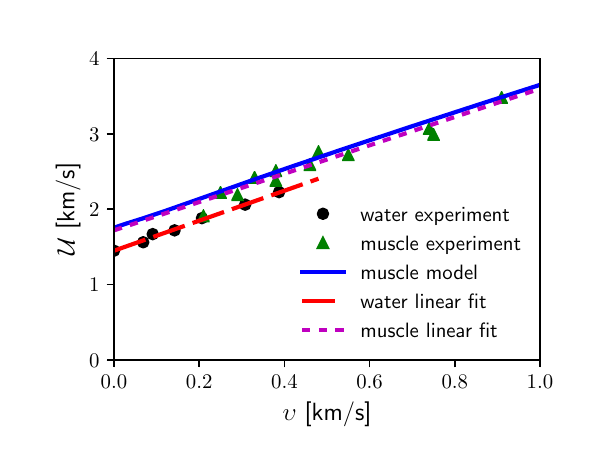} \label{fig2b}} 
\end{subfigure} \\
\begin{subfigure}
 \, (c) {\includegraphics[width = 0.36\textwidth]{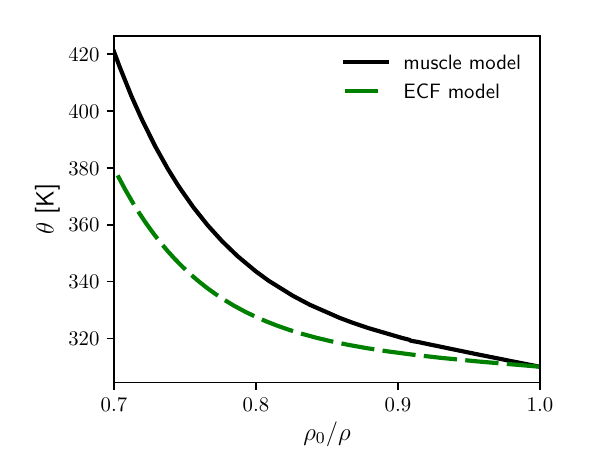} \label{fig2c}} 
\end{subfigure}
\caption{\label{fig2} Model results and/or shock compression data for (a) Hugoniot stress vs.~mass density, (b) shock velocity vs.~particle velocity, and (c) temperature. Data in (a,b) are for water \cite{nagayama2002} human blood \cite{nagoya1995}, and porcine skeletal muscle \cite{wilgeroth2012}. }
\end{figure}

All static and dynamic shear properties are not fully established from Ref.~\cite{wilgeroth2012}, so
order-of-magnitude estimates are used based on literature values \cite{martins1998,song2007,holz2009,morrow2010,ito2010,gultekin2016} for skeletal, and in some cases cardiac,
muscle.
A standard scalar measure \cite{claytonNEIM2019} of shear stress $\tau$ of the mixture for uniaxial shock compression is the first of  
\begin{align}
\label{eq:taumix}
\tau = {\textstyle{\frac{3}{4}}} (P - p); \quad
\tau = - {\textstyle{\frac{3}{4}}} \sum_\alpha [ (\sigma^\alpha)^1_1 + p^\alpha].
\end{align}
If a material is isotropic, $\tau$ is
half the von Mises stress under uniaxial strain.
The second expression in \eqref{eq:taumix} specializes the first.  Both phases $\alpha = 1,2$
contribute to $p$ via each EOS; only the tissue phase contributes to $\tau$ via
deviatoric matrix and fiber, elastic and viscoelastic, stresses. 
Low-rate data \cite{martins1998,song2007,holz2009,morrow2010,ito2010,gultekin2016} suggest $\mu^1_{\rm S}$ and $\mu^1_1$ should be in the kPa range, with fiber exponential stiffening $k^1_1$ on the order of 10.
Define the sums of glassy viscoelastic stiffening factors  $\beta^1_{\rm S} = \sum_m \beta^1_{{\rm S} m}$ 
and ${\beta}^1_{{\Phi}1} = \sum_n {\beta}^1_{{\Phi}1,n}$.
Low- to moderate-rate data on cardiac tissue \cite{gultekin2016} suggest values up to the order of $10^3$.
Dynamic compression data on porcine muscle \cite{song2007} show von Mises stresses in the MPa range
for strain rates on the order of $10^3$/s.  Extrapolating, $\tau$ is anticipated up to the order of 10 MPa
for shock loading, where strain rates appear on the order of  $10^5$/s given rise times under 1 $\mu$s \cite{wilgeroth2012}.  Accordingly, $\beta^1_{\rm S}$ and ${\beta}^1_{{\Phi}1}$ are estimated for shock compression as
$10^5$, probing very high-frequency modes.

For fracture cohesive energy in \eqref{eq:fracen}, the usual phase-field description
is invoked for matrix and fibers: $\bar{E}^1_{\rm C} = \bar{\Upsilon}^1 /\bar{l}^1 =
  \bar {\mathsf J}^1_{\rm C} / (2 \bar{l}^1)$
and ${E}^1_{{\rm C} 1} = {\Upsilon}^1_1 /{l}_1 =   {\mathsf J}^1_{{\rm C}1} / (2 {l}^1_1)$.
Toughness ${\mathsf J}_{\rm C}$ of muscle is known only for the whole tissue \cite{ary2022}, so here
the simplest physical choice $ n_0^1 \bar{\mathsf J}^1_{\rm C} = n_0^1 {\mathsf J}^1_{{\rm C}1} = \frac{1}{2}{\mathsf J}_{\rm C}$ is used.
Similarly, length constants for each mechanism are both set equal to a value
calibrated later for modeling tensile damage: $\bar{l}^1 = l^1_1 = l^1$, on the order of 10-15 single-fiber diameters \cite{wilgeroth2012}.
Values are about $20 \times$ those used elsewhere for modeling skin \cite{claytonSYMM2023}.
Standard phase-field choices $\bar{\vartheta}^1 = \vartheta^1_1 = \vartheta^1 = 2$ \cite{gultekin2019,claytonSYMM2023} are used for degradation functions \eqref{eq:degf1} and \eqref{eq:degf2}.
Regarding generalized Finsler metrics, $\bar{r}^1 = \tilde{r}^1_1 = 2$ is adopted from prior work on skin \cite{claytonSYMM2023}, and remanent microstructure strain factors are set equal:  $\bar{\epsilon}^1 = n_0^1 \bar{\kappa}^1 / \bar{r}^1 = 
\tilde{\epsilon}^1_1 = n_0^1 \tilde{\kappa}^1_1 / \tilde{r}^1_1 = \epsilon^1_{\rm r}$.
Experimental data on vascular tissue \cite{maher2012} furnish $\epsilon_{\rm r} = 0.2$, set
positive (dilative) here, as in other soft tissues \cite{rubin2002,claytonSYMM2023}. Under uniaxial-stress compression \cite{maher2012}, axial shortening is overcompensated by radial and circumferential expansion: the arterial wall tissue is residually stretched. Arterial data  \cite{maher2012} suggest $\epsilon^1_{\rm r}$ is higher (lower) for tissues with more collagen (less elastin), but experimental data on skeletal muscle components do not exist to justify different choices of $\epsilon^1_{\rm r}$ for matrix and fibers.

Shown in Fig.~\ref{fig2}(a) for skeletal muscle is mixture Hugoniot stress $P$ versus mixture density ratio $\rho/ \rho_0$. Experimental data on muscle \cite{wilgeroth2012}, blood \cite{nagoya1995}, and water \cite{nagayama2002} are shown for comparison, along with model predictions for the ECF in isolation.
The mixture theory captures most of the shock data well, exceptions being several anomalous 
points in the domain $\rho_0 / \rho \in [0.82,0.87]$. Similar statements
apply for mixture $\mathcal{U}$ versus $\upsilon$ data \cite{wilgeroth2012} and model results
in Fig.~\ref{fig2}(b).  Muscle is stiffer than blood, ECF, and water.
Hugoniot $\theta$ predictions in Fig.~\ref{fig2}(c) show a substantial temperature rise,
with higher temperatures in muscle than ECF in isolation.
This could cause burn damage, not modeled here.
\begin{figure}[t]
\begin{subfigure}
 (a){\includegraphics[width = 0.32\textwidth]{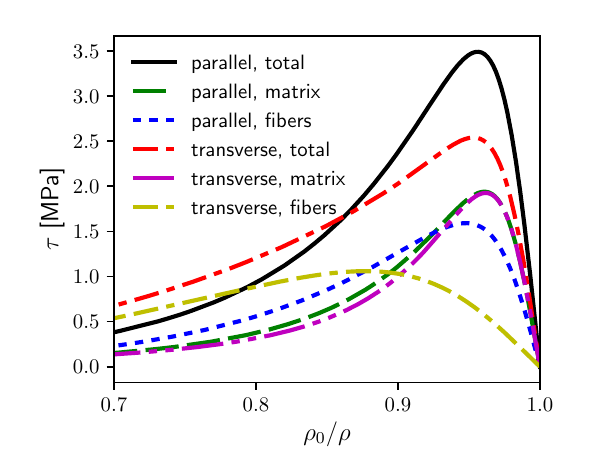} \label{fig3a}}
\end{subfigure} \\
\begin{subfigure}
  (b) {\includegraphics[width = 0.32\textwidth]{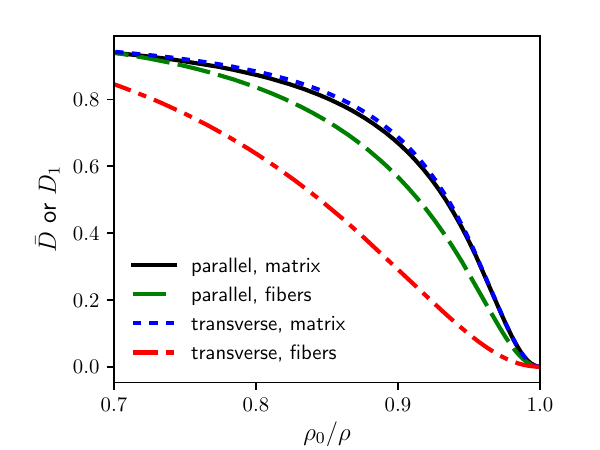} \label{fig3b}} 
\end{subfigure} \\
\begin{subfigure}
  (c) {\includegraphics[width = 0.32\textwidth]{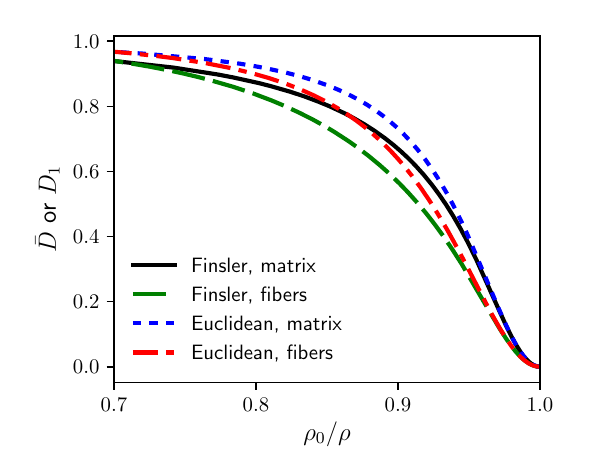} \label{fig3c}} 
\end{subfigure} \\
\begin{subfigure}
 (d) {\includegraphics[width = 0.32\textwidth]{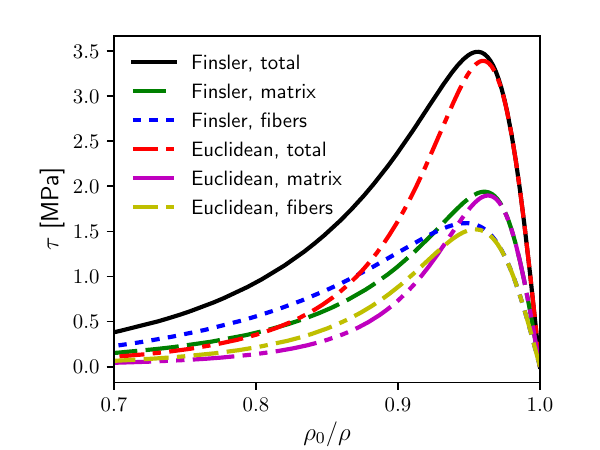} \label{fig3d}} 
\end{subfigure}
\caption{\label{fig3} Predictions for parallel and transverse fiber orientations and Finsler metrics: (a) shear stress vs.~mass density and (b) damage. Model results for Finsler and Euclidean metrics and parallel orientation: (c) damage and (d) shear stress. }
\end{figure}

Results in Fig.~\ref{fig2} are for shock compression parallel to the fiber direction $ {\bm \iota}$.
Shear stress of the mixture (supplied only by the tissue phase) $\tau$ is predicted in Fig.~\ref{fig3}(a) for tissue shocked parallel and transverse to the fiber direction. Contributions of matrix and fiber deviatoric stresses are delineated; these simply sum to give $\tau$.
For parallel compression, matrix and fibers contribute similarly in magnitude. For transverse compression,
fibers support less load, and $\tau$ is lower. In both arrangements,
$\tau$ is at most on the order of $10^{-2}P$, so orientation does not discernibly influence the Hugoniot stress
that is dominated by $p = P - \frac{4}{3} \tau$.
For $\rho_0 / \rho \lesssim 0.95$, damage in matrix and fibers causes a reduction in strength, leading to
reduced $\tau$ at high compressions.
Order parameters $\bar{D}$ (matrix) and $D_1$ (muscle fibers) at Hugoniot states are shown in Fig.~\ref{fig3}(b).
For parallel loading, degradation occurs similarly for matrix and fibers.
For transverse loading, less degradation occurs in the fibers; their strain energy is lower in this
arrangement, giving smaller elastic driving force in $\pi^1_{{\rm D} 1}$.
Shock-recovered samples \cite{wilgeroth2012aip} show microstructure changes indicative of damage
in fibers (myofibrils) and slippage at cellular interfaces, implying matrix damage.
Other experiments, including microscopy and histology after static crushing of muscle,
show shear-induced damage in fibers, interfaces, and extracellular matrix \cite{sanderson1975,winkler2011}.
Predictions agree with these trends.

Model results for muscle in Figs.~\ref{fig2} and \ref{fig3}(a,b) use the generalized Finsler metric with remnant strain $\epsilon^1_{\rm r} = 0.2$ (Table~\ref{table1}).
Predictions in Fig.~\ref{fig3}(c,d) compare aforementioned results for damage and $\tau$ with those obtained
when $\epsilon^1_{\rm r} = 0$, producing Euclidean metrics.
Recall $\epsilon^1_{\rm r} > 0$ depicts dilatation of the material manifold $\mathfrak{M}^\alpha = \mathfrak{M}^1$ as tearing commences and internal surfaces enlarge.
Under shock compression, shear-induced dilatation can occur in solid phases as fracture surfaces slide and open \cite{claytonNEIM2019}. As a result, area of free surfaces increases, leading to an increase in total fracture surface energy in the model at fixed damage order parameters. As corroborated by Fig.~\ref{fig3}(c), damage is suppressed (i.e., more diffuse tearing and rupture) at large deformation when a Finsler metric is used relative to a Euclidean metric. Higher energetic cost of fracture in \eqref{eq:fracen} for the former ($\hat{G}^\alpha > 1$) explains this. Conversely, with higher values of $\bar{D}$ and $D_1$, $\tau$ decays more rapidly with increasing compression in Fig.~\ref{fig3}(d) when a Euclidean metric ($\hat{G}^\alpha = 1$) is used.

For shock stability, $P > 0$, ${\rm d} P/{\rm d} J < 0$, and ${\rm d}^2 P / {\rm d} J^2 > 0$ \cite{chen1971}.
These, and complementary conditions along isentropes
plus $\partial P / \partial \eta > 0$ \cite{coleman1966}, were verified for $J \in [0.7,1]$ for all cases in Fig.~\ref{fig3}.
Damage reduces the tangent shear modulus, but this is more than offset by increasing
tangent elastic bulk and shear moduli under compression.

%% file: s4b.tex
\subsection{\label{sec4b} Static and dynamic uniaxial stress response}

\noindent {\bf Liver.}
The theory is implemented to model uniaxial-stress compression
of bovine liver across a wide range of strain rates as studied experimentally
\cite{pervin2011}, demonstrating efficacy of the model's viscoelastic and damage kinetics. 
Liver parenchyma is comprised of cells (hepatocytes), blood vessels (sinusoids), lymphatic vessels, bile ducts,
and fibrous extracellular matrix (ECM). The organ is encased in a membrane (peritoneum) and
 connective tissue (Glisson's capsule).
In vivo, the liver is internally pressurized, expanded, and perfused with blood, with a fluid volume fraction on the order of 0.5 \cite{ricken2010}.
Most experimental characterizations, including those modeled here \cite{pervin2011},
consider excised samples of the parenchyma, initially at ambient pressure (i.e., not perfused), 
excluding the peritoneum, Glisson's capsule, and major vessels and ligaments.
In these cases, initial blood volume is substantially lower, and the response is
usually isotropic.

The material is depicted as a mixture of two phases ($N = 2$):
a solid tissue phase ($\alpha =1$) and a fluid phase ($\alpha = 2$) consisting
of blood. The EOS used in Sec.~\ref{sec4a} is reinvoked, with properties in
Table~\ref{table1}, any differences between bovine and human blood ignored. In the non-perfused state, the initial fluid fraction is $n_0^2 = 0.12$ \cite{bonfig2010}, the solid fraction $n_0^1 = 0.88$.
Effects of intracellular and extracellular fluids other than blood are encompassed by the EOS of the
first phase, with free energy of \eqref{eq:EOSf1} and \eqref{eq:EOSf2}. In addition, free energy of the solid phase ($\alpha = 1$) consists of matrix deviatoric elastic ($\Psi^1_{\rm S}$) and viscoelastic ($\Psi^1_{\Gamma}$)
terms, fiber elastic ($\Psi^1_{{\rm F} 1}$) and viscoelastic  ($\Psi^1_{{\Phi} 1}$) terms, and damage to
matrix and fibers ($\Psi^1_{\rm D}$).  A single fiber family is sufficient ($k =1$),
fully dispersed with $\kappa^\alpha_k \rightarrow \kappa^1_1 = \frac{1}{3}$ in \eqref{eq:structens} for isotropy. 
Damage order parameters for matrix and fibers,  $\bar{D}^\alpha \rightarrow \bar{D}^1 = \bar{D}$  and $D^\alpha_k \rightarrow D^1_1 = D_1$, reduce deviatoric stress in \eqref{eq:totstatstress} and
furnish surface energy in \eqref{eq:fracen}.
For loading rates up to the order of $10^3$/s and viscosities $\hat{B}^\alpha, \hat{\mu}^\alpha$ in Table~\ref{table1}, viscous stress from blood should not exceed tens of Pa.
This is negligible relative to total stresses in the kPa to MPa range \cite{pervin2011}, and thus ignored.
For compression, cavitation damage in the fluid is irrelevant.

The sample is a cylindrical annulus \cite{pervin2011}, deformed uniformly in the longitudinal (i.e., axial) direction
to a stretch of $F^1_1(t) = \lambda(t) = 1 - \dot{\epsilon} t \leq 1$ at  constant
``engineering strain'' rate of $\dot{\epsilon}$.
A Cartesian coordinate frame defines the axial 1-direction and orthogonal (radial) 2- and 3-directions.
Longitudinal velocity history $(\upsilon^\alpha)^1(t)$ and axial deformation gradient $(F^\alpha)^1_1(t)$ are identical in each phase.
In the initial state, partial pressure in each phase is equilibrated to reference ambient
pressure: $p^\alpha_0 = n_0^\alpha p^\alpha_{{\rm R} 0}$ with $p^\alpha_{{\rm R} 0} = 1$ atm.
At low rates ($\dot{\epsilon} \lesssim 10^2$/s), isothermal conditions apply:
$\theta^\alpha = \theta_0 = 310$ K. For high rates ($\dot{\epsilon} \gtrsim 10^2$/s), 
macroscopically adiabatic conditions apply: ${\bf q}^\alpha = {\bf 0}$. Interphase mass transfer is excluded: $\hat{c}^\alpha = 0$.

Two different boundary conditions are considered for transverse (i.e., radial) stress and deformation.
First is a ``drained'' condition, whereby each phase expands or contracts independently
to maintain equilibrium with external atmosphere:
$(\sigma^\alpha)^2_2 = (\sigma^\alpha)^3_3= p^\alpha_0$.
Transverse velocities $(\upsilon^1)^2 = (\upsilon^1)^3$ and $(\upsilon^2)^2 = (\upsilon^2)^3$
are not necessarily equal in each phase, so transverse diffusion velocities $(\mu^\alpha)^2,(\mu^\alpha)^3$
need not vanish. Hydraulic conductivity is assumed infinite as a limiting case, so $\lambda^{\alpha \beta} \rightarrow 0$ and ${\bf h}^\alpha \rightarrow {\bf 0}$ in \eqref{eq:hdef}.
For the drained case, each $\alpha$ likewise maintains its own temperature $\theta^\alpha(t)$, with $\omega^{\alpha \beta} \rightarrow 0$  in \eqref{eq:epsdef} as a similarly limiting case, so $\epsilon^\alpha \rightarrow 0$.
Temperatures are updated by solving \eqref{eq:temprate} separately for $\alpha = 1,2$.

Second is an ``undrained'' or ``tied'' condition, whereby each phase expands or contracts radially
with the same transverse velocity history.  All diffusion velocities vanish: ${\bm \mu}^\alpha = {\bf 0}$.
Transverse deformation is obtained by equilibrating the total transverse stress of \eqref{eq:stresstot} to atmospheric pressure: $\sigma^2_2 = \sum_{\alpha =1,2} (\sigma^\alpha)^2_2 = p^\alpha_{{\rm R} 0} = 1$ atm.
Consistently, for high-rate loading, each phase has the same temperature: $\theta^1(t) = \theta^2(t) = \theta(t)$, updated by integrating the sum of \eqref{eq:temprate} over $\alpha =1,2$. Undrained conditions are consistent with limiting very high, yet still finite, $\lambda^{\alpha \beta} \tilde{\rightarrow} \infty$
and $\omega^{\alpha \beta} \tilde{\rightarrow} \infty$, so ${\bf h}^\alpha \rightarrow {\bf 0}$ and $\epsilon^\alpha \rightarrow 0$.

Axial deformation $\lambda(t)$ is imposed over small time steps $\Delta t$.
Thermomechanical responses of
each phase and the mixture as a whole are obtained by solution and integration of
the constitutive (i.e., stress-deformation-temperature) equations, \eqref{eq:temprate} if high-rate loading, and kinetic laws for viscoelasticity and damage. For viscoelasticity, two relaxation modes are sufficient for the deviatoric
matrix ($m = 1,2$ in \eqref{eq:Q1}) and one for fiber ($n=1$ in \eqref{eq:Q1f}) contributions to stress,
with volumetric (pressure) contributions omitted for reasons explained in Sec.~\ref{sec4a}.
The algorithm of Refs.~\cite{holz1996b,gultekin2016} is used to solve \eqref{eq:Q4} and \eqref{eq:Q3f}.
Damage is absent in the fluid and spatially homogeneous in the solid: $\nabla \bar{D} =
\nabla D_1 = {\bf 0}$, ensuring stress fields are homogeneous within each phase, consistent
with momentum conservation in the absence of acceleration waves. Gradient energies in \eqref{eq:fracen} and conjugate forces $\bar{\bm \zeta}^\alpha_{\rm D}, {\bm \zeta}^\alpha_{{\rm D} k}$ in \eqref{eq:zetabarD}, \eqref{eq:zetafiber} vanish.
Order parameters and dissipated energies are obtained by integrating \eqref{eq:TDGLbar}, \eqref{eq:DDis}, \eqref{eq:TDGLk}, and \eqref{eq:DDisk} over the load history with nonzero fracture viscosities
$\bar{\nu}^1_{\rm D}$ and $\nu^1_{ {\rm D} 1}$.

\begin{table}[]
\caption{\label{table2}%
Physical properties or model parameters ($\alpha = k = 1$) for bovine liver ($n_0^1 = 0.88$) and rabbit muscle ($n_0^1 = 0.9$)} 
\begin{ruledtabular}
\begin{tabular}{lcc}
Property & Bovine Liver & Rabbit Muscle \\
\hline
$\rho^\alpha_{{\rm R} 0}$ [g/cm$^3$]  & 1.06 & 1.10 \\
$B^\alpha_\eta$ [GPa] & 2.67 & 3.28  \\
$c^\alpha_\epsilon$ [J/g$\cdot$K] & 3.51 & 3.25 \\
$\gamma_0^\alpha$ [-]  & 0.114 & 0.313 \\
$B^\alpha_{\eta {\rm p}}$ [-]  & 8.0 & 8.0 \\
$k^\alpha_{\rm V}$ [-]  & 6.0 & 6.0 \\
$\mu^\alpha_{\rm S}$ [kPa]  & 1.0 &  1.0\\
$\mu^\alpha_k$ [kPa]  & 100 & 600 \\
$k^\alpha_k$ [-] &  $1.0 \times 10^{-6}$ & 2.1 \\
$\beta^\alpha_{{\rm S 1}}$ [-] & 20  & 900 \\
$\beta^\alpha_{{\rm S 2}}$ [-] &  150 & $\ldots$ \\
$\beta^\alpha_{\Phi k, 1}$ [-] &  1.0 & 0.1 \\
$\tau^\alpha_{{\rm S 1}}$ [s] & 0.05  & 0.05 \\
$\tau^\alpha_{{\rm S 2}}$ [s] & $1.0 \times 10^{-3}$  & $\ldots$ \\
$\tau^\alpha_{\Phi k, 1}$ [s] & $1.0 \times 10^{-3}$ & 0.05 \\
$\hat{\nu}^\alpha_{\rm D}$ [s]  & 0.05 & 0 \\
${\mathsf J}_{\rm C} $ [kJ/m$^2$]  &  0.08 & 0.84 \\
$l^\alpha$ [mm]   &  1.00 & 0.88 \\
$\vartheta^\alpha$ [-] &  2.0 & 2.0 \\
${\epsilon}^\alpha_{\rm r}$ [-] & 0.2 & 0.2 \\
$\bar{r}^\alpha = \tilde{r}^\alpha_k $ [-] & 2.0 & 2.0 \\
 \end{tabular}
\end{ruledtabular}
\end{table}

\begin{figure}[]
\begin{subfigure}
 (a){\includegraphics[width = 0.36\textwidth]{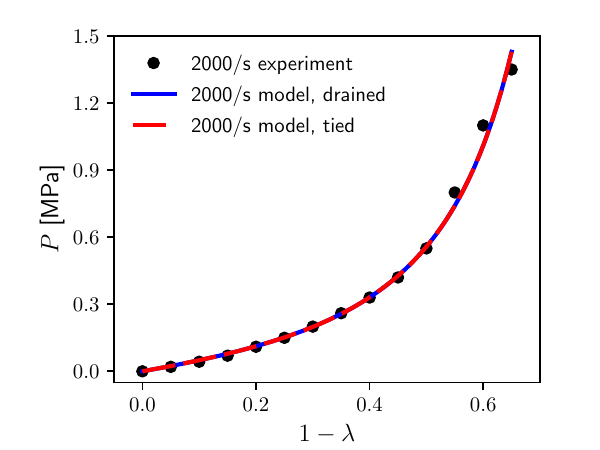} \label{fig4a}}
\end{subfigure} \\
\begin{subfigure}
  (b) {\includegraphics[width = 0.36\textwidth]{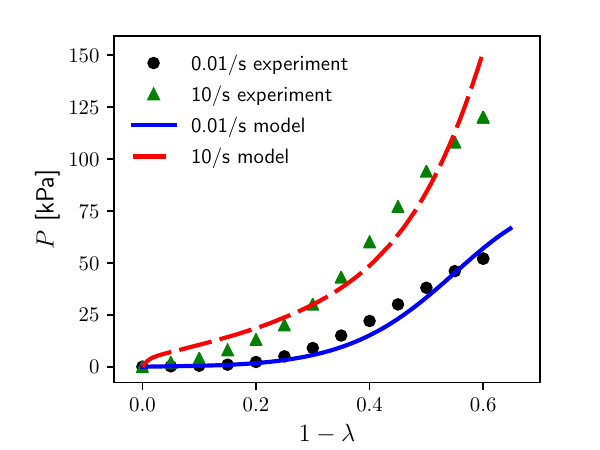} \label{fig4b}} 
\end{subfigure} \\
\begin{subfigure}
  \quad (c) {\includegraphics[width = 0.36\textwidth]{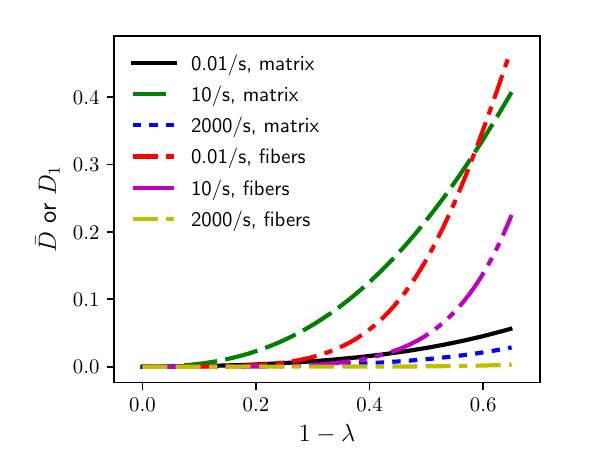} \label{fig4c}} 
\end{subfigure}
\caption{\label{fig4} Model results and experimental data \cite{pervin2011} for
bovine liver compressed to stretch $\lambda$: (a) axial stress $P$ at strain rate $\dot{\epsilon} = 2000$/s, (b) $P$ at $\dot{\epsilon} = 0.01$/s and $\dot{\epsilon} = 10$/s, and (c) predicted matrix and fiber damage $\bar{D}$ and $D_1$ at all three rates}
\end{figure}

Properties for the isolated solid tissue phase ($\alpha = 1$) of bovine liver are given in Table~\ref{table2}.
EOS properties, namely $\rho_{ {\rm R} 0}^\alpha$, $B^\alpha_\eta$, $\gamma_0^\alpha$, and $c^\alpha_\epsilon$, are calculated from mixture rules in \eqref{eq:muscprops1}--\eqref{eq:muscpropsfin} using
known values for the fluid ($\alpha = 2$) phase (i.e., blood in Table~\ref{table1}), $n_0^1 = 0.88$ \cite{bonfig2010}, and available properties for the liver as a whole (solid + fluid) \cite{gao2010,sour2010,jones2018}.
Bulk nonlinear stiffening coefficients $k^\alpha_{\rm V}$ and $B^\alpha_{\eta {\rm p}} = B^\alpha_{\theta {\rm p}}$
are assumed identical to those of skeletal muscle in Table~\ref{table1} since high-pressure data are not available for their determination.  Values are inconsequential for pressures obtained here under uniaxial-stress compression,
wherein $|J^\alpha - 1| < 10^{-4}$.
Shear moduli $\mu^\alpha_{\rm S}$ and $\mu^\alpha_{k}$, stiffening $k^\alpha_k$, viscoelastic strength factors $\beta^\alpha_{ {\rm S} m}$ and $\beta^\alpha_{ {\Phi}k,n}$,
and relaxation times $\tau^\alpha_{ {\rm S} m}$ and $\tau^\alpha_{ {\Phi}k,n}$ 
for $\alpha = k=1$, $m = 1,2$ and $n = 1$ are fit to data \cite{pervin2011} in Fig.~\ref{fig4}(a,b) at rates $\dot{\epsilon} = 0.01$/s, $\dot{\epsilon} = 10$/s, and $\dot{\epsilon} = 2000$/s.
Total first Piola-Kirchhoff or ``engineering'' stress magnitude for this purpose, noting $J^1 \approx 1$ and $ (\sigma^\alpha)^1_1 \leq - n_0^\alpha p^\alpha_{{\rm R} 0}$, is measured relative to ambient pressure $ p^\alpha_{{\rm R} 0}$:
\begin{equation}
\label{eq:PK1}
P = \frac{ J^1 |(\sigma^1_1 + p^1_{{\rm R} 0})|}{(F^1)^1_1}
 = \frac{J^1}{ \lambda} \bigr{ \lvert} \sum_\alpha [(\sigma^\alpha)^1_1 + n_0^\alpha p^\alpha_{{\rm R} 0}] \bigr{\rvert}.
\end{equation}

Fracture toughness of the mixture, ${\mathsf J}_{\rm C}$, is obtained from Ref.~\cite{azar2008}, presumed similar for porcine and bovine liver.
Procedures of Sec.~\ref{sec4a} 
give  $ n_0^1 \bar{\mathsf J}^1_{\rm C} = n_0^1 {\mathsf J}^1_{{\rm C}1} = \frac{1}{2}{\mathsf J}_{\rm C}$.
Length constants for matrix and fibers are set equal to the value
 in Table~\ref{table2} to best represent data in Fig.~\ref{fig4}(a,b): $\bar{l}^1 = l^1_1 = l^1 = 1$ mm.
Recalling $\bar{E}^1_{\rm C} = \bar{\Upsilon}^1 /\bar{l}^1 =
  \bar {\mathsf J}^1_{\rm C} / (2 \bar{l}^1)$
and ${E}^1_{{\rm C} 1} = {\Upsilon}^1_1 /{l}_1 =   {\mathsf J}^1_{{\rm C}1} / (2 {l}^1_1)$,
for homogeneous damage, \eqref{eq:fracen} and evolution of order parameters
depend only on the ratio of toughness to length (i.e., cohesive energies $\bar{E}^1_{\rm C}$,
 ${E}^1_{{\rm C} 1}$) and not toughness and length independently. 
If gradient regularization lengths $\bar{l}^1_{\rm R}, l^1_{ {\rm R} 1}$ must be chosen in \eqref{eq:fracen} based on mesh size constraints rather than physical observations (e.g., the smaller value of 0.1 mm used for arterial rupture in Ref.~\cite{gultekin2019}), $\bar{\alpha}^1 = \bar{l}^1_{\rm R} / \bar{l}^1$ and
 $\alpha^1_1 = l^1_{ {\rm R} 1}/l^1_{1}$ can be invoked independently
without affecting the cohesive energy.
Standard values $\bar{\vartheta}^1 = \vartheta^1_1 = \vartheta^1 = 2$ \cite{gultekin2019,claytonSYMM2023} enter \eqref{eq:degf1} and \eqref{eq:degf2}.
The same rate dependence of damage,  $\hat{\nu}^\alpha_{\rm D}$, normalized by cohesive energy $E^\alpha_{\rm C}  = \bar{E}^\alpha_{\rm C} = E^\alpha_{{\rm C} k}$ and with units of time, is used for matrix and fibers ($\alpha = k = 1$): $\bar{\nu}_{\rm D}^\alpha = \nu^\alpha_{{\rm D} k} = E^\alpha_{\rm C} \hat{\nu}^\alpha_{\rm D}$. The value in Table~\ref{table2} produces credible results in the context of Fig.~\ref{fig4}(c). The same parameters for generalized Finsler metrics ${\bf G}^\alpha$ are used for liver and muscle, explained in Sec.~\ref{sec4a}.

The high-rate response of compressed liver is shown in Fig.~\ref{fig4}(a).
Model results assume adiabatic conditions, but predicted temperature change is negligible.
Damage order parameters attained small maxima at high rates (i.e., $\bar{D} \lesssim 0.03$ and $D_1 \lesssim 0.003$ in Fig.~\ref{fig4}(c))
due to the viscosity $\hat{\nu}^\alpha_{\rm D}$ preventing notable degradation over the brief time period of deformation. This is consistent with  data \cite{pervin2011} that show continuously increasing stiffness
at this loading rate, with no evidence of material failure. Differences between drained and undrained conditions are indiscernible because the solid tissue is nearly incompressible.

Low- and moderate-rate stress histories are reported in Fig.~\ref{fig4}(b). Model results are isothermal and for drained conditions only; results for undrained, not shown, are nearly identical.
Fits to data are reasonable but not as close as those for high-rate loading in Fig.~\ref{fig4}(a). At low rates, data show reduction in the degree of stiffening at large compression, indicative
of strength degradation \cite{fung1993,yang2015,mitsuhashi2018}.
This phenomenon is captured by damage kinetics, Fig.~\ref{fig4}(c).
Damage increases monotonically with compressive strain $1-\lambda$, and at moderate- to high-rates is more severe in matrix than fibers.
Histological analysis after dynamic blunt impact \cite{malec2021} showed fractures in liver parenchyma
avoid fibers and interlobular septa and more often propagate along interfaces, consistent with higher levels of ``matrix'' damage $\bar{D}$ relative to fiber damage $D_1$ in Fig.~\ref{fig4}(c).
Conversely, at the lowest strain rate (0.01/s), fiber damage overtakes matrix damage at large compression ($\lambda \lesssim 0.77$) , and
is more pronounced due to longer load times for relaxation to ensue.
\\

\noindent {\bf Skeletal muscle.}
The theory is now implemented to study uniaxial-stress tensile behavior of
rabbit skeletal muscle at low and moderate strain rates, with and without activation
from electrical stimulation. Model results seek to depict experiments reported in Refs.~\cite{taniguchi2003,ito2010}.

Calculations proceed in the same manner as just described for modeling liver, with a few exceptions.
First, tension is modeled rather than compression, with stretch ratio 
$F^1_1(t) = \lambda(t) = 1+ \dot{\epsilon} t \geq 1$ at two
rates \cite{taniguchi2003}: $\dot{\epsilon} = 0.17$/s and $\dot{\epsilon} = 15$/s.
Engineering tensile stress is $P$ of \eqref{eq:PK1}, where now  $ (\sigma^\alpha)^1_1 \geq -n_0^\alpha p^\alpha_{{\rm R} 0}$.  Both drained and undrained conditions are considered,
all isothermal.
Second, the data do not indicate any consistent rate sensitivity of damage or failure stretch \cite{ito2010},
so equilibrium equations \eqref{eq:mDkinhug} used in Sec.~\ref{sec4a} for porcine muscle still apply.
 These correspond to \eqref{eq:TDGLbar} and \eqref{eq:TDGLk} with null viscosities $\bar{\nu}_{\rm D}^\alpha = \nu^\alpha_{{\rm D} k} \rightarrow 0$, giving zero dissipation in \eqref{eq:DDis} and \eqref{eq:DDisk}. 
Lastly, active tension, irrelevant for liver, is considered for muscle.
The form of free energy $\Psi^\alpha_{\rm A}$ and Cauchy stress term $ {\bm \sigma}^\alpha_{\rm A}$ in \eqref{eq:Psiact}
and \eqref{eq:PsiAsig} are adapted directly from Ref.~\cite{ito2010} since
they reproduce the over-stress from activation recorded in isometric experiments \cite{taniguchi2003}:
\begin{align}
\label{eq:PsiA1}
& \Psi^1_{\rm A} = {\textstyle{\frac{2}{3}}} \Delta_{\rm A} \mu_{\rm A} (\lambda_{{\rm A} 1} - \lambda_{{\rm A} 0}) 
[ \bar{\Lambda}^{ p_{\rm A}} /  p_{\rm A} - \bar{\Lambda}^{ r_{\rm A}} /  r_{\rm A} ],
\\ \label{eq:SigA1}
&  {\bm \sigma}^1_{\rm A} =  {\textstyle{\frac{2}{3}}}  \frac{\rho^\alpha}{\rho^\alpha_{ {\rm R} 0} } 
 \frac{ \Delta_{\rm A} \mu_{\rm A} }{\lambda^1_1} \bar{\Lambda}^{ p_{\rm A}-1}[1 - \bar{\Lambda}^{ q_{\rm A}-1}]
 \tilde{\bf h}^1_1,
 \\  \label{eq:barLam}
&  \bar{\Lambda} = \begin{cases}  
&  \frac{\lambda^1_1 - \lambda_{{\rm A} 0}}{ \lambda_{{\rm A} 1} - \lambda_{{\rm A} 0}} \quad \forall \, \lambda^1_1 \in 
(\lambda_{{\rm A} 0},\lambda_{{\rm A} 1}), \\
 & 0 \qquad \qquad \qquad (\text{otherwise}).
 \end{cases}
  \end{align}
Active tension vanishes for $\bar{\Lambda}$ outside domain $[0,1]$. Recall fiber stretch obeys $(\lambda^\alpha_k)^2 = {I^\alpha_k} = {\tilde{\bf C}^\alpha:{\bf H}^\alpha_k}$ with $\alpha = k = 1$.
  Parameters are $\mu_{\rm A}$ (stress units) and dimensionless set $(\lambda_{{\rm A} 0},\lambda_{{\rm A} 1}, p_{\rm A},q_{\rm A})$,
  with  $r_{\rm A} = p_{\rm A} + q_{\rm A} - 1$.
  A dimensionless internal variable for activation is $\{ \Delta_k^\alpha (t) \} \rightarrow \Delta_1^1(t)  = \Delta_{\rm A} $.
 Only discrete states are considered:  $\Delta_{\rm A} = 1$ in the fully active state and $\Delta_{\rm A} = 0$ in the fully passive state.  Transient switching between states and partial activation are not addressed here or in the experiments \cite{taniguchi2003,ito2010}.
 Energy $\chi^\alpha_k$ in \eqref{eq:Psiact} and kinetic law \eqref{eq:actkin} need not be prescribed; no contribution to dissipation arises since  $D_t^1 \Delta_{\rm A} = 0$ in \eqref{eq:dissiA}.
 The $\frac{2}{3}$ in \eqref{eq:PsiA1} is omitted in Ref.~\cite{ito2010} where compressibility is ignored.

As prescribed in Sec.~\ref{sec4a} for porcine muscle, rabbit muscle consists of solid tissue $\alpha = 1$ and ECF ($\alpha =2$), where parameters for ECF are in Table~\ref{table1}. The initial solid volume fraction remains $n_0^1 = 0.9$. Parameters for rabbit skeletal muscle are compiled in Table~\ref{table2}. 
Thermophysical properties entering the EOS are identical to those for porcine tissue in Table~\ref{table1}.
Shear properties $\mu^\alpha_{\rm S}$, $\mu^\alpha_{k}$, and $k^\alpha_k$ are calibrated to the data \cite{taniguchi2003,ito2010}, with $k = 1$ sufficient.  
Fibers are fully aligned, $\kappa^\alpha_k \rightarrow \kappa^1_1 = 0$ in \eqref{eq:structens},
giving anisotropic response. The glassy viscoelastic assumption used for modeling shocks in Sec.~\ref{sec4a} is inappropriate for low and moderate strain rates.
Instead, viscoelastic strength factors $\beta^\alpha_{ {\rm S} m}$ and $\beta^\alpha_{ {\Phi}k,n}$
and relaxation times $\tau^\alpha_{ {\rm S} m}$ and $\tau^\alpha_{ {\Phi}k,n}$ are fit to  experimental data; here,
a single mode suffices: $m = n = 1$.
The activation parameters in \eqref{eq:PsiA1} are verbatim from Ref.~\cite{ito2010}:
$\mu_{\rm A} = 962$ kPa, $\lambda_{{\rm A} 0} = 0.9$, $\lambda_{{\rm A} 1} = 1.32$, $p_{\rm A}= 1.65$, and $q_{\rm A}= 2.0$. 
Assumptions for properties modulating matrix and fiber damage are the same as those explained for porcine tissue in Sec.~\ref{sec4a}, with matching toughness $\mathsf{J}_{\rm C}$ \cite{ary2022}.
Length $l^1 = \bar{l}^1 = l^1_1$ provides cohesive energies $E_{\rm C} = \bar{E}_{\rm C} = E_{{\rm C} 1}$ in \eqref{eq:fracen}.  The value in Table~\ref{table2} best fits softening and failure in test data \cite{taniguchi2003,ito2010} at the lowest loading rate. Finsler metric parameters in Table~\ref{table2} match those of porcine muscle in Table~\ref{table1}; none are adjusted when fitting the data.
 
 For tensile loading, cavitation of the fluid ($\alpha =2$) is not impossible. 
 Calibration of the model for water under isentropic expansion to its 8.7 MPa
 cavitation stress \cite{boteler2004} gives $\bar{E}^\alpha_{ {\rm C}} = 0.1818$ MPa.
 Use of the same cohesive energy for ECF gives a cavitation stress of 8.9 MPa, and
 $J^\alpha \gtrsim 1.001$ ($\alpha =2$) is needed to initiate detectable damage $\bar{D}^\alpha$. Such expansion is never achieved in the present modeling of muscle response: damage in the ECF is negligible here.
 
 \begin{figure}[]
\begin{subfigure}
 (a){\includegraphics[width = 0.36\textwidth]{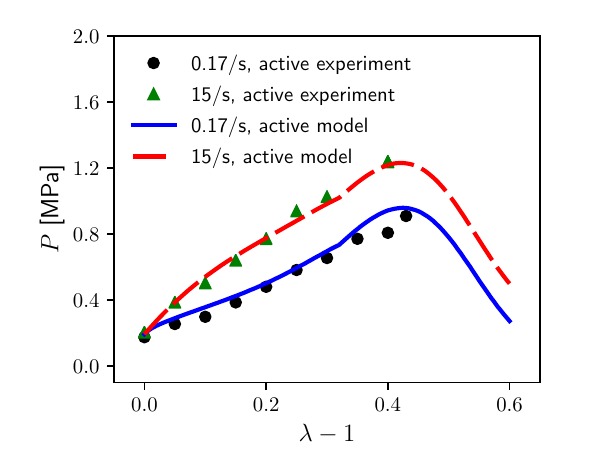} \label{fig5a}}
\end{subfigure} \\
\begin{subfigure}
  (b) {\includegraphics[width = 0.36\textwidth]{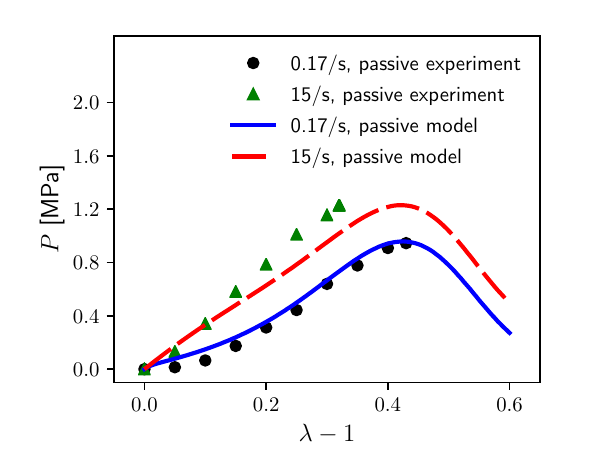} \label{fig5b}} 
\end{subfigure} \\
\begin{subfigure}
  \quad (c) {\includegraphics[width = 0.36\textwidth]{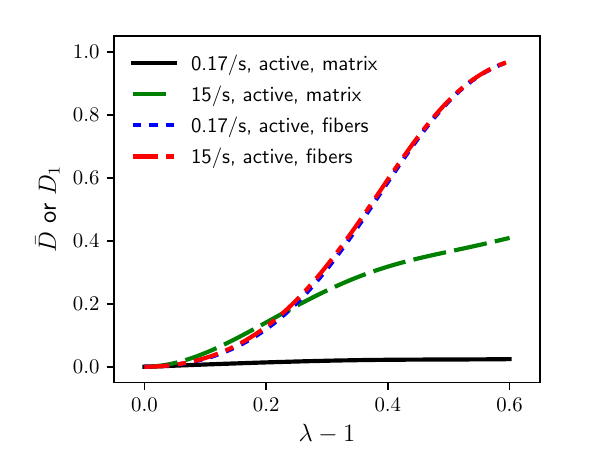} \label{fig5c}} 
\end{subfigure}
\caption{\label{fig5} Model results and experimental data \cite{taniguchi2003,ito2010} for
rabbit skeletal muscle to tensile stretch $\lambda$ at rates of $\dot{\epsilon} = 0.17$/s and  $\dot{\epsilon} = 15$/s: (a) axial stress $P$ in active state ($\Delta_{\rm A} = 1$), (b) $P$ in passive state ($\Delta_{\rm A} = 0$), and (c) predicted matrix and fiber damage $\bar{D}$ and $D_1$, active state (passive nearly identical)}
\end{figure}
 
 Model outcomes and experimental stress-stretch data are compared in Fig.~\ref{fig5}(a) for active
 states and Fig.~\ref{fig5}(b) for passive states,
 at engineering strain rates of $\dot{\epsilon} = 0.17$/s and $\dot{\epsilon} = 15$/s. 
 The fiber direction ${\bm \iota} = {\bm \iota}^1_1$ is aligned with the direction of elongation.
Model results in Fig.~\ref{fig5} are
 for drained lateral boundaries; predictions for undrained conditions are nearly indiscernible from drained and thus not shown.
 For active and passive states, the material is stiffer at the higher rate, with larger peak (failure) stress. Predicted failure stretch is similar at both strain rates.
 The material supports larger $P$ in the active state over domain $1 \leq \lambda \lesssim 1.32$, including
 an initial stress of $P = 0.192$ MPa at $\lambda(t = 0) = 1$ that matches experiments.
 
  The model closely depicts the majority of data points \cite{taniguchi2003,ito2010}, an exception under-prediction
  of large $\lambda$-$P$ data at $\dot{\epsilon} = 15$/s for the passive state in Fig.~\ref{fig5}(b). 
  An over-prediction of peak stress for active loading at the higher rate was obtained from
   a phenomenological model \cite{ito2010}.
   That model, however, contained five adjustable parameters, whereas only the length parameter $l^1$ was
   adjusted for damage modeling in application of the present theory.
   More elaborate coupling among viscoelastic and damage kinetics, with more parameters, could improve agreement, but such an exercise is unjustified for closer fitting of relatively few data points. Unlike results for compression in Sec.~\ref{sec4a} where the matrix and fibers 
   supported similar stress, here, under tensile loading, the fibers bear the majority of the load $P$,
   with the ratio of fiber to matrix stress increasing as stretch increases and rate decreases.
   The correspondingly larger strain energy in the fibers provides a larger driving force for fiber damage $D_1$ than
   matrix damage $\bar{D}$, which is nearly negligible at $\dot{\epsilon} = 0.17$/s, as shown in Fig.~\ref{fig5}(c).
   As discussed in Ref.~\cite{lamsfuss2022}, under tensile loading at low rates, damage to muscle fibers
   is prominently observed over delamination and damage to the endomysium (i.e., the matrix including connective tissue in which fibers are embedded).
   The current predictions concur with these observations.
   At the higher rate of $\dot{\epsilon} = 15$/s, viscoelastic energy of the matrix is sufficient to induce matrix damage,
   though it remains less severe than fiber damage for $\lambda \gtrsim 1.23$, which is nearly the same at both rates.
   The damage model is decoupled from $\Psi_{\rm A}$,
   so order parameters $\bar{D}$ and $D_1$ have indistinguishable histories for active versus passive states.

%% file: s4c.tex
\subsection{\label{sec4c} Shock evolution}

\noindent {\bf Analytical solution.}
Growth and decay of planar shock waves are studied, with shock fronts treated
as singular surfaces per theory of Sect.~\ref{sec2b}. An analytical solution is
derived for solid-fluid mixtures with viscoelastic and damage mechanisms, extending
Ref.~\cite{claytonIJES2022} that considered nonlinear elastic solid-fluid mixtures without
internal variables and Ref.~\cite{chen1971} that considered fluids with internal state variables. To streamline notation, let internal variables
$\{ {\bm \xi}^\alpha \} \rightarrow ({\bm{\mathsf{a}}}^\alpha, {\bm{\mathsf {b}}}^\alpha )$. 
Generic internal variable class ${\bm{\mathsf{a}}}^\alpha$ obeys kinetic laws
of form \eqref{eq:isvdot} specialized to
\begin{equation}
\label{eq:adot}
D_t^\alpha {\bm{\mathsf{a}}}^\alpha = D_t^\alpha {\bm{\mathsf{a}}}^\alpha ({\bf F}^\alpha,
{\eta}^\alpha({\bf F}^\alpha,
{\theta}^\alpha, {\bm{\mathsf{a}}}^\alpha, {\bm{\mathsf{b}}}^\alpha), {\bm{\mathsf{a}}}^\alpha, {\bm{\mathsf{b}}}^\alpha).
\end{equation}
Generic class ${\bm{\mathsf{b}}}^\alpha$ obeys equilibrium conditions of form
\begin{equation}
\label{eq:bequil}
{\bm \pi}^\alpha_{\mathsf b} = \rho^\alpha \partial u^\alpha / \partial {\bm{\mathsf{b}}}^\alpha = {\bf 0}.
\end{equation}
Type ${\bm{\mathsf{a}}}^\alpha$ include dissipative variables for viscoelasticity and
order parameters for rate-dependent fracture; ${\bm{\mathsf{b}}}^\alpha$
include rate-insensitive order parameter(s). Now excluding gradient regularization and explicit ${\bf X}^\alpha$-dependence, 
\begin{equation}
\label{eq:unograd}
u^\alpha = u^\alpha ({\bf F}^\alpha,\eta^\alpha, {\bm{\mathsf{a}}}^\alpha, {\bm{\mathsf{b}}}^\alpha ), \, 
 \theta^\alpha = \theta^\alpha  ({\bf F}^\alpha,\eta^\alpha, {\bm{\mathsf{a}}}^\alpha, {\bm{\mathsf{b}}}^\alpha)
\end{equation}
are internal energy and temperature of \eqref{eq:legendre} and \eqref{eq:legendre2}.
Unless $\llbracket {\bm \xi}^\alpha \rrbracket = {\bf 0}$, gradient regularization yields infinite energy density in the shock front as its width approaches zero.  It also introduces complexity in Rankine-Hugoniot equations via 
${\bm \zeta}^\alpha$, precluding
analytical solutions without undue assumptions on shock structure \cite{claytonIJF2017} not used here. 

From \eqref{eq:ghatrelate}, 
 $\hat{g} = \hat{G}^\alpha \Rightarrow g/G^\alpha = 1$. Then 
1-D kinematics, continuum balance
 laws, and entropy production are, with $\hat{c}^\alpha = 0$, ${\bf q}^\alpha = {\bf 0}$, $r^\alpha = 0$, $h^\alpha = {\bf h}^\alpha \cdot {\bf n}$, and $P^\alpha = - t^\alpha$,
 \begin{align}
 \label{eq:evolcont0}
 & F^\alpha = \partial \chi^\alpha / \partial X^\alpha = J^\alpha, \qquad D_t^\alpha J^\alpha = 
 \partial \upsilon^\alpha / \partial X^\alpha,  
 \\
 \label{eq:evolcont1}
& \rho_0^\alpha = J^\alpha \rho^\alpha , \quad 
 \rho_0^\alpha D_t^\alpha \upsilon^\alpha = -(\partial P^\alpha / \partial X^\alpha) + J^\alpha h^\alpha,  \\
 \label{eq:evolcont2}
 & \rho_0^\alpha D_t^\alpha u^\alpha = - P^\alpha (\partial \upsilon^\alpha / \partial X^\alpha) + J^\alpha \epsilon^\alpha,
 \\ 
 \label{eq:evolcont3}
 & 
 \rho_0^\alpha \theta^\alpha D_t^\alpha \eta^\alpha = J^\alpha ( \epsilon^\alpha -  {\bm \pi}^\alpha_{\mathsf a} \cdot D_t^\alpha {\bm{\mathsf{a}}}^\alpha),
 \\ 
 \label{eq:evolcont4}
 & \sum_\alpha ({J^\alpha }/{\theta^\alpha})  (\epsilon^\alpha - {\bm \pi}^\alpha_{\mathsf a} \cdot D_t^\alpha {\bm{\mathsf{a}}}^\alpha + c^\alpha \theta^\alpha \eta^\alpha) \geq 0.
 \end{align}
 
 A single shock propagates in the ($x,X^\alpha$)-direction at Lagrangian speed $\mathcal{U}^\alpha$. Ahead of the shock front $\Sigma^\alpha$, each phase $\alpha$ obeys equilibrium and uniformity conditions:
 \begin{align}
 \label{eq:plusstate}
 & J^{\alpha+} = 1, \quad \upsilon^{\alpha +} = 0, \quad \theta^{\alpha +} = \theta_0, \quad
 \eta^{\alpha+} = {\rm constant}
 \nonumber
 \\
 &  \Rightarrow \, \rho^{\alpha +} = \rho_0^\alpha, \quad
 h^{\alpha +} = 0, \quad \epsilon^{\alpha +} = 0, \quad {\mathcal U^\alpha} = {\mathcal U};
 \\  \label{eq:plusisv}
 &
 {\bm{\mathsf{a}}}^{\alpha +} = {\bm{\mathsf{a}}}^{\alpha}_0 = {\rm constant},  
 \quad
 {\bm{\mathsf{b}}}^{\alpha +} = {\bm{\mathsf{b}}}^{\alpha}_0 = {\rm constant}.
 \end{align}
 Rankine-Hugoniot equations ($P^\alpha \! > 0 \! \Leftrightarrow \! {\rm compression}$) are 
 \begin{align}
 \label{eq:evolRH1}
 & \llbracket \upsilon^\alpha \rrbracket = - {\mathcal U} \llbracket J^\alpha \rrbracket, \qquad  \llbracket P^\alpha \rrbracket = -\rho_0^\alpha {\mathcal U}^2 \llbracket J^\alpha \rrbracket, 
 \\
 \label{eq:evolRH2}
 & \rho_0^\alpha \llbracket u^\alpha \rrbracket = - \langle P^\alpha \rangle \llbracket J \rrbracket, \qquad
 \sum_\alpha \rho_0^\alpha \llbracket \eta^\alpha \rrbracket \geq 0. 
 \end{align}
 To avoid infinite dissipation in the shock front \cite{chen1971,morro1980,morro1980b},
 \begin{align}
 \label{eq:jumpevola}
 \llbracket  {\bm{\mathsf{a}}}^\alpha \rrbracket = {\bf 0} \, \Rightarrow \, {\bm{\mathsf{a}}}^{\alpha-}
 = {\bm{\mathsf{a}}}^\alpha_0.
 \end{align}
 
  Jumps $ \llbracket  {\bm{\mathsf{b}}}^\alpha \rrbracket $ in non-dissipative variables can be nonzero across $\Sigma^\alpha$ so long as \eqref{eq:bequil} holds. However, it is assumed that
 \eqref{eq:bequil} can be solved, at least implicitly, at any $(X^\alpha, t)$ with the first of each of \eqref{eq:unograd}, \eqref{eq:evolcont1}, and then via \eqref{eq:jumpevola},
 \begin{equation}
 \label{eq:bsoln}
  {\bm{\mathsf{b}}}^\alpha  = \bar{\bm{\mathsf{b}}}^\alpha (J^\alpha, \eta^\alpha, {\bm{\mathsf{a}}}^\alpha), \, \, \, 
   {\bm{\mathsf{b}}}^{\alpha -} = \bar{\bm{\mathsf{b}}}^{\alpha -} (J^{\alpha -}, \eta^{\alpha-}, {\bm{\mathsf{a}}}^\alpha_0).
 \end{equation}
 Now with \eqref{eq:jumpevola} and \eqref{eq:bsoln} at a given $(\cdot)^+$ state, 
the first of \eqref{eq:evolRH2} can be written
 $H(J^{\alpha-},\eta^{\alpha-},\bar{\bm{\mathsf{b}}}^{\alpha-} (J^{\alpha -}, \eta^{\alpha-})) = 0$. Again, at least implicitly, this can be solved for entropy along the Hugoniot and then the other state variables:
 \begin{align}
 \label{eq:evolhug}
& \llbracket \eta^\alpha \rrbracket = \eta^\alpha_{\rm H} (\llbracket J^\alpha \rrbracket),  \quad
  \llbracket  {\bm{\mathsf{b}}}^\alpha  \rrbracket =  {\bm{\mathsf{b}}}^\alpha_{\rm H} 
  (\llbracket J^\alpha \rrbracket),
 \\ \label{eq:evolhug2}
 &  \llbracket P^\alpha \rrbracket = P^\alpha_{\rm H} (\llbracket J^\alpha \rrbracket), 
 \quad
   \llbracket u^\alpha \rrbracket = u^\alpha_{\rm H} (\llbracket J^\alpha \rrbracket).
 \end{align}
 Hugoniot states do not depend explicitly on $h^{\alpha-}$ or $\epsilon^{\alpha -}$.
 From \eqref{eq:consteq3} with ${\bf F}^\alpha = {\rm diag}(J^\alpha,1,1)$, note, then define
 \begin{align}
 \label{eq:PJ}
 & P^\alpha = - \rho^\alpha_0 \partial u^\alpha / \partial J^\alpha, \quad
 \theta^\alpha = \partial u^\alpha / \partial \eta^\alpha;
 \\ & 
 \label{eq:Ctan}
 {\mathsf C}^\alpha = - \partial P^\alpha / \partial J^\alpha = \rho^\alpha_0 \partial^2 u^\alpha / 
 \partial (J^\alpha)^2,    \\
 \label{eq:Gtan}
 & 
  {\mathsf G}^\alpha = - \partial P^\alpha / \partial \eta^\alpha = 
  - \rho^\alpha_0 J^\alpha \theta^\alpha (\gamma^\alpha)^1_1,
   \\
 \label{eq:ABtan}
 & {\bm{\mathsf{A}}}^\alpha =   - \partial P^\alpha / \partial  {\bm{\mathsf{a}}}^\alpha,
 \quad
   {\bm{\mathsf{B}}}^\alpha =   - \partial P^\alpha / \partial  {\bm{\mathsf{b}}}^\alpha,
    \\
\label{eq:btan}
& {\bm{\mathsf{b}}}'^{\alpha} = \partial \bar{\bm{\mathsf{b}}}^{\alpha}/ \partial J^\alpha ,
\quad
 {\bm{\mathsf{b}}}''^{\alpha} = \partial^2 \bar{\bm{\mathsf{b}}}^{\alpha}/ (\partial J^\alpha)^2, 
   \\
\label{eq:Ctanhat}
& \hat {\mathsf C}^\alpha = {\mathsf C}^\alpha +  {\bm{\mathsf{B}}}^{\alpha} \cdot {\bm{\mathsf{b}}}'^{\alpha},
\quad
{\bm{\mathsf{B}}}'^{\alpha} = \partial {\bm{\mathsf{B}}}^{\alpha}/ \partial J^\alpha,
  \\
 \label{eq:Ctanprime}
& {\mathsf C'}^\alpha =  \partial {\mathsf C} ^\alpha / \partial J^\alpha = \rho^\alpha_0 \partial^3 u^\alpha / 
 \partial (J^\alpha)^3, \\
\label{eq:Ctanhatprime}
& \hat {\mathsf C}'^\alpha = {\mathsf C}'^\alpha +  {\bm{\mathsf{B}}}'^{\alpha} \cdot {\bm{\mathsf{b}}}'^{\alpha} + {\bm{\mathsf{B}}}^{\alpha} \cdot {\bm{\mathsf{b}}}''^{\alpha},
\\
\label{eq:hatG}
& \hat{\mathsf G}^\alpha = {\mathsf G}^\alpha + {\bm{\mathsf{B}}}^{\alpha} \cdot {\bm{\mathsf{b}}}_\eta^{\alpha}.
 \end{align}
 
 Weak shocks are analyzed in the limit $\llbracket J^\alpha \rrbracket \rightarrow 0$.
Applying theorems \cite{coleman1966} relating isentropic and tangent moduli,
\eqref{eq:evolhug} and \eqref{eq:evolhug2} are expanded from a $(\cdot)^+$ state:
\begin{align}
\label{eq:taylorP}
& \llbracket P^\alpha \rrbracket = -\hat{\mathsf C}^{\alpha+}  \llbracket J^\alpha \rrbracket
-  {\textstyle{\frac{1}{2}}} \hat{\mathsf C}'^{\alpha+} \llbracket J^\alpha \rrbracket^2 + O(\llbracket J^\alpha \rrbracket^3),
 \\
 \label{eq:tayloru}
& \llbracket \rho_0^\alpha u^\alpha \rrbracket = -P^{\alpha+}  \llbracket J^\alpha \rrbracket
+ {\textstyle{\frac{1}{2}}} \hat{\mathsf C}^{\alpha+} \llbracket J^\alpha \rrbracket^2 + O(\llbracket J^\alpha \rrbracket^3),
 \\
  \label{eq:taylorT}
& \llbracket \rho_0^\alpha \theta^\alpha \rrbracket = \hat{\mathsf G}^{\alpha+}  \llbracket J^\alpha \rrbracket
+ {\textstyle{\frac{1}{2}}} (\textstyle{ \frac{\partial \hat{\mathsf G}}{ \partial J }})^{\alpha +} 
 \llbracket J^\alpha \rrbracket^2 + O(\llbracket J^\alpha \rrbracket^3),
 \\
  \label{eq:taylorb}
& \llbracket  {\bm{\mathsf{b}}}^\alpha \rrbracket = {\bm{\mathsf{b}}}'^{\alpha+}  \llbracket J^\alpha \rrbracket
+ {\textstyle{\frac{1}{2}}} {\bm{\mathsf{b}}}''^{\alpha+}
 \llbracket J^\alpha \rrbracket^2 + O(\llbracket J^\alpha \rrbracket^3),
 \\
\label{eq:taylorent}
& \llbracket \eta^\alpha \rrbracket = {\textstyle{\frac{1}{12}}} \{ \hat{\mathsf C}'^{\alpha+}/(\rho_0^\alpha \theta_0) \} \llbracket J^\alpha \rrbracket^3 + O(\llbracket J^\alpha \rrbracket^4). 
\end{align}
From \eqref{eq:taylorent}, $\llbracket \eta^\alpha \rrbracket$ is of order three
in $\llbracket J^\alpha \rrbracket$, and $\llbracket J^\alpha \rrbracket \leq 0$ when $\hat{\mathsf C}'^{\alpha+} < 0$
to satisfy the second of \eqref{eq:evolRH2} for a single $\alpha$.
Interactions ${\bf h}^\alpha$ and $\epsilon^\alpha$ are respectively
odd and even functions of $\upsilon^\beta$ \cite{bowen1974,claytonIJES2022}, so from \eqref{eq:plusstate}, \eqref{eq:evolRH1}, and \eqref{eq:taylorT},
\begin{align}
\label{eq:taylorh}
& 
 h^{\alpha - } = - {\mathcal U} \sum_\beta
(\partial h^\alpha / \partial \upsilon^\beta)^+ \llbracket J^\beta \rrbracket 
+ O(\llbracket J^\beta \rrbracket^2),
\\
\label{eq:tayloreps}
& \epsilon^{\alpha -} =  \sum_\beta
(\partial \epsilon^\alpha / \partial \theta^\beta)^+ ( \hat{ \mathsf G}^\beta /\rho^\beta_0)^{+}  \llbracket J^\beta \rrbracket 
+ O(\llbracket J^\beta \rrbracket^2).
\end{align}
From \eqref{eq:evolRH2} and \eqref{eq:taylorP}, $\mathcal U^\beta$ approaches the sound speed:
\begin{align}
\label{eq:Usound}
&  ({\mathcal U}^\beta)^2 = ({\mathcal C}^\beta)^2 + 
 {\textstyle{\frac{1}{2}}} (\hat{\mathsf C}'^\beta / \rho^\beta_0)^{ +}
  \llbracket J^\beta \rrbracket + O(\llbracket J^\beta \rrbracket^2),  \nonumber  
 \\ 
 & \mathcal{C}^\beta = \sqrt{ \hat{\mathsf C}^{\beta +} / \rho_0^\beta }.
\end{align}
Considering $\alpha \neq \beta$ and ${\mathcal C}^\alpha \neq {\mathcal C}^\beta$, \eqref{eq:Usound} implies \cite{bowen1974,claytonIJES2022}
\begin{align}
\label{eq:Umult1}
& \{ ({\mathcal C}^\alpha)^2 - ({\mathcal C}^\beta)^2 \} \llbracket J^\beta \rrbracket 
+ O  (\llbracket J^\alpha \rrbracket \llbracket J^\beta \rrbracket)
+ O  (\llbracket J^\beta \rrbracket^2) = 0 \nonumber 
\\ & \Rightarrow
\, { \mathcal U}^2  = ({\mathcal C}^\alpha)^2 + O  (\llbracket J^\alpha \rrbracket)
\quad  ({\rm for \, one} \, \llbracket J^\alpha \rrbracket \neq 0), \nonumber \\
& \quad \llbracket J^\beta \rrbracket = 0 \, \, (\forall \, \beta  = 1,2,\ldots,\alpha-1,\alpha+1,\ldots,N).
\end{align}
Each distinct wave speed ${\mathcal C}^\alpha$ corresponds to isolated jump $\llbracket J^\alpha \rrbracket$. From \eqref{eq:taylorP}--\eqref{eq:taylorent}, a weak shock in phase $\alpha$ does not induce jumps $\llbracket P^\beta \rrbracket$, $\llbracket u^\beta \rrbracket$, $\llbracket \theta^\beta \rrbracket$, etc.~in other phases $\beta$. Since $\llbracket J^\beta \rrbracket = 0$ for $\beta \neq \alpha$, \eqref{eq:taylorh} and \eqref{eq:tayloreps} become
\begin{align}
\label{eq:taylorh2}
& 
 h^{\alpha - } = - {\mathcal C}^\alpha 
(\partial h^\alpha / \partial \upsilon^\alpha)^+ \llbracket J^\alpha \rrbracket 
+ O(\llbracket J^\alpha \rrbracket^2),
\\
\label{eq:tayloreps2}
& \epsilon^{\alpha -} = 
(\partial \epsilon^\alpha / \partial \theta^\alpha)^+ ( \hat{ \mathsf G}^\alpha /\rho^\alpha_0)^{+}  \llbracket J^\alpha \rrbracket 
+ O(\llbracket J^\alpha \rrbracket^2),
\end{align}
that also apply for $h^{\beta-}, \epsilon^{\beta-}$, $\beta \neq \alpha$ with
$(\partial h^\alpha / \partial \upsilon^\alpha)^+ \rightarrow (\partial h^\beta / \partial \upsilon^\alpha)^+$ and $(\partial \epsilon^\alpha / \partial \theta^\alpha)^+ \rightarrow
(\partial \epsilon^\beta / \partial \theta^\alpha)^+$ \cite{bowen1974}.

Resuming analysis of the nonlinear (i.e., strong-shock) regime,
denote by $f^\alpha$ any of ($J^\alpha$, $P^\alpha$, $\eta^\alpha$, $\theta^\alpha$, $\rho^\alpha$, $\upsilon^\alpha$, $u^\alpha$, ${\bm{\mathsf{b}}}^\alpha $).
Recall that across surface $\Sigma^\alpha$, $f^\alpha$, $D_t^\alpha f^\alpha$, and $\nabla^\alpha_0 f^\alpha = \partial f^ \alpha / \partial X^\alpha$ can be discontinuous.
From \eqref{eq:jumpevola}, ${\bm{\mathsf{a}}}^\alpha $ is continuous; however, $ D_t^\alpha {\bm{\mathsf{a}}}^\alpha$ and $\nabla^\alpha_0 {\bm{\mathsf{a}}}^\alpha$ need not be so. Applying \eqref{eq:deltaderiv} with the last of \eqref{eq:plusstate} gives \cite{chen1971}
\begin{align}
\label{eq:deltajump}
& \delta_t \llbracket f^\alpha \rrbracket = \llbracket D_t^\alpha f^\alpha \rrbracket + {\mathcal U}
 \llbracket \nabla_0^\alpha f^\alpha \rrbracket, \\
\label{eq:deltaa}
&
\llbracket D_t^\alpha {\bm{\mathsf{a}}}^\alpha \rrbracket = - {\mathcal U}
 \llbracket \nabla_0^\alpha {\bm{\mathsf{a}}}^\alpha \rrbracket.
\end{align}
Recall \eqref{eq:velgrad} in 1-D is $D_t^\alpha J^\alpha = \nabla_0^\alpha \upsilon^\alpha$ via \eqref{eq:evolcont0}. 
Using \eqref{eq:deltajump} on $\llbracket J^\alpha \rrbracket$
and $\llbracket \upsilon^\alpha \rrbracket$ and \eqref{eq:deltaderiv} on ${\mathcal U}$ in the first of \eqref{eq:evolRH1} gives
\begin{align}
\label{eq:sdec1}
2 {\mathcal U} \delta_t \llbracket J^\alpha \rrbracket + \llbracket J^\alpha \rrbracket \delta_t {\mathcal U}
= {\mathcal U}^2 \llbracket \nabla_0^\alpha J^\alpha \rrbracket - \llbracket D_t^\alpha \upsilon^\alpha \rrbracket.
\end{align}
From \eqref{eq:evolcont1},  \eqref{eq:evolcont3}, \eqref{eq:plusstate}, \eqref{eq:evolRH1}, and \eqref{eq:PJ}--\eqref{eq:ABtan},
\begin{align}
\label{eq:Dtv}
& \rho_0^\alpha \llbracket D_t^\alpha \upsilon^\alpha \rrbracket = 
{\mathsf C}^{\alpha -} \llbracket \nabla_0^\alpha J^\alpha \rrbracket + 
{\mathsf G}^{\alpha -} \llbracket \nabla_0^\alpha \eta^\alpha \rrbracket  + J^{\alpha -} \llbracket h^\alpha \rrbracket \nonumber \\
& \qquad \qquad \quad \, \, +
{\bm{\mathsf{A}}}^{\alpha -} \cdot \llbracket \nabla_0^\alpha   {\bm{\mathsf{a}}}^\alpha \rrbracket + 
{\bm{\mathsf{B}}}^{\alpha -} \cdot \llbracket \nabla_0^\alpha   {\bm{\mathsf{b}}}^\alpha \rrbracket , 
\\
\label{eq:Dteta}
& \rho_0^\alpha \theta^{\alpha -} \llbracket D_t^\alpha \eta^\alpha \rrbracket = J^{\alpha -}( \llbracket \epsilon^\alpha \rrbracket - {\bm \pi}^{\alpha -}_{\mathsf a} \cdot \llbracket D_t^\alpha {\bm{\mathsf{a}}}^\alpha \rrbracket ),
\end{align}
where ${\bm \pi}^\alpha_{\mathsf a} = \rho^\alpha \partial u^\alpha / \partial {\bm{\mathsf{a}}}^\alpha$.
With $f^\alpha \rightarrow \eta^\alpha$, putting \eqref{eq:Dteta} into \eqref{eq:deltajump}  gives
$\llbracket \nabla_0^\alpha \eta^\alpha \rrbracket$ in terms of $\delta_t \llbracket \eta^\alpha \rrbracket$
and jumps on the right side of \eqref{eq:Dteta}. This $\llbracket \nabla_0^\alpha \eta^\alpha \rrbracket$ is substituted for the second term on the right in \eqref{eq:Dtv}. The fourth term on the right of \eqref{eq:Dtv} is $- \frac{1}{\mathcal U} {\bm{\mathsf{A}}}^{\alpha -} \cdot
 \llbracket D_t^\alpha {\bm{\mathsf{a}}}^\alpha \rrbracket$ via \eqref{eq:deltaa}. Using this result,
\eqref{eq:btan}, and \eqref{eq:Dteta}, the fifth term includes
\begin{align}
\label{eq:fifthterm}
& 
\llbracket \nabla_0^\alpha   {\bm{\mathsf{b}}}^\alpha \rrbracket =
{\bm{\mathsf{b}}}'^{\alpha -} \llbracket \nabla_0^\alpha J^\alpha \rrbracket
- (1/ {\mathcal U}) \, {{\bm{\mathsf{b}}}_{\mathsf a}^{\alpha -}} \cdot \llbracket D_t^\alpha {\bm{\mathsf{a}}}^\alpha 
\rrbracket
\nonumber 
\\ &
+ \frac{{\bm{\mathsf{b}}}_\eta^{\alpha -}}{\mathcal U} \biggr{\{} \delta_t \llbracket \eta^\alpha \rrbracket
- \frac{J^{\alpha -} (\llbracket \epsilon^\alpha \rrbracket -
{\bm \pi}^{\alpha -}_{\mathsf a} \cdot \llbracket D_t^\alpha {\bm{\mathsf{a}}}^\alpha \rrbracket ) }
{\rho_0^\alpha \theta^{\alpha - }} \biggr{\}},
\\ 
& {\bm{\mathsf{b}}}_{\mathsf a}^{\alpha} = \partial \bar{\bm{\mathsf{b}}}^\alpha / \partial {\bm{\mathsf{a}}}^\alpha, 
\qquad {\bm{\mathsf{b}}}_\eta^{\alpha} = \partial \bar{\bm{\mathsf{b}}}^\alpha / \partial \eta^\alpha.
\end{align}
Putting \eqref{eq:fifthterm} into \eqref{eq:Dtv}, the latter is
inserted into \eqref{eq:sdec1}:
\begin{align}
\label{eq:Jdec1}
& \delta_t \llbracket J^\alpha \rrbracket = \frac{1}{2 {\mathcal U}} \biggr{\{} 
\biggr{(} {\mathcal U}^2 - \frac{ \hat{\mathsf C}^{\alpha -}}{\rho_0^\alpha} \biggr{)} \llbracket \nabla_0^\alpha J^\alpha \rrbracket 
- \frac{ \hat{\mathsf G}^{\alpha -}}{\rho_0^\alpha {\mathcal U}}
\delta_t \llbracket \eta^\alpha \rrbracket
\nonumber
\\ & 
\qquad \qquad - \delta_t {\mathcal U} \llbracket J^\alpha \rrbracket  
- \frac{J^{\alpha -}  }{\rho_0^\alpha}  \llbracket h^\alpha \rrbracket 
+ \frac{J^{\alpha -}  \hat{\mathsf G}^{\alpha -} }{(\rho^\alpha_0)^2 {\mathcal U} \theta^{\alpha -} } \llbracket \epsilon^\alpha \rrbracket
\nonumber
\\ &
\qquad \qquad + {\bm{\mathsf{L}}}^{\alpha -} \cdot \llbracket D_t^\alpha {\bm{\mathsf{a}}}^\alpha \rrbracket \biggr{\}},
\\ 
\label{eq:Ldef}
& {\bm{\mathsf{L}}}^\alpha = \frac{1}{ \rho_0^\alpha {\mathcal U}} \biggr{\{}
{\bm{\mathsf{A}}}^{\alpha } + {\bm{\mathsf{B}}}^{\alpha } \cdot  {\bm{\mathsf{b}}}_{\mathsf a}^{\alpha}
- \frac{J^{\alpha}  \hat{\mathsf G}^\alpha
} {\rho_0^\alpha \theta^{\alpha }}
 {\bm \pi}^{\alpha }_{\mathsf a} \biggr{\}}.
\end{align}
From \eqref{eq:bequil}, \eqref{eq:evolRH1}, \eqref{eq:evolRH2}, \eqref{eq:PJ}, \eqref{eq:Ctanhat}, \eqref{eq:deltaa}, and \eqref{eq:hatG},
\begin{align}
\label{eq:deltau}
& \rho_0^\alpha \delta_t \llbracket u^\alpha \rrbracket = 
-(P^{\alpha -} + \rho_0^\alpha {\mathcal U}^2 \llbracket J^\alpha \rrbracket ) \delta_t \llbracket J^\alpha \rrbracket \nonumber 
\\ & \qquad \qquad \qquad -  \llbracket J^\alpha \rrbracket \delta_t P^{\alpha -}
- \rho_0^\alpha {\mathcal U}  \llbracket J^\alpha \rrbracket^2 \delta_t {\mathcal U}, 
\\ & 
\rho_0^\alpha \delta_t \llbracket u^\alpha \rrbracket = - P^{\alpha -} \delta_t \llbracket J^\alpha \rrbracket + \rho_0^\alpha \theta^{\alpha -} \delta_t \llbracket \eta^\alpha \rrbracket,
\\
\label{eq:delPminus}
& \delta_t P^{\alpha -} = \delta_t \llbracket P^\alpha \rrbracket = - \hat{C}^{\alpha -} \delta_t \llbracket J^\alpha \rrbracket
- \hat{G}^{\alpha -} \delta_t \llbracket \eta^\alpha \rrbracket .
\end{align}
Using \eqref{eq:deltau}--\eqref{eq:delPminus} to eliminate $\delta_t \llbracket u^\alpha \rrbracket$ and
$ \delta_t P^{\alpha -}$, then differentiating \eqref{eq:evolRH1} produces the following:
\begin{align}
\label{eq:uelim}
&  \rho_0^\alpha \llbracket J^\alpha \rrbracket^2 {\mathcal U} \delta_t {\mathcal U}
 = {\mathsf G}^{\alpha -}  \llbracket J^\alpha \rrbracket ( 1 - 1/ \hat{\zeta}^\alpha) \delta_t \llbracket \eta^\alpha \rrbracket \nonumber \\ & 
  \qquad \qquad  \qquad \quad + \hat{\mathsf C}^{\alpha -} ( 1 - \hat{\xi}^\alpha) \llbracket J^\alpha \rrbracket,
\\
\label{eq:zetaxi}
&  \hat{\xi}^\alpha = \rho_0^\alpha {\mathcal U}^2 / \hat{\mathsf C}^{\alpha -}, \quad \hat{\zeta}^\alpha = \hat {\mathsf G}^{\alpha -}  \llbracket J^\alpha \rrbracket / (\rho^\alpha_0 \theta^{\alpha -});
\\
\label{eq:deltaUU}
& 2 \rho_0^\alpha \llbracket J^\alpha \rrbracket {\mathcal U} \delta_t {\mathcal U} =
\hat{\mathsf C}^{\alpha -} (1 - \hat{\xi}^\alpha ) \delta_t \llbracket J^\alpha \rrbracket
+ \hat{G}^{\alpha -} \delta_t \llbracket \eta^\alpha \rrbracket.
\end{align}
Solving \eqref{eq:uelim} and \eqref{eq:deltaUU} for $\delta_t {\mathcal U}$ and $\delta_t \llbracket \eta^\alpha \rrbracket$, then insertion in \eqref{eq:Jdec1} with
\eqref{eq:plusstate} and \eqref{eq:delPminus} yields the fully nonlinear shock evolution equations for $\delta_t \llbracket J^\alpha \rrbracket$ and $\delta_t \llbracket P^\alpha \rrbracket$:
\begin{align}
\label{eq:deltaUfin}
& 
\delta_t {\mathcal U} = {\mathcal U}  \frac{(1- \hat{\xi}^\alpha)}{ (2 - \hat{\zeta}^\alpha)  \hat{\xi}^\alpha \llbracket J^\alpha \rrbracket } \delta_t \llbracket J^\alpha \rrbracket ,
\\ 
\label{eq:deltaetafin}
& 
\delta_t \llbracket \eta^\alpha \rrbracket = 
\frac{ \hat{\mathsf C}^{\alpha -}} { \hat{\mathsf G}^{\alpha -}} 
\frac { \hat{\zeta}^\alpha (1- \hat{\xi}^\alpha)} { (2 - \hat{\zeta}^\alpha) } \delta_t \llbracket J^\alpha \rrbracket,
\\ 
\label{eq:deltaJfin}
& 
\delta_t \llbracket J^\alpha \rrbracket = 
{\mathcal U}  \frac{(1- \hat{\xi}^\alpha) (2 - \hat{\zeta}^\alpha)  \{ \Lambda^\alpha - (\nabla_0^\alpha J^{\alpha})^{-} \} }
{ (3 \hat{\xi}^\alpha +1 ) - \hat{\zeta}^\alpha (3 \hat{\xi}^\alpha -1 ) },
\\
\label{eq:deltaPfin}
& 
\delta_t \llbracket P^\alpha \rrbracket = 
- \hat{\mathsf C}^{\alpha -} {\mathcal U}  \frac{ \{ 3 - \hat{\xi}^\alpha (1+ \hat{\zeta}^\alpha) \}
 \{ \Lambda^\alpha - (\nabla_0^\alpha J^{\alpha})^{-} \} }
{ (3 \hat{\xi}^\alpha +1 ) - \hat{\zeta}^\alpha (3 \hat{\xi}^\alpha -1 ) },
\\
\label{eq:Lamfin}
&
\Lambda^\alpha = 
\frac{ 1 + \llbracket J^{\alpha} \rrbracket } {(1- \hat{\xi}^\alpha)  \hat{\mathsf C }^{\alpha -} }
\bigr{\{}
\frac{\rho_0^\alpha}{J^{\alpha -}} [ {\bm{\mathsf{L}}}^{\alpha -}
\cdot (D_t^\alpha  {\bm{\mathsf{a}}}^{\alpha})^{-} ] \nonumber 
\\ & \qquad \qquad \qquad  \quad 
 - h^{\alpha -} 
+ [ {\hat{\mathsf G}^{\alpha -}}/({ \rho_0^\alpha  {\mathcal U} \theta^{\alpha -}})] \epsilon^{\alpha -}  \bigr{\}}.
\end{align}
When $(\nabla_0^\alpha J^{\alpha})^{-}$ equals a critical strain gradient $\Lambda^\alpha$ (a function of $(\cdot)^{-}$ conditions immediately behind the wave front), \eqref{eq:deltaUfin}--\eqref{eq:deltaPfin} vanish so the shock is steady. 

Preceding derivations are for phase $\alpha$. Now let $\mathcal U$ for phase $\alpha$ be imposed simultaneously on $\beta \neq \alpha$. The trivial solution to \eqref{eq:evolRH1} is $\llbracket J^\beta \rrbracket = 0$.
Noting \eqref{eq:deltaUU} and \eqref{eq:deltaetafin} still apply with $\alpha \rightarrow \beta$, substitution of \eqref{eq:deltaUfin} into the former gives a non-trivial solution (i.e., shock interaction law):
\begin{align}
\label{eq:interactlaw}
& \delta_t \llbracket J^\beta \rrbracket = 
\frac{(2 - \hat{\zeta}^\beta) \hat{\xi}^\beta ( 1- \hat{\xi}^\alpha)}
{(2 - \hat{\zeta}^\alpha) \hat{\xi}^\alpha ( 1- \hat{\xi}^\beta) }
\frac{ \llbracket J^\beta \rrbracket}{ \llbracket J^\alpha \rrbracket} 
 \delta_t \llbracket J^\alpha \rrbracket .
\end{align}

In the weak limit, $\mathcal{U} = {\mathcal U}^\alpha  \rightarrow {\mathcal C}^\alpha = {\rm constant}$ and shock evolution depends only on $(\cdot)^+$ states.
As $\llbracket J^\alpha \rrbracket \rightarrow 0$, from \eqref{eq:evolRH1}, 
\eqref{eq:evolRH2}, \eqref{eq:taylorP}, \eqref{eq:taylorT}, \eqref{eq:taylorh}, \eqref{eq:tayloreps}, \eqref{eq:deltaJfin}, and \eqref{eq:Lamfin},
\begin{align}
\label{eq:weakxi}
& \hat{\xi}^\alpha = 1 - {\textstyle{\frac{1}{2}}} ( \hat{\mathsf C}'^{\alpha +} /
\hat{\mathsf C}^{\alpha + }) \llbracket J^\alpha \rrbracket + O ( \llbracket J^\alpha \rrbracket ^2), 
\\
& \hat{\zeta}^\alpha = \{ \hat{\mathsf G}^{\alpha +} / (\rho_0^\alpha \theta_0) \} 
 \llbracket J^\alpha \rrbracket + O ( \llbracket J^\alpha \rrbracket ^2),
 \\
 \label{eq:weaklam}
 & \Lambda^\alpha = - 4 \,  {\mathcal C}^\alpha  \rho_0^\alpha \omega^\alpha / \hat{\mathsf C}'^{\alpha +}
 + O ( \llbracket J^\alpha \rrbracket ), 
 \end{align}
 \begin{align}
  \label{eq:weakomega}
 & \omega^\alpha = - \frac{1}{2 \rho_0^\alpha} \biggr{\{} 
 \frac{1}{({\mathcal C}^{\alpha})^2} \biggr{[}
{\bm{\mathsf{A}}}^{\alpha +} + {\bm{\mathsf{B}}}^{\alpha + } \cdot  {\bm{\mathsf{b}}}_{\mathsf a}^{\alpha +}
- \frac{ \hat{\mathsf G}^{\alpha +}
} {\rho_0^\alpha \theta_0}
 {\bm \pi}^{\alpha +}_{\mathsf a} \biggr{]} 
 \nonumber
 \\
 \nonumber
 & \qquad \qquad \quad
  \cdot \left[ \biggr{(} \frac{ \partial (D_t^\alpha {\bm{\mathsf{a}}}^\alpha)}{\partial J^\alpha} \biggr{)}^+ + 
  \biggr{(} \frac{ \partial (D_t^\alpha {\bm{\mathsf{a}}}^\alpha)}{\partial {\bm{\mathsf{b}}}^\alpha } \cdot  {\bm{\mathsf{b}}}'^\alpha \biggr{)}^+
    \right] 
 \\
 &
 \qquad \quad + \left( \frac{\partial h^\alpha}{\partial \upsilon^\alpha} \right)^+
 +\frac{1}{\theta_0} \left( \frac{ \hat{ \mathsf G}^{\alpha +}}{\rho_0^\alpha {\mathcal C}^\alpha } \right)^2 \left( \frac{\partial \epsilon^\alpha}{\partial \theta^\alpha} \right)^+ \biggr{\}} ,
 \\ 
 \label{eq:weakJjump}
  & \delta_t \llbracket J^\alpha \rrbracket = -\omega^\alpha \llbracket J^\alpha \rrbracket 
  + O( \llbracket J^\alpha \rrbracket  \llbracket \nabla^\alpha_0 J^\alpha \rrbracket ; \llbracket J^\alpha \rrbracket ^2).
 \end{align}
 Omitting higher-order products on the right of \eqref{eq:weakJjump} \cite{bowen1974}, 
 \begin{align}
 \label{eq:Jdecaylin}
  \llbracket J^\alpha \rrbracket(t) = \Delta J^\alpha_0 \exp (-\omega^\alpha t),
  \, \, \, \Delta J^\alpha_0 =  \llbracket J^\alpha \rrbracket (t=0).
 \end{align}
 For small $\llbracket J^\alpha \rrbracket$, if $\nabla_0^\alpha J^\alpha$ remains negligible behind wave front $\Sigma^\alpha$, 
 shock amplitude evolves at a rate determined by $\omega^\alpha = {\rm constant}$.
 Jumps $\llbracket P^\alpha \rrbracket$, $\llbracket u^\alpha \rrbracket$, $\llbracket \theta^\alpha \rrbracket$, and $\llbracket  {\bm{\mathsf{b}}}^\alpha \rrbracket$  evolve proportionally via \eqref{eq:taylorP}--\eqref{eq:taylorb}; $\llbracket \eta^\alpha \rrbracket \rightarrow 0$ by \eqref{eq:taylorent}.
  
Derivations of Sec.~\ref{sec4c} apply trivially for a single-phase material ($\alpha = N = 1$, 
 $h^\alpha \rightarrow 0$, $\epsilon^\alpha \rightarrow 0$).
They can also describe a shock moving with velocity $\mathcal U$ through the mixture as a whole per Sec.~\ref{sec2c}. Assuming
$\upsilon^\alpha = \upsilon$ and $\theta^\alpha = \theta$ for all constituents, then
$J^\alpha = J$, diffusion velocities ${ \bm \mu}^\alpha = {\bm 0}$, and thus $h^\alpha = 0$ and $\epsilon^\alpha = 0$.
Then from Sec.~\ref{sec2c}, mixture quantities include $\psi = u - \theta \eta$ and
\begin{align}
\label{eq:mixshock}
& P = \sum_\alpha P^\alpha, \quad \rho_0 = \sum_\alpha n_0^\alpha \rho^\alpha_{{\rm R} 0}, \nonumber
\quad \rho_0 u = \sum_\alpha \rho_0^\alpha u^\alpha, 
\\
& \hat{\mathsf C} =  \sum_\alpha \hat{\mathsf C}^\alpha,
\quad
 \hat{\mathsf G} =  \sum_\alpha \frac{\partial P^\alpha}{ \partial \eta},
 \quad
 \eta = - \frac{\partial \psi}{\partial \theta}.
\end{align}
Since $u^\alpha$ is independent of $({\bm{\mathsf{a}}}^\beta, {\bm{\mathsf {b}}}^\beta) \, \forall \, \beta \neq \alpha$ per \eqref{eq:unograd}, 
\begin{align}
\label{eq:Lsum}
& {\bm{\mathsf{L}}} \cdot \llbracket D_t^\alpha {\bm{\mathsf{a}}} \rrbracket  = \sum_\alpha {\bm{\mathsf{L}}}^{\alpha} \cdot \llbracket D_t^\alpha {\bm{\mathsf{a}}}^\alpha \rrbracket .
\end{align}

%% file: s4c2.tex
\noindent {\bf Biologic tissue.}
Quantities entering $\Lambda^\alpha$ and $\omega^\alpha$ of \eqref{eq:weaklam} and \eqref{eq:weakomega} are evaluated for constitutive frameworks of Sec.~\ref{sec3}.
First, consider an ideal gas. Variables (${\bm{\mathsf{a}}}^\alpha,{\bm{\mathsf{b}}}^\alpha$) are irrelevant, $\rho_0^\alpha = n_0^\alpha \rho_{ {\rm R} 0}^\alpha$, and
\eqref{eq:idealeos} and \eqref{eq:idealu} give
\begin{align}
\label{eq:idealvars}
& P^{\alpha+} = n_0^\alpha p_{ {\rm R} 0}^\alpha, \quad
\hat{\mathsf C}^{\alpha + } = n_0^\alpha p_{ {\rm R} 0}^\alpha (1+ \gamma_0^\alpha),
\\
\label{eq:idealvars2}
& 
\hat{\mathsf C}'^{\alpha + } = - n_0^\alpha p_{ {\rm R} 0}^\alpha (1+ \gamma_0^\alpha)(2+ \gamma_0^\alpha),
\\
\label{eq:idealvars3} 
& \hat{\mathsf G}^{\alpha + } = - \rho_0^\alpha \theta_0 \gamma^\alpha_0,
\quad
{\mathcal C}^\alpha = [  n_0^\alpha p_{ {\rm R} 0}^\alpha (1+ \gamma_0^\alpha) / \rho^\alpha_0 ]^{1/2},
\\
\label{eq:Gprime}
& (\partial \hat{\mathsf G}^\alpha / \partial J^\alpha)^+ = - \hat{\mathsf G}^{\alpha +}
(1 + \gamma^\alpha_0).
\end{align}

Next, consider a compressible fluid obeying the EOS in \eqref{eq:EOSi1}--\eqref{eq:tEOSi}.
With null cavitation for compression, (${\bm{\mathsf{a}}}^\alpha$, ${\bm{\mathsf{b}}}^\alpha$) are again irrelevant, $\rho_0^\alpha = n_0^\alpha \rho_{ {\rm R} 0}^\alpha$, \eqref{eq:Gprime} holds, and
\begin{align}
\label{eq:EOSvars}
& P^{\alpha+} = n_0^\alpha p_{ {\rm R} 0}^\alpha, \quad
\hat{\mathsf C}^{\alpha + } = n_0^\alpha 
( p_{ {\rm R} 0}^\alpha+ B_\eta^\alpha),
\\
\label{eq:EOSvars2}
& 
\hat{\mathsf C}'^{\alpha + } = -n_0^\alpha \{ 2 p^\alpha_{{\rm R} 0} + B^\alpha_\eta 
(1 + B_{\eta {\rm p}}^\alpha) \},
\\
\label{eq:EOSvars3} 
& \hat{\mathsf G}^{\alpha + } = - \rho_0^\alpha \theta_0 \gamma^\alpha_0,
\quad
{\mathcal C}^\alpha = [  n_0^\alpha 
( p_{ {\rm R} 0}^\alpha+ B_\eta^\alpha) / \rho^\alpha_0 ]^{1/2}.
\end{align}

Last, a solid tissue with EOS \eqref{eq:EOSi1}--\eqref{eq:tEOSi},
viscoelastic matrix, viscoelastic fibers, matrix- and fiber-damage is addressed.
Bulk and shear viscoelasticity and
fiber family $k = 1$ furnish internal variables
${\bm{\mathsf{a}}}^\alpha \rightarrow \{ {\bm \Gamma}^\alpha_{{\rm V} l},
{\bm \Gamma}^\alpha_{{\rm S} m}, {\bm \Gamma}^\alpha_{ {\Phi} k,n} \}$ 
with initial conditions
 $\{ {\bm \Gamma}^{\alpha +}_{{\rm V} l} = 
{\bm \Gamma}^{\alpha +}_{{\rm S} m} =  {\bm \Gamma}^{\alpha +}_{ {\Phi} k,n} = {\bf 0} \}$.
Rate-insensitive damage supplies order parameters 
${\bm{\mathsf{b}}}^\alpha \rightarrow \{ \bar{D}^\alpha, D^\alpha_k \}$
with initial conditions $ \{ \bar{D}^{\alpha +} = D^{\alpha+}_k =0 \}$.
Fibers are either aligned, $\kappa^\alpha_k = 0$ parallel to ($x,X^\alpha$), or isotropic, $\kappa^\alpha_k = \frac{1}{3}$.
Constant active tension is permitted, affecting energy, stress $(\sigma_{\rm A})^1_1$, and stiffness
via \eqref{eq:Psiact} and \eqref{eq:PsiAsig}; since $\Delta^\alpha_k = {\rm constant}$,  
this does not affect dissipation nor need enter ${\bm{\mathsf{a}}}^\alpha$.
As usual, $\rho_0^\alpha = n_0^\alpha \rho_{ {\rm R} 0}^\alpha$
and $({\mathcal C}^\alpha)^2 = { \hat{\mathsf C}^{\alpha +}/ \rho^\alpha_0}$. Lengthy, yet routine, derivations yield 
\begin{align}
\label{eq:sol1}
& P^{\alpha +} = n_0^\alpha \{ p^\alpha_{{\rm R} 0} - (\sigma^\alpha_{\rm A})^{1+}_1 \}, 
\quad {\mathsf G}^{\alpha + } = - \rho_0^\alpha \theta_0 \gamma^\alpha_0; 
\\
& (\partial \hat{\mathsf G}^\alpha / \partial J^\alpha)^+ = - \hat{\mathsf G}^{\alpha +}
(1 + \gamma^\alpha_0), 
 \\
 \label{eq:sol1b}
& \hat{\mathsf C}^{\alpha + } =  {\mathsf C}^{\alpha + }, \quad
\hat{\mathsf C}'^{\alpha + } =  {\mathsf C}^{\prime \alpha + }, \quad
\hat{\mathsf G}^{\alpha + } = {\mathsf G}^{\alpha + },
\\
\label{eq:sol2}
& {\mathsf C}^{\alpha + } = {\mathsf C}^{\alpha + }_{ \rm V} + 
{\mathsf C}^{\alpha + }_{ \rm S} +  {\mathsf C}^{\alpha + }_{ \Gamma} +
   {\mathsf C}^{\alpha + }_{ \rm A} +
  {\mathsf C}^{\alpha + }_{ \rm F} +  {\mathsf C}^{\alpha + }_{ \Phi} ,
   \\
   \label{eq:sol3}
   & {\mathsf C}'^{\alpha + } = {\mathsf C}'^{\alpha + }_{ \rm V} + 
 {\mathsf C}'^{\alpha + }_{ \rm S} +  {\mathsf C}'^{\alpha + }_{ \Gamma} +
   {\mathsf C}'^{\alpha + }_{ \rm A} + 
 {\mathsf C}'^{\alpha + }_{ \rm F} +  {\mathsf C}'^{\alpha + }_{ \Phi} ;  
     \\
   \label{eq:sol4}
   &  {\mathsf C}^{\alpha + }_{ \rm V}  =  n_0^\alpha 
( p_{ {\rm R} 0}^\alpha+ B_\eta^\alpha),
 \quad {\mathsf C}^{\alpha + }_{ \rm S} = {\textstyle{\frac{4}{3}}} n_0^\alpha \mu_{\rm S}^\alpha ,
     \\
     \label{eq:sol5}
     & {\mathsf C}^{\alpha + }_{ \Gamma} = n_0^\alpha \bigr{(} B_\theta^\alpha \sum_l \beta^\alpha_{ {\rm V} l}  
     +  {\textstyle{\frac{4}{3}}}  \mu^\alpha_{\rm S} \sum_m \beta^\alpha_{{\rm S} m} \bigr{)},
      \\ 
      &  {\mathsf C}^{\alpha + }_{ \rm A} = n^\alpha_0 
      (\partial^2 \Psi^\alpha_{{\rm A}k} / \partial (J^\alpha)^2)^+   \qquad (k= 1),
      \\
     \label{eq:sol6}
     & {\mathsf C}^{\alpha + }_{ \rm F} =  \begin {cases}
       \frac{8}{9} n_0^\alpha \mu^\alpha_k & (\kappa^\alpha_k = 0), \\
     0 & (\kappa^\alpha_k = \frac{1}{3}), \\
     \end{cases}
     \\ &  {\mathsf C}^{\alpha + }_{ \Phi} =
     \begin {cases}
   \frac{8}{9} n_0^\alpha  \mu^\alpha_k \sum_{n} \beta^\alpha_{\Phi k,n}   &  (\kappa^\alpha_k = 0), \\
    0 & (\kappa^\alpha_k = \frac{1}{3}); \\
     \end{cases} 
       \\
   \label{eq:sol7}
   &  {\mathsf C}^{ \prime \alpha + }_{ \rm V}  =  -n_0^\alpha \{ 2 p^\alpha_{{\rm R} 0} + B^\alpha_\eta 
(1 + B_{\eta {\rm p}}^\alpha) \}, 
\\
&
 {\mathsf C}^{ \prime \alpha + }_{ \rm S} = -{\textstyle{\frac{28}{9}}} n_0^\alpha \mu_{\rm S}^\alpha,
     \\
     \label{eq:sol8}
     & {\mathsf C}^{ \prime \alpha + }_{ \Gamma} =  - 3 n_0^\alpha \bigr{(} 
     B_\theta^\alpha\sum_l \beta^\alpha_{ {\rm V} l}  
     +  {\textstyle{\frac{28}{27}}}  \mu^\alpha_{\rm S} \sum_m \beta^\alpha_{{\rm S} m} \bigr{)}, 
     \\
     & {\mathsf C}^{ \prime \alpha + }_{ \rm A} = n^\alpha_0  (\partial^3 \Psi^\alpha_{{\rm A}k} / \partial (J^\alpha)^3)^+  \qquad (k= 1),
      \\
     \label{eq:sol9}
     & {\mathsf C}^{ \prime \alpha + }_{ \rm F} =  \begin {cases}
       \frac{32}{27} n_0^\alpha   \mu^\alpha_k   & (\kappa^\alpha_k = 0), \\
     0 & (\kappa^\alpha_k = \frac{1}{3}), \\
     \end{cases}
     \\
     & {\mathsf C}^{ \prime \alpha + }_{ \Phi} =
     \begin {cases}
      \frac{32}{27} n_0^\alpha   \mu^\alpha_k \sum_{n}  \beta^\alpha_{\Phi k,n}    &  (\kappa^\alpha_k = 0), \\
    0  & (\kappa^\alpha_k = \frac{1}{3}); \\
     \end{cases} 
     \\ 
     \label{eq:sol10}
     &       {\bm{\mathsf{B}}}^{\alpha +}  \rightarrow  {\bf 0}, \quad 
     {\bm{\mathsf{b}}}^{\alpha +}_{\mathsf a} \rightarrow  {\bf 0}, \quad
       {\bm{\mathsf{b}}}'^{\alpha +} \rightarrow  {\bf 0}, \quad
        {\bm{\mathsf{b}}}^{\alpha +}_{\eta} \rightarrow  {\bf 0},
         \\ 
     \label{eq:sol11}
     &       {\bm{\mathsf{A}}}^{\alpha +}  \rightarrow
      -n_0^\alpha  \biggr{\{}
      \frac{\partial {\bf Q}^\alpha_{ {\rm V} l}}{ \partial J^\alpha},
       \frac{\partial {\bf Q}^\alpha_{ {\rm S} m}}{ \partial J^\alpha},
        \frac{\partial {\bf Q}^\alpha_{ \Phi k,n}}{ \partial J^\alpha}
      \biggr{\}}^+, 
      \\
      & 
      \biggr{(} \frac{ \partial (D_t^\alpha {\bm{\mathsf{a}}}^\alpha)}{\partial J^\alpha} \biggr{)}^+
       \rightarrow
     \biggr{\{}
     \frac{ \partial {\bf Q}^\alpha_{ {\rm V} l} /  \partial J^\alpha} {\beta^\alpha_{ {\rm V} l} B_\theta^\alpha \tau^\alpha_{ {\rm V} l}},
          \frac{ \partial {\bf Q}^\alpha_{ {\rm S} m} /  \partial J^\alpha}
          {\beta^\alpha_{ {\rm S} m} \mu_{\rm S}^\alpha \tau^\alpha_{ {\rm S} m}}, \cdots
          \nonumber
          \\
          &  
           (\partial {\bf Q}^\alpha_{ \Phi k,n} /  \partial J^\alpha) / ( \beta^\alpha_{ \Phi k,n} \mu_{k}^\alpha \tau^\alpha_{ \Phi k,n}) \bigr{\}}
           , \quad \,
{\bm \pi}^{\alpha +}_{\mathsf a } \rightarrow {\bf 0}, \quad             
                      \\
   &           {\bm{\mathsf{A}}}^{\alpha +} \cdot 
                \bigr{(} { \partial (D_t^\alpha {\bm{\mathsf{a}}}^\alpha)}/{\partial J^\alpha} \bigr{)}^+
                 \rightarrow
               {\mathcal A}^\alpha_{\Gamma {\rm V}} +   
         {\mathcal A}^\alpha_{\Gamma {\rm S}}     +
         {\mathcal A}^\alpha_{\Phi} ,
         \label{eq:sol13}
    \\
     &
     {\mathcal A}^\alpha_{\Gamma {\rm V}} = - 3 n_0^\alpha B_\theta^\alpha \sum_l \frac{\beta^\alpha_{{\rm V}l}}{\tau^\alpha_{{\rm V}l}},
     \quad
      {\mathcal A}^\alpha_{\Gamma {\rm S}}  = -{\textstyle{\frac{8}{3}}} n_0^\alpha \mu^\alpha_{\rm S}
      \sum_m \frac{\beta^\alpha_{{\rm S}m}}{\tau^\alpha_{{\rm S}m}}, \nonumber
      \\
    &   {\mathcal A}^\alpha_{\Phi }  =  
    \begin{cases}  - \frac{32}{27} n_0^\alpha  \mu^\alpha_k \sum_n {\beta^\alpha_{\Phi k,n}}/{\tau^\alpha_{\Phi k,n}} & (\kappa_k^\alpha = 0), \\
    0 & (\kappa_k^\alpha = \frac{1}{3}).
    \end{cases}
     \end{align}

\begin{table*}[]
\caption{\label{table3}%
Shock evolution parameters for rabbit muscle (ECF and solid, $n_0^1 = 0.9$), bovine liver (blood and solid, $n_0^1 = 0.88$) and canine lung (air and solid, $n_0^1 = 0.336$).  ``Mixture'' invokes same shock simultaneously to each phase.
$(\Delta P)^\alpha_0$, $(\Delta \theta)^\alpha_0$, $(\Delta \bar{D})^\alpha_0$,
and $(\Delta D_1)^\alpha_0$
are initial stress, temperature, and matrix/fiber damage jumps for initial strain change
$ \Delta J^\alpha_0 =  \llbracket J^\alpha \rrbracket (t = 0) = -0.1$.}
\begin{ruledtabular}
\begin{tabular}{lccc|ccc|ccc}
Property or &  & \underline{Muscle} & & & \underline{Liver} & & & \underline{Lung}  & \\
model prediction & ECF & Solid & Mixture & Blood & Solid & Mixture & Air & Solid & Mixture \\
\hline
$\rho^\alpha_{ 0}$ [g/cm$^3$]  & 0.103 & 0.990 & 1.093 & 0.127 & 0.933 & 1.060 & 7.56$\times 10^{-4}$  & 0.337 & 0.338 \\
$\hat{\mathsf C}^{\alpha +}$ [GPa]  & 0.220 & 2.954 & 3.174 & 0.317 & 2.350  & 2.667 & 
 9.42$\times 10^{-5}$ & 9.26$\times 10^{-5}$ & 1.87$\times 10^{-4}$ \\
$\hat{\mathsf C}'^{\alpha+}$ [GPa]  & -1.751 & -26.57 & -28.32 & -4.118 & -21.15  & -25.27 & -2.26$\times 10^{-4}$  & -2.42$\times 10^{-4}$  & -4.68$\times 10^{-4}$  \\
${\mathcal C}^{\alpha}$ [km/s]  & 1.462 & 1.727 & 1.704 & 1.578 & 1.587  & 1.586 & 0.353 & 1.66$\times 10^{-2 }$ & 2.35$\times 10^{-2 }$  \\
$\hat{\mathsf G}^{\alpha +}$ [g$\cdot$K/cm$^3$]  & -4.215 & -96.06 & -99.15 & -6.309 & -32.97  & -39.31 & -9.38$\times 10^{-2 }$ & -11.91 & -11.95 \\
${\omega}^{\alpha}$ [1/s]  & 8.69$\times 10^{9}$ & 9.05$\times 10^{8}$ & 7.01$\times 10^{-3}$ & 1.89$\times 10^{7}$ & 2.57$\times 10^{6}$  & 6.62$\times 10^{-2}$ 
& 2.97$\times 10^{4} $ & 1.13$\times 10^{2}$ & 2.95$\times 10^{-2}$ \\
${\Lambda}^{\alpha}$ [1/m]  & 2.99$\times 10^6$ & 2.33$\times 10^5$ & 1.84$\times 10^{-6}$ & 3.68$\times 10^3$ & 7.21$\times 10^2$  & 1.76$\times 10^{-5}$ & 1.40$\times 10^2$ & 1.04$\times 10^1$ & 2.01$\times 10^{-3}$ \\
$(\Delta P)^\alpha_0$ [MPa] & 30.8 &  428 & 459 & 52.3 & 341 & 393 & 1.06$\times 10^{-2}$ & 1.05$\times 10^{-2}$ & 2.10$\times 10^{-2}$ \\
$(\Delta \theta)^\alpha_0$ [K] & 4.32 & 10.3 & 9.66 & 5.25 & 3.73 & 3.92 & 13.3 & 3.73 & 3.74 \\
$(\Delta \bar{D})^\alpha_0$ [-] matrix & $\ldots$ & 0.045 & 0.045 &$\ldots$ & 0.100  & 0.100 & $\ldots$ & 5.03$\times 10^{-3}$ & 5.03$\times 10^{-3}$ \\
$(\Delta D_1)^\alpha_0$ [-] fibers & $\ldots$ & 0.022 & 0.022 &$\ldots$ & 0  & 0 & $\ldots$ & 0 & 0 \\
 \end{tabular}
\end{ruledtabular}
\end{table*}

For rate-insensitive damage, from $\bar{\pi}^\alpha_{{\rm D} } = 0$ and $\pi^\alpha_{{\rm D}k } = 0$ via \eqref{eq:bequil} and using \eqref{eq:taylorb} and \eqref{eq:sol10}
with $\bar{\vartheta}^\alpha = \vartheta^\alpha_k  =2$,
\begin{align}
\label{eq:damjump}
& \bar{D}^{\alpha -} = 
 \frac{1}{n_0^\alpha \bar{E}_{ \rm C}^\alpha} ( {\mathsf C}^{\alpha +}_{\rm S} + {\mathsf C}^{\alpha +}_{\Gamma})
\llbracket J^\alpha \rrbracket^2 + O(\llbracket J^\alpha \rrbracket^3),
\\
\label{eq:damjump2}
& {D}^{\alpha -}_k = 
 \frac{1}{n_0^\alpha {E}_{ {\rm C} k}^\alpha} ( {\mathsf C}^{\alpha +}_{\rm F} + {\mathsf C}^{\alpha +}_{\Phi})
\llbracket J^\alpha \rrbracket^2 + O(\llbracket J^\alpha \rrbracket^3).
\end{align}
Notice ${D}^{\alpha -}_k \rightarrow 0$ in \eqref{eq:damjump2} if $\kappa^\alpha_k = \frac{1}{3}$. 
  To at least $O(\llbracket J^\alpha \rrbracket^2)$, \eqref{eq:damjump} and \eqref{eq:damjump2}
are unaffected by Finsler versus Euclidean metrics
$({\bf g},{\bf G}^\alpha)$
for current prescriptions
$\bar{r}^\alpha \geq 2$ and $\tilde{r}^\alpha_k \geq 2$.
From  ${\bm{\mathsf{b}}}^{\alpha +} \rightarrow  {\bf 0}$ and \eqref{eq:sol10},
rate-insensitive fractures do not affect weak-shock evolution \eqref{eq:weaklam}--\eqref{eq:Jdecaylin}.

If fractures are rate dependent, then  ${\bm{\mathsf{b}}}^\alpha \rightarrow { \bf 0} $ and
 ${\bm{\mathsf{a}}}^\alpha \rightarrow \{ {\bm \Gamma}^\alpha_{{\rm V} l},
{\bm \Gamma}^\alpha_{{\rm S} m}, {\bm \Gamma}^\alpha_{ {\Phi} k,n}, \bar{D}^\alpha, D^\alpha_k  \}$. Then \eqref{eq:jumpevola} yields $\bar{D}^{\alpha -} =0$ and
$D^{\alpha -}_k = 0$ in lieu of \eqref{eq:damjump} and \eqref{eq:damjump2}. From
\eqref{eq:TDGLbar} and \eqref{eq:TDGLk}, damage kinetics do not
contribute to ${\bm{\mathsf{A}}}^{\alpha +}$ or ${\bm \pi}^{\alpha +}_{\mathsf a }$ nor $\Lambda^\alpha$ or $\omega^\alpha$ in \eqref{eq:weaklam}, \eqref{eq:weakomega}; \eqref{eq:sol11}--\eqref{eq:sol13} are unchanged.
Importantly, damage, regardless of rate (in)dependence, can still affect strong-shock evolution
in \eqref{eq:deltaUfin}--\eqref{eq:Lamfin}.

Phase interactions affect $\Lambda^\alpha$ and $\omega^\alpha$, from \eqref{eq:hdef}--\eqref{eq:omegaab}, 
\begin{align}
\label{eq:phsinter1}
\biggr{(} \frac{\partial h^\alpha}{ \partial \upsilon^\alpha} \biggr{)}^+ \! = - \sum_{\beta \neq \alpha} \bar{\lambda}^{\alpha \beta +},  \, 
\biggr{(}  \frac{\partial \epsilon^\alpha}{ \partial \theta^\alpha} \biggr{)}^+ \! = - \sum_{\beta \neq \alpha} \bar{\omega}^{\alpha \beta +} \! .
\end{align}
Consider now a two-phase mixture of solid ($\alpha = 1 \rightarrow {\rm s}$) and fluid ($\alpha = 2 \rightarrow {\rm f}$). Adopting physics in Refs.~\cite{regueiro2014,suh2021},
\begin{align}
\label{eq:lam12reg}
& \bar{\lambda}^{12+} = (n^{\rm f}_0)^2 \hat{\mu}^{\rm f} / \Xi, \qquad
\bar{\omega}^{12+} = \alpha_{\rm v} {\kappa}^{\rm fs},
\end{align}
with fluid viscosity $\hat{\mu}^{\rm f}$, system permeability $\Xi$, 
interfacial area per unit volume $\alpha_{\rm v}$, and heat transfer coefficient
${\kappa}^{\rm fs}$.
Although macroscopic Newtonian viscosity and Fourier conduction are excluded for singular shocks
\cite{morro1980,morro1980b}, microscopic $h^\alpha$ and $\epsilon^\alpha$ include viscosity and heat transfer.

Recall from \eqref{eq:Umult1} that the weak-shock solution \eqref{eq:Jdecaylin} for a multi-phase material corresponds to strain jump $ \Delta J^\alpha_0 $ and resulting discontinuities in $P^\alpha$ and $\theta^\alpha$ applied as a loading condition for one phase $\alpha$, with all other
phase $\beta = 1,2, \ldots \alpha-1,\alpha+1,\ldots,N$ witnessing no discontinuities
in $J^\beta$, $P^\beta$, or $\theta^\beta$.  This shock moves through all phases at
speed ${\mathcal C}^\alpha$; phase interactions $h^\alpha$ and $\epsilon^\alpha$ induce decay in amplitude $\llbracket J^\alpha \rrbracket(t)$ so long as $\bar{\lambda}^{\alpha \beta} > 0$ and $\bar{\omega}^{\alpha \beta} > 0$.
Velocities $\upsilon^\beta$ and temperatures $\theta^\beta$ ($\beta \neq \alpha$) can evolve continuously in
space-time behind the wave front from such interactions, their values indeterminate.

\begin{figure}[]
\begin{subfigure}
 (a){\includegraphics[width = 0.36\textwidth]{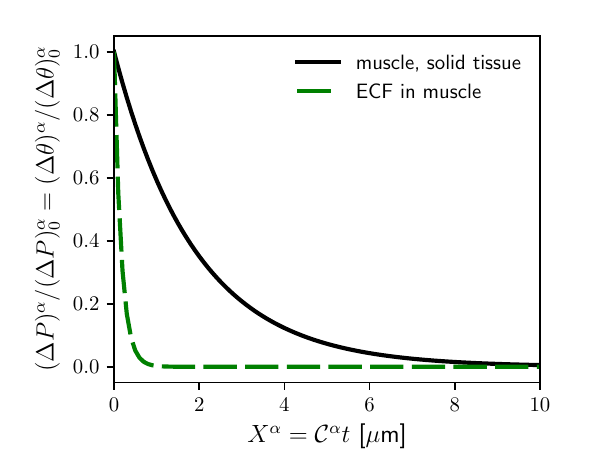} \label{fig6a}}
\end{subfigure} \\
\begin{subfigure}
  (b) {\includegraphics[width = 0.36\textwidth]{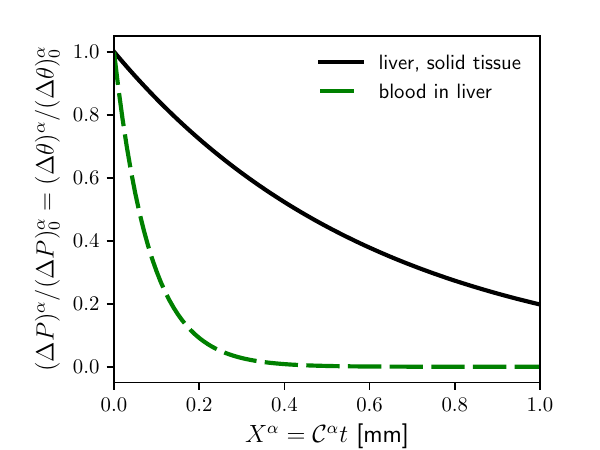} \label{fig6b}} 
\end{subfigure} \\
\begin{subfigure}
\, (c) {\includegraphics[width = 0.36\textwidth]{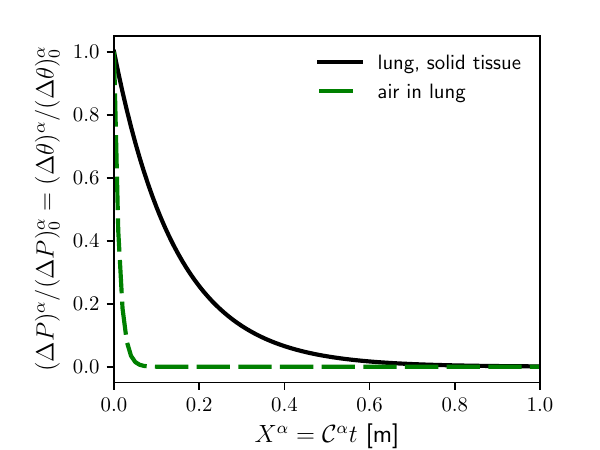} \label{fig6c}} 
\end{subfigure}
\caption{\label{fig6} Predicted ratio of jump in shock stress $(\Delta P)^\alpha$ or temperature $(\Delta \theta)^\alpha$ normalized by initial magnitude $ (\Delta P)^\alpha_0$ or $( \Delta \theta)^\alpha_0$, for shock amplitude $\Delta J^\alpha_0$ applied individually to phase $\alpha$, vs.~propagation distance $X^\alpha$ in the weak shock limit: (a) rabbit muscle, (b) bovine liver, and (c) canine lung. Wave speed is $\mathcal{C}^\alpha$; different scales (i.e., $\mu$m, mm, or m) used for $X^\alpha$.}
\end{figure}

In contrast, if the mixture is idealized as homogeneous with matching $\upsilon^\alpha$ and $\theta^\alpha$, then $h^\alpha$ and $\epsilon^\alpha$ do not explicitly affect shock evolution. In this ``tied'' case, $ \Delta J^\alpha_0$ is applied simultaneously at $t=0$
to all phases $\alpha = 1,\ldots,N$ as a loading condition.
Speed $\mathcal C  = (\hat{\mathsf C} / \rho_0)^{1/2}$ results from stiffness and density of the whole mixture in
 \eqref{eq:mixshock}, and
\begin{align}
\label{eq:solmix}
\hat{\mathsf C}^{\prime +} = \sum_\alpha \hat{\mathsf C}^{\prime \alpha +}, \quad
\hat{\mathsf G}^{+} = -\rho_0 \theta_0 \frac {\sum_\alpha \gamma_0^\alpha \rho_0^\alpha c_\epsilon^\alpha }{
\sum_\alpha \rho_0^\alpha c_\epsilon^\alpha }.
\end{align}
\noindent {\bf Predictions.}
The analytical solution for weak shock evolution, \eqref{eq:weaklam}--\eqref{eq:Jdecaylin}, is
 applied to three biologic systems at $\theta_0 = 310\,$K, each comprised of one solid tissue phase and one fluid: skeletal muscle with interstitial fluid, liver with blood, and lung with air.
 Properties, $\omega^\alpha$, and $\Lambda^\alpha$ are given in Table~\ref{table3} for each component, 
and for the homogeneous idealization of \eqref{eq:mixshock}, \eqref{eq:Lsum}, and
 \eqref{eq:solmix} labeled ``Mixture''. The lower four rows contain initial jumps in
 stress \eqref{eq:taylorP}, temperature \eqref{eq:taylorT}, matrix- \eqref{eq:damjump}, and fiber-damage \eqref{eq:damjump2}, each to $O(\llbracket J^\alpha \rrbracket^2$) for initial strain jump $\Delta J^\alpha_0 = -0.1$.
 If damage is rate-dependent, its jumps are zero instead. Normalized exponential decay of $\llbracket P^\alpha \rrbracket$ and $\llbracket \theta^\alpha \rrbracket$
 arising from \eqref{eq:Jdecaylin} is shown for each component in Fig.~\ref{fig6}.
 When mixtures are shocked uniformly, decay from viscoelastic dissipation alone manifests over much larger distances (not shown).
 
 Constitutive and metric parameters are those for rabbit skeletal muscle and bovine lung
 of Sec.~\ref{sec4b}, Table~\ref{table1} (ECF, blood) and Table~\ref{table2} (solid phases).
 For muscle, active tension from \eqref{eq:SigA1} affects initial stress but does not appreciably change tabulated results.
 Air is modeled as an ideal gas with ${\mathfrak R}^\alpha = 287\,$J/kg$\cdot$K
 and $\gamma_0^\alpha = 0.4$ \cite{claytonIJES2022}. Viscosity for air in \eqref{eq:lam12reg} is $\hat{\mu}^{\rm f} = 18.3 \,\mu{\rm Pa}\cdot{\rm s}$
 \cite{regueiro2014}.
 Solid properties are for canine lung \cite{claytonMOSM2020,claytonAIP2020,claytonIJES2022}
 with bulk and shear viscoelasticity ($l = m = 1$), isotropic fibers ($\kappa^\alpha_k = \frac{1}{3}$),
solid fraction $n_0^1 = 0.336$, $B^\alpha_\eta =164\,$kPa, $\mu_{\rm S}^\alpha = 2.98\,$kPa, $ \beta^\alpha_{ {\rm V} l} = 0.009$,  $ \beta^\alpha_{ {\rm S} m} = 1.5$,
$ \tau^\alpha_{ {\rm V} l} = \tau^\alpha_{ {\rm S} m} =0.5\,$s.
Values $\gamma_0^\alpha =0.114$ and $\bar{E}_{\rm C}^\alpha = 22.7\,$kPa, unmeasured for lung parenchyma, are borrowed from liver parenchyma.  
This value of $\gamma_0^\alpha$ for lung is $\approx 2 \times$ that of Ref.~\cite{claytonIJES2022}, with the latter estimate $10^3 \times$ that of
classical thermodynamics \cite{claytonNMC2011} using the low $B^\alpha_\theta$ of the highly porous structure. 

In \eqref{eq:lam12reg}, $\Xi =6.7 \times 10^{-18}{\rm m}^2$ for muscle \cite{wang2020},
$\Xi = 1.5 \times 10^{-14} {\rm m}^2$ for liver \cite{zheng2021}, and $\Xi = 1.83 \times 10^{-10} {\rm m}^2$ for lung \cite{regueiro2014}. Regarding heat transfer, $\kappa^{\rm fs} = 6 \,$W/m$^2 \cdot$K for muscle and liver \cite{chato1980}, and $\kappa^{\rm fs} = 41.2 \,$W/m$^2 \cdot$K for lung \cite{saha2022}. From idealized microstructure geometries \cite{wang2020,bonfig2010,fung1990}, 
a contact area estimate for muscle is $\alpha_{\rm v} = \pi/R_0 $ with $R_0 = 30 \, \mu$m the fiber radius \cite{wilgeroth2012}, for
 liver $\alpha_{\rm v} = 2 n_0^{\rm f} /R_0$ with $R_0 = 4 \, \mu$m the capillary radius \cite{chato1980}, and for
 lung  $\alpha_{\rm v} = \pi/ (2R_0) $ with $R_0 = 30 \, \mu$m the alveolar radius \cite{claytonMOSM2020}.
 
 From Table~\ref{table3} and Fig.~\ref{fig6}, when $\Delta J^\alpha_0$ is applied to one phase alone, then $|\Lambda^\alpha|$ and $|\omega^\alpha|$ are large, with $\omega^\alpha > 0$ for transient decay.
 For ${\rm d} P^\alpha_{\rm H} / {\rm d} \llbracket J^\alpha \rrbracket \approx -\hat{\mathsf C}^{\alpha -} < 0$ and
 since $\Lambda^\alpha > 0$, a negative stress gradient $\nabla_0 P^{\alpha -}$
 is needed for a steady shock: $P^\alpha$ should decrease steeply as $\Sigma^\alpha$ is approached from the $(\cdot)^-$ side.
 For muscle and liver, $(\partial h^\alpha / \partial \upsilon^\alpha)^+$
 dominates $\omega^\alpha$ due to relatively large $\Xi$ and $\hat{\mu}^{\rm f}$ in
 \eqref{eq:lam12reg}.  For lung, both 
 $(\partial h^\alpha / \partial \upsilon^\alpha)^+$ and $(\partial \epsilon^\alpha / \partial \theta^\alpha)^+$ have significant influence on shock decay, the latter
 due to a comparatively low ${\mathcal C}^\alpha$.
 Viscoelastic dissipation embodied in ${\mathcal A}_{\Gamma {\rm V}}$, ${\mathcal A}_{\Gamma {\rm S}}$, and ${\mathcal A}_{\Phi}$ has relatively small effects,
 negligible compared to interphase drag and heat exchange.
 The near-incompressibility of these tissues, also typical of soft polymers at
 high rates \cite{khand2023}, leads to small glassy shear moduli relative to bulk moduli,
 the latter (bulk) modeled here having little or no viscoelastic relaxation.
 
From Fig.~\ref{fig6}, decay distance is shortest in muscle and
 longest in lung.  In each system, a shock applied to the fluid decays over a much shorter distance than one applied to the solid.
  From Table~\ref{table3}, stress rises $(\Delta P)^\alpha_0$ are highest in solid phases of muscle and liver having largest tangent moduli.  Temperature $(\Delta \theta)^\alpha_0$ is highest in air in the lung.
 Damage is greatest in liver and smallest in lung; fiber damage is absent in both organs in the weak-shock limit as fibers are isotropic.
 In anisotropic muscle, fiber damage $(\Delta {D}_k)^\alpha_0$ is $\approx \frac{1}{2} \times$ matrix damage $(\Delta \bar{D})^\alpha_0$,
 in contrast to (strong-shock) Hugoniot solutions of Fig.~\ref{fig3}(b,c) whereby these damage mechanisms evolve more similarly.
  
 When $\Delta J^\alpha_0$ is applied to both phases at once and the homogeneous  idealization is invoked, 
 $|\Lambda^\alpha|$ and $|\omega^\alpha|$ are deemed small, with 
 contributions only from viscoelasticity. For these conditions, weak shocks should be
 steady with near-uniform stress-strain states trailing their wave fronts.
 Weak-shock experiments are needed to confirm this prediction; known data \cite{wilgeroth2012} are for strong shocks.
 

%% file: s5.tex
\section{\label{sec5} Concluding remarks}

A theoretical framework is posited  
for modeling multi-phase soft biologic materials over wide
ranges of loading rate and pressure. Results for
uniaxial-stress and shock loading
agree with experimental data on biologic fluids, skeletal muscle,
and liver.  Damage, represented by order parameters
for matrix and fiber fractures, is found to be rate insensitive in muscle but rate dependent
in liver. An analytical solution has been derived for shock evolution including
phase interactions, viscoelasticity, and tissue damage.
Predictions for weak shock decay reveal dominance of interphase momentum and energy exchange over viscoelastic dissipation and damage/fracture kinetics.